\documentclass[journal,10pt,a4paper,twocolumn,twoside]{IEEEtran}

\IEEEoverridecommandlockouts

\usepackage[active]{srcltx} 
\usepackage{graphicx}
\usepackage{balance}
\usepackage{cite}
\usepackage{amssymb}
\usepackage[cmex10]{amsmath}
\usepackage{amsfonts}
\usepackage{amsthm}

\usepackage{eurosym}

\usepackage{booktabs}
\usepackage{hyperref} 

\usepackage{multirow}

\usepackage{algorithm}
\usepackage{algorithmicx}
\usepackage{algpseudocode}

\usepackage{subcaption}

\usepackage{braket}
\usepackage{tikz}
\usetikzlibrary{quantikz}
\usepackage{adjustbox}

\usepackage{rotating}

\usepackage{placeins}

\begin{document}

\newcommand{\meterCom}{\scalebox{.5}{\begin{tikzcd} \meter{} \end{tikzcd}}}
\newcommand{\cwCom}{\scalebox{.5}{\begin{tikzcd} \pgfmatrixnextcell \cw \end{tikzcd}}}

\newcommand{\outputgrouph}[6]{\POS"#1,#3"."#2,#3"."#1,#3"."#2,#3", \POS"#1,#3"."#2,#3"."#1,#3"."#2,#3"*!C!<#6,#4>=<0em>{#5}}
\newcommand{\eqdef}{\stackrel{\triangle}{=}}


\newtheorem{theo}{Theorem}
\newtheorem{theor}{Theorem}
\newtheorem{cor}{Corollary}
\newtheorem{lem}{Lemma}
\newtheorem{prop}{Proposition}
\newtheorem{ins}{Insight}

\theoremstyle{remark}

\theoremstyle{definition}
\newtheorem{defin}{Definition}
\newtheorem{ass}{Assumption}
\newtheorem{rem}{Remark}

\renewcommand{\qed}{$\blacksquare$}

\renewcommand{\algorithmiccomment}[1]{// #1}

\title{When Entanglement meets Classical Communications: Quantum Teleportation for the Quantum Internet\\
\vspace{5mm}
\large{(Invited Paper)}}
\author{Angela~Sara~Cacciapuoti,~\IEEEmembership{Senior Member,~IEEE}, Marcello~Caleffi,~\IEEEmembership{Senior Member,~IEEE}, Rodney Van Meter, Lajos Hanzo,~\IEEEmembership{Fellow,~IEEE}
	\thanks{A.S. Cacciapuoti and M. Caleffi are with DIETI, University of Naples Federico II, Naples, 80125 Italy (e-mail: angelasara.cacciapuoti@unina.it; marcello.caleffi@unina.it). They are also with the Laboratorio Nazionale di Comunicazioni Multimediali (CNIT), Naples, 80126 Italy.}
	\thanks{Rodney Van Meter is with the Faculty of Environment and Information Studies, Keio University, Fujisawa, Kanagawa 252-0882, Japan (e-mail: rdv@sfc.wide.ad.jp).}
	\thanks{Lajos Hanzo is with with the School of Electronics and Computer Science, University of Southampton, Southampton SO17 1BJ, UK (e-mail: lh@ecs.soton.ac.uk).}
}


\maketitle

\begin{abstract} 
\textit{Quantum Teleportation} is the key communication functionality of the Quantum Internet, allowing the ``transmission'' of qubits without either the physical transfer of the particle storing the qubit or the violation of the quantum mechanical principles. Quantum teleportation is facilitated by the action of \textit{quantum entanglement}, a somewhat counter-intuitive physical phenomenon with no direct counterpart in the classical word. As a consequence, the very concept of the classical communication system model has to be redesigned to account for the peculiarities of quantum teleportation. This re-design is a crucial prerequisite for constructing any effective quantum communication protocol. The aim of this manuscript is to shed light on this key concept, with the objective of allowing the reader: i) to appreciate the fundamental differences between the transmission of classical information versus the teleportation of quantum information; ii) to understand the communications functionalities underlying quantum teleportation, and to grasp the challenges in the design and practical employment of these functionalities; iii) to acknowledge that quantum information is subject to the deleterious effects of a noise process termed as quantum decoherence. This impairment has no direct counterpart in the classical world; iv) to recognize how to contribute to the design and employment of the Quantum Internet.
\end{abstract}


\section{Introduction}
\label{sec:1}

The interconnection of quantum devices via the \textit{Quantum Internet} -- i.e. through a network enabling quantum communications among remote quantum nodes -- represents a disruptive technology \cite{CacCalTaf-18,Han-12,CalCacBia-18,NguBabAla-18,VanMet-14}. Indeed, the Quantum Internet is capable of supporting functionalities with no direct counterpart in the classical world \cite{PirBra-16,Gib-16,Sim-17,DurLamHeu-17,Cas-18}, such as secure communications \cite{WehElkHan-18}, blind computing \cite{Bro-09}, exponential increase of the quantum computing power \cite{VanMet-14,CalCacBia-18} and advanced quantum sensing techniques \cite{Deg-17}. These functionalities have the potential of fundamentally changing markets and industries -- such as commerce, intelligence and military affairs.

At a first sight, the design of the Quantum Internet might sound like a trivial task. After all, the number of devices interconnected by the conventional Internet exceeds 17 billion \cite{IOT-19}, hence connecting few extra quantum devices might not seem like a `big deal'. However, the laws of quantum mechanics impose unusual constraints on the design of the Quantum Internet. 

Specifically, the Quantum Internet facilitates quantum communications among remote nodes by transmitting quantum bits (qubits) -- which differ fundamentally from classical bits -- or by creating distributed, entangled quantum states with no classical equivalent.
A classical bit encodes one of two mutually exclusive states, being in only one state at any time. In contrast, a qubit can be in a \textit{superposition} of the two basis states (see Sec.~\ref{sec:2}). Hence, while $n$ classical bits are only ever in one of the $2^n$ possible states at any given moment, an $n$-qubit register can be in a \textit{superposition} of all of the possible states \cite{VanMet-14,CacCalTaf-18}.

Unfortunately, quantum mechanics does not allow an unknown qubit to be copied or observed/measured \cite{NieChu-11,RiePol-11}. Hence, although we can map a qubit to the spin of an electron or to a photon and it can also be directly transmitted to a remote node via a fiber link, if the traveling photon is corrupted, the original quantum information stored within the qubit is definitely destroyed. As a consequence, the direct transmission of qubits via photons is not readily feasible -- unless the network applications can tolerate the loss of information and/or having a low transmission success rate, as in Quantum Key Distribution (QKD) networks \cite{WehElkHan-18}.

Thankfully, \textit{quantum teleportation}, originally proposed in \cite{BenWie-92,BenBraCre-93}, constitutes an astonishing strategy for ``\textit{transmitting}'' qubits within a quantum network, without either the physical transfer of the particle storing the qubit or the violation of the quantum mechanical principles \cite{CacCalTaf-18,CalCacBia-18,GyoImrNgu-18,VanMet-14}. Quantum teleportation has been experimentally verified over substantial distances, such as 1200 kilometers \cite{RenXuYon-17}, which exploits the weird quantum phenomenon represented by the quantum entanglement \cite{NieChu-11}. Specifically, to realize quantum teleportation a pair of parallel resources are needed. One of these resources is classical: two bits must be transmitted from the source to the destination. The other resource is quantum: an entangled pair of qubits must be generated and shared between the source and the destination. As a consequence, quantum teleportation requires two parallel communication links, a classical link for transmitting the pair of classical bits and a quantum link for entanglement generation and distribution \cite{NieChu-11}.

With this in mind, it appears plausible that the very concept of the classical communication system model, as originally proposed by Shannon in his pioneering contribution \cite{Sha-48}, has to be redesigned to account for these peculiarities of quantum mechanics. This re-design is a crucial prerequisite for conceiving quantum communication protocols.

The aim of this treatise is to shed light on this key concept, with the objective of allowing the reader:
\begin{itemize}
	\item[i)] to appreciate the fundamental differences between the transmission of classical information versus the teleportation of quantum information;
	\item[ii)] to understand the communications functionalities underlying quantum teleportation and to highlight the challenges of turning the vision of the Quantum Internet into reality.
\end{itemize}

In Sec.~\ref{sec:2} we commence by introducing the preliminaries of quantum mechanics, required for appreciating the fundamental difference between classical and quantum networks.

Then, in Sec.~\ref{sec:3}, we review the quantum teleportation process, by providing the rudimentary mathematical preliminaries, followed by describing some representative schemes conceived for practical entanglement generation and distribution. These basics are crucial for understanding the communication system model proposed in Sec.~\ref{sec:4} to account for the peculiarities of noiseless quantum teleportation. Then, in Sec.~\ref{sec:5}, we introduce realistic imperfections into the quantum teleportation process.

Specifically, similarly to classical communications, quantum communications are subject to the impairments imposed by the deleterious effects of the environment. These imperfections are termed as \textit{decoherence} -- a type of quantum noise with no direct counterpart in the classical world. Understanding decoherence is pivotal for the design of efficient quantum communication techniques and protocols. Hence, in Sec.~\ref{sec:5} we highlight the theoretical framework of quantum noise modeling from a communications engineering perspective to allow the reader:
\begin{itemize}
	\item[iii)] to recognize that the quantum-domain noise is \textit{multiplicative} rather than being additive and it exhibits an asymmetric behavior\footnote{As detailed in Sec.~\ref{sec:5}, decoherence may impose different types of errors on a qubit, such as bit-flip errors, phase-flip errors, as well as simultaneous bit- and phase-flip errors, while, in the classical domain, only bit-flips may occur. Given the nature of the quantum-domain impairments, the probability of bit-flips and phase-flips tends to be different, regardless of the specific material representing the qubits, as seen in Table~1 of \cite{Han-16} and Fig.~6 of \cite{BotAlaBab-18}, which the authors succinctly refer to as an `asymmetric' property. In this paper we adopt this terminology but in a broader sense, as it will be clarified in Sec.~\ref{sec:5}.} with respect to the three Cartesian coordinates\footnote{As introduced in Sec.~\ref{sec:2.6} and further detailed in Sec.~\ref{sec:5}, there exists a one-to-one mapping between a qubit and a Cartesian vector $\mathbf{r}=[r_x,r_y,r_z] \in \mathcal{R}^3$, known as \textit{Bloch vector}.} representing a qubit.
\end{itemize}

Indeed, as it will be detailed in Sec.~\ref{sec:5}, decoherence is not the only source of impairments in the quantum teleportation process. In fact, quantum teleportation relies on a sequence of operations applied to the quantum states, as it will be detailed in the following sections. The imperfections of these operations aggravate the impairments affecting the quantum teleportation. However, the imperfections accumulated throughout the quantum operations strongly depend on the particular technology adopted for representing a qubit. Hence, in Sec.~\ref{sec:6} we will report on the results of an extensive campaign of teleportation experiments carried out by using the IBM Q quantum processor \cite{Cas-17}, with the aim of gaining experimental insights into the cumulative impairments affecting the teleported qubit at the destination in order to confirm the modeling of the quantum decoherence detailed in Sec.~\ref{sec:5}.

Finally, Sec.~\ref{sec:7} concludes the paper by summarizing the results obtained and by providing a long-term perspective on the design of the Quantum Internet.

\section{Background}
\label{sec:2}
In this section, we introduce some preliminaries on quantum mechanics.

\subsection{The Hilbert Space}
\label{sec:2.1}
According to one of the quantum mechanics postulates, any isolated or closed quantum physical system is associated with a complex Hilbert space\footnote{In the finite-dimensional complex vector spaces encountered in quantum computation and information processing, a Hilbert space is equivalent to a vector space with inner product.}. This complex vector space is known as the state space of the system. The system is completely described by its state vector, which is a unit-vector in the system's state space \cite{NieChu-11,RiePol-11}.

The simplest quantum mechanical system is the quantum bit (qubit), whose state space is two-dimensional. To characterize a quantum state in the state space, a basis that is orthonormal to this state space has to be chosen. In the following, we adopt the conventional \textit{bra-ket} notation\footnote{The \textit{bra-ket} notation (also known as Dirac's notation) is a standard notation describing quantum states. In a nutshell, a ket $\ket{\cdot}$ represents a column vector, whereas a bra $\bra{\cdot} = \ket{\cdot}^{\dag}$ represents the conjugate transpose of the corresponding ket.} for denoting a qubit \cite{NieChu-11,VanMet-14,BotAlaBab-18,Han-QERC-18}.
The most commonly used basis is the standard (or computational) basis, which corresponds to the convention:
\begin{equation}
	\label{eq:2.1}
	\ket{0} \equiv \begin{bmatrix} 1 \\ 0 \end{bmatrix}, \quad 
		\ket{1} \equiv \begin{bmatrix} 0 \\ 1 \end{bmatrix}.
\end{equation}

Given the vector space framework postulated by quantum mechanics, the state $\ket{\psi}$ of a qubit can be expressed as a linear combination of the basis states chosen:
\begin{equation}
	\label{eq:2.2}
	\ket{\psi} = \alpha \ket{0} + \beta \ket{1},
\end{equation}
where $\alpha$ and $\beta \in \mathbb{C}$ are complex numbers, known as the amplitudes of the state $\ket{\psi}$. Equation \eqref{eq:2.2} portrays the state $\ket{\psi}$ in a \textit{superposition} of the two basis states. The condition of $\ket{\psi}$ being a unit-vector, which can be formulated as $\bra{\psi}\ket{\psi}=1$, is therefore equivalent to $|\alpha|^2+|\beta|^2=1$. This condition is also known as the normalization condition of the state vectors \cite{NieChu-11,RiePol-11}. 

Intuitively, the states $\ket{0}$ and $\ket{1}$ are analogous to the values 0 and 1 that a bit may assume. However, a qubit differs from a classical bit, since the superpositions of the two basis states in \eqref{eq:2.2} can also exist \cite{NieChu-11,RiePol-11}. Consequently, a classical bit encodes one of two mutually exclusive states, being in only one state at any time. Conversely, a qubit can be in a superposition of the two basis states.

Indeed, the meaning of the amplitudes of the state $\ket{\psi}$ in \eqref{eq:2.2} is much deeper. In fact, according to the quantum measurement postulate, although a qubit may be in a superposition of two orthogonal states, when we want to observe or measure its value, it collapses into one of the two orthogonal states \cite{NieChu-11,VanMet-14,BotAlaBab-18}. More explicitly, $|\alpha|^2$ and $|\beta|^2$ uniquely determine the probabilities of obtaining $\ket{0}$ or $\ket{1}$, respectively, by measuring the qubit state on the basis $\{\ket{0}, \ket{1}\}$\footnote{The measurement of a qubit state may also be carried out in a basis different from that in which the qubit was prepared in \cite{NieChu-11,RiePol-11}. In the above description, for the sake of clarity, we assumed the standard basis also for the measurement.}. Hence the normalization condition may also be further interpreted in the light of $|\alpha|^2$ and $|\beta|^2$ being probabilities.

After its measurement/observation, the original quantum state collapses to the measured state. Hence, the measurement irreversibly alters the original qubit state. For instance, if the outcome of measuring a superposed qubit is the state zero, the qubit collapses into this specific state and any further measurement will always give the state zero outcome, regardless of the original contribution of the state one to the superposed state \cite{CacCalTaf-18}. 

\begin{figure}[t!]
	\centering
	\includegraphics[width=.9\columnwidth]{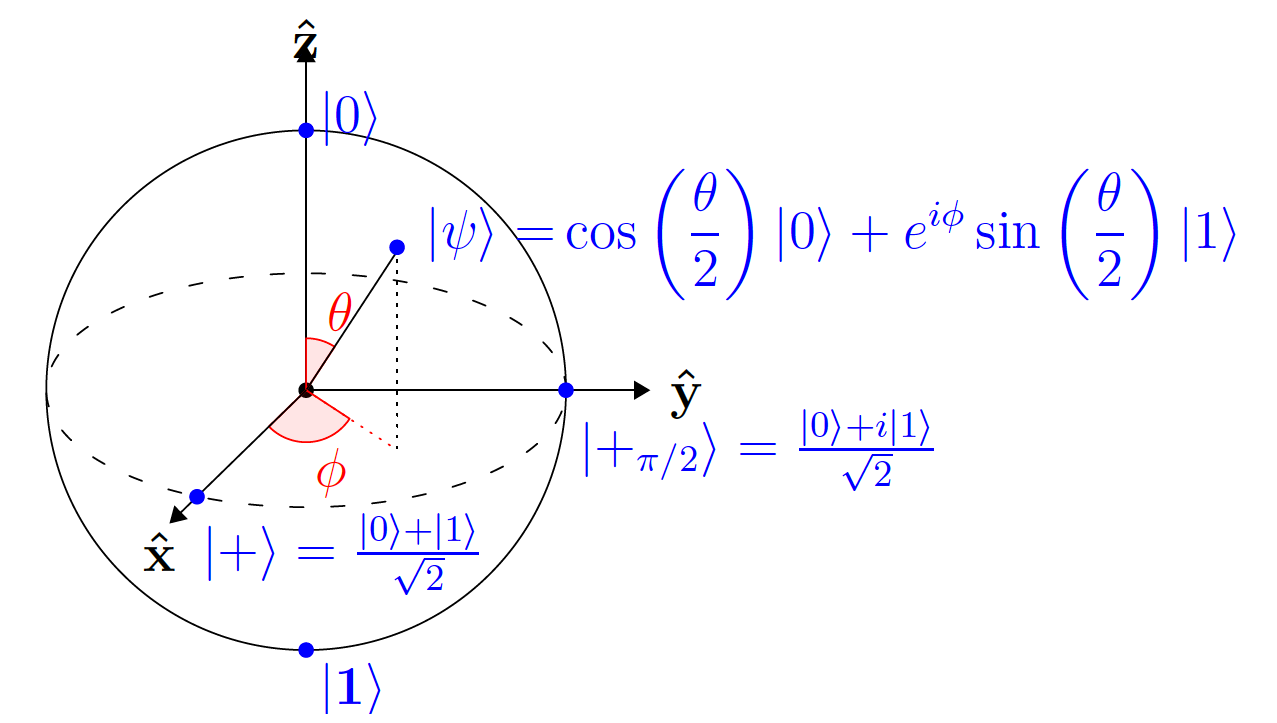}
	\caption{Bloch sphere: geometrical representation of a qubit in spherical coordinates. A pure state $\ket{\psi} = \alpha \ket{0} + \beta \ket{1}$ is represented by a point on the sphere surface, with $\alpha = \cos{\frac{\theta}{2}}$ and $\beta = e^{i \phi} \sin{\frac{\theta}{2}}$.}
	\label{Fig:01}
\end{figure}

To elaborate a little further, the state of a qubit is often represented geometrically by the Bloch sphere, which is depicted in Fig.~\ref{Fig:01} and surveyed in more depth in \cite{VanMet-14,NieChu-11,RiePol-11,BotAlaBab-18}. Specifically, any pure quantum state is represented by a point on the sphere's surface, with $\theta$ and $\phi$ denoting the spherical coordinates. Unfortunately, visualizing the state of more than one qubit is more complicated, since the state space grows exponentially with the number of qubits, as described in the next subsection. 
%
%

\subsection{Phase}
\label{sec:2.1.bis}
The angle $\phi$ mentioned above is referred to as the \textit{phase} of the quantum state \cite{VanMet-14}. To be more rigorous, the difference between the \textit{global phase} and \textit{relative phase} has to be clarified.

Specifically, we say that the state $e^{i\gamma} \ket{\psi}$ is equal to $\ket{\psi}$ up to the \textit{global phase factor} $e^{i\gamma}$, with $\gamma$ being the global phase. In fact, the statistics of measurement predicted for these two states are the same, being $|e^{i\gamma}|=1$ \cite{NieChu-11,VanMet-14,RiePol-11}. Therefore, from an observational point of view, these two states are identical. Hence, the global phase factors are neglected, since they are irrelevant to the observed properties of the physical system.

However, the \textit{relative phase} $\phi$ cannot be neglected, and indeed it is critical to quantum computation. Specifically, the relative phase (in the standard basis) of a superposition $\ket{\psi} = \alpha \ket{0} + \beta \ket{1}$ is a measure -- in the complex plane -- of the angle between $\alpha$ and $\beta$, i.e, $\phi : \beta/\alpha=e^{i\phi} |\beta|/|\alpha|$ \cite{RiePol-11}. A pair of superpositions
$\ket{\psi} = \alpha \ket{0} + \beta \ket{1}$ and $\ket{\psi'} = \alpha' \ket{0} + \beta' \ket{1}$, whose amplitudes have the same magnitudes but differ in the relative phase\footnote{For instance, $\ket{+} \equiv \frac{\ket{0} + \ket{1}}{\sqrt{2}}$ and $\ket{+_{\pi/2}} \equiv \frac{\ket{0} + i \ket{1}}{\sqrt{2}}$ have the same magnitudes, i.e., $|\alpha|=|\alpha'|$ and $|\beta| = |\beta'|$, but they differ by the \textit{relative phase} of $\pi/2$.}, represent different states. Furthermore, the role of the relative phase is pivotal in creating the interference patterns exploited for instance in the construction of quantum algorithms. In fact, the state of a quantum system is a wave function that matches Schr$\ddot{o}$dinger's equation. Similar to classical wave mechanics, a pair of waves can interfere, either constructively or destructively, depending on the relative phases of the waves. When the resultant interference is constructive, it enhances the amplitude (hence, probability) of a particular state. By contrast, when it is destructive, it reduces the probability. Since the phase of a state is actually complex-valued, the sum of phases is also complex \cite{VanMet-14}.

\subsection{Composite Quantum System}
\label{sec:2.2}
In classical physics, the legitimate states of a system of $n$ objects, whose individual states can be described by a vector in a two-dimensional vector space, can be described by vectors in a vector space of $2n$ dimensions, i.e. the classical state spaces combine through the direct sum. By contrast, in quantum mechanics, the state space of a composite quantum system made up of $n$ quantum systems, each having states modeled by two-dimensional vectors, is much larger. Indeed, the vector spaces associated with the constituent quantum systems can be combined using their tensor product, which is denoted by $\otimes$, resulting in a vector space of $2^n$ dimensions. Hence, if the systems are numbered $0$ through $n-1$, and the system having the index $i$ is prepared in the state $\ket{\psi_i}$, then the joint state $\ket{\psi}$ of the resultant composite system is \cite{RiePol-11,NieChu-11}:
\begin{equation}
	\label{eq:2.3}
	\ket{\psi}= \ket{\psi_0} \otimes \ket{\psi_1} \otimes \ldots \ket{\psi_{n-1}}.
\end{equation} 
A more compact and readable notation uses $\ket{\psi_0 \psi_1 \ldots \psi_{n-1}}$ to represent $\ket{\psi_0} \otimes \ket{\psi_1} \otimes \ldots \ket{\psi_{n-1}}$.

By exploiting this notation, if $V$ and $W$ are vector spaces corresponding to a qubit, each having the standard basis of $\{\ket{0}, \ket{1}\}$, then the composite two-qubit system $V \otimes W$ has the basis:
\begin{equation}
	\label{eq:2.4}
	\{\ket{00}, \ket{01}, \ket{10}, \ket{11}\}.
\end{equation}
Just like a single qubit system, a possible state of a two-qubit system can be in a superposition of the basis states:
\begin{equation}
	\label{eq:2.5}
	\ket{\psi} = \alpha_0 \ket{00} + \alpha_1 \ket{01} + \alpha_2 \ket{10} + \alpha_3 \ket{11} =
		\begin{bmatrix} \alpha_0 \\ \alpha_1 \\ \alpha_2 \\ \alpha_3 \end{bmatrix},
\end{equation}
with $\alpha_0, \alpha_1, \alpha_2, \alpha_3 \in \mathbb{C}: |\alpha_0|^2+|\alpha_1|^2+|\alpha_2|^2+|\alpha_3|^2=1$. 

Upon further generalizing this procedure, a state of an n-qubit system can be in a superposition of all the $2^n$ basis states, which is formulated as:
\begin{equation}
	\label{eq:2.6}
	\ket{\psi} = \sum_{i=0}^{2^n-1} \alpha_i \ket{i},
\end{equation}
with $\alpha_i \in \mathbb{C}: \sum_{i=0}^{2^n-1} |\alpha_i|^2=1$. 

Hence, while $n$ classical bits are only ever in one of the $2^n$ possible states at any given moment, an $n$-qubit register can be in a \textit{superposition} of all of the possible states. Quantum algorithms manipulate the amount of the total system's quantum wave function and the phase to build \textit{interference patterns}, affecting the probability of measuring particular values to execute the algorithms mentioned above.

\subsection{Entanglement}
\label{sec:2.3}
The vast majority of $n$-qubit states cannot be written as the tensor product of $n$ single-qubit states, even though they are all linear combinations of the basis states of the composite n-qubit system. 
The particular states that cannot be written as the tensor product of $n$ single-qubit states are termed as \textit{entangled states} \cite{RiePol-11,NieChu-11}.

In a nutshell, the entanglement may be interpreted as a special case of the superposition of multiple qubits, where the combined state cannot be decomposed into the tensor product of individual states. 
More formally, given a state $\ket{\psi}$ of a composite quantum system associated with the vector space $V$ and a tensor decomposition of $V$, namely $V=V_0\otimes V_1\otimes \ldots \otimes V_{n-1}$, the state $\ket{\psi}$ is \textit{separable}, or unentangled -- with respect to that decomposition -- if it can be written as
$\ket{\psi}=\ket{v_0} \otimes \ket{v_1}\otimes \ldots \otimes \ket{v_{n-1}}$, where $\ket{v_i}$ belongs to $V_i$. Otherwise, $\ket{\psi}$ is entangled with respect to this particular decomposition, but may be unentangled with other decompositions into subsystems\footnote{The entanglement is not an absolute property of a quantum state, but depends on the particular decomposition of the composite system into subsystems under consideration; states entangled with respect to the single-qubit decomposition may be unentangled with respect to other decompositions into subsystems. Hence it must be specified or
clear from context which of the many legitimate tensor decompositions of $V$ is under consideration \cite{RiePol-11}.}.

Let us consider for example the \textit{Bell states}, called also \textit{EPR pairs}, in honor of an article written by Einstein, Podolsky, and Rosen in 1935 \cite{EinPodRos-35}:
\begin{align}
	\label{eq:2.7}
	\ket{\Phi^+} &= \frac{1}{\sqrt{2}} \big( \ket{00} + \ket{11} \big)\\
	\label{eq:2.8}
	\ket{\Phi^-} &= \frac{1}{\sqrt{2}} \big( \ket{00} - \ket{11} \big)\\
	\label{eq:2.9}
	\ket{\Psi^+} &= \frac{1}{\sqrt{2}} \big( \ket{01} + \ket{10} \big)\\
	\label{eq:2.10}
	\ket{\Psi^-} &= \frac{1}{\sqrt{2}} \big( \ket{01} - \ket{10} \big),
\end{align}
which represents four maximally entangled 2-qubit states. The Bell states cannot be decomposed, since it is impossible to find $a_0, a_1, b_0, b_1$ for assuring that:
\begin{align}
	\label{eq:2.11}
	\left(a_0\ket{0}+ b_0\ket{1}\right) \otimes \left(a_1\ket{0}+ b_1\ket{1}\right) = \ket{\Phi^{\pm}}
\end{align}
or equivalently, that:
\begin{align}
	\label{eq:2.12}
	\left(a_0\ket{0}+ b_0\ket{1}\right) \otimes \left(a_1\ket{0}+ b_1\ket{1}\right) = \ket{\Psi^{\pm}}.
\end{align}
To better understand the entanglement concept, let us consider for example $\ket{\Phi^+}$. By measuring each of the two qubits forming the EPR pair independently, one obtains a random distribution of zero and one outcomes with equal probability. However, if the results of the two independent measurements are compared, we find that every time the measurement of a qubit yielded zero so did the measurement of the other qubit, and the same happened with the outcome one. Indeed, according to quantum mechanics, as soon as one of the two qubits is measured, the state of the other also becomes instantaneously determined \cite{NieChu-11,RiePol-11,VanMet-14,CacCalTaf-18}. This quantum entanglement behavior led Einstein and his colleagues to the so-called EPR paradox: the measurement of a qubit instantaneously changes the state of the second qubit, regardless of the distance dividing the two qubits. This seems to involve information being transmitted faster than light, violating the Relativity Theory. But the paradox is illusory, since entanglement does not allow the transmission of information faster than light, as will be clarified in Sec.~\ref{sec:3}.

\subsection{Quantum State Transformations}
\label{sec:2.4}

Closed quantum systems evolve in time according to deterministic, reversible unitary operations \cite{NieChu-11,RiePol-11,VanMet-14,Sch-07}. Hence Nature does not allow arbitrary transformations of a quantum system.
That is, the state $\ket{\psi(t)}$ of the system at time $t$ is related to the state $\ket{\psi(0)}$ of the system at an initial time instant $0$ through a unitary operator $U(\cdot)$, which depends only on the time instants $t$ and $0$:
\begin{equation}
	\label{eq:2.13}
	\ket{\psi(t)}=U(\ket{\psi(0)})
\end{equation}

\begin{rem}
Any linear operator $A: V \rightarrow W$ between vector spaces $V$ and $W$ admits a matrix representation that is completely equivalent to the operator $A(\cdot)$. Hence, the matrix representation and the operator are interchangeable. Consequently, in the following, we will use the same symbol to denote both of them without any loss of generality.
\end{rem}
\begin{rem}
Unitary operators are special linear operators having unitary matrix representations, i.e. we have $U^\dag U=I$. A unitary operator is an invertible operator satisfying $U^{-1}=U^\dag$ \cite{HorJoh-12}.
\end{rem}

The unitary operators (or equivalently the quantum transformations) can be regarded as gates in the circuit model usually adopted both in quantum computation and in quantum information processing. Hence, although experimentalists usually describe the behavior of a quantum system by emphasizing the temporal nature of the evolution, in the circuit model perspective the temporal dependence is hidden within input-output relationships (e.g., \eqref{eq:2.13} and it can be re-written as $\ket{\psi}_{\text{out}}=U(\ket{\psi}_{\text{in}})$) \cite{NieChu-11,RiePol-11,VanMet-14}. As a consequence, unless explicitly stated, in the following we will not emphasize the temporal dependence of a quantum gate.

Please note that the expressions ``quantum transformation'' or ``quantum operator'' refer to unitary operators applied to the state space, not to measurement operators. Geometrically speaking, all quantum state transformations may also be interpreted as rotations of the complex vector space associated with the quantum state space.

In Table~\ref{Tab:01}, we summarize the most popular quantum gates. Naturally, the Identity operation $I$ leaves the quantum state unchanged. The Pauli-X operation imposes a bit-flip, the Pauli-Z a phase-flip, while Pauli-Y represents a joint bit- and phase-flip. The Hadamard operation maps the basis states into superpositions through a combination of Pauli-X and Pauli-Z operations. Finally, the Controlled-Not (CNOT) operation performs the Pauli-X operation on the second qubit whenever the first qubit is $\ket{1}$, and otherwise leaves it unchanged. It is important to further highlight that the Pauli-Z gate reported in Table~\ref{Tab:01} changes the relative phase of a superposition in the standard basis, and hence it is critical to quantum computation, as mentioned in Sec.~\ref{sec:2.1.bis}.

An important consequence of the unitary nature of the quantum transformations is the \textit{no-cloning theorem} \cite{Par-70,WotZur-82,Die-82}: unknown quantum states cannot be copied or cloned \cite{NieChu-11,RiePol-11}. Indeed, the no-cloning theorem has a deep and complex impact on the design of quantum communications, as it will be discussed in detail in the following sections. The corruption of the transmitted classical information by the noise does not imply the total loss of the information, since a copy of the original information can be stored at the source. By contrast, the corruption of the transmitted quantum information by decoherence implies the irreversible loss of information.

\begin{table*}
	\caption{Popular Quantum Gates}
	\label{Tab:01}
	\centering
	\begin{tabular}{|l|c|c|c|c|c|c|c|}
		\toprule
		\textbf{Gate} & Identity & 
			 \begin{tabular}{@{}c@{}}
				Pauli-X \\ (NOT)
			\end{tabular}
			& Pauli-Y & Pauli-Z & Hadamard &
			 \begin{tabular}{@{}c@{}}
				Controlled-NOT \\ (CNOT)
			\end{tabular}
			\\
		\midrule
		\textbf{Symbol} &
			\begin{tikzcd} & \gate{I} & \qw \end{tikzcd}	 &
			 \begin{tabular}{@{}c@{}}
				\begin{tikzcd} & \gate{X} & \qw \end{tikzcd} \\
				or equivalently\\
				\begin{tikzcd} & \targ{} & \qw \end{tikzcd} 
			\end{tabular} &
			\begin{tikzcd} & \gate{Y} & \qw \end{tikzcd} &
			\begin{tikzcd} & \gate{Z} & \qw \end{tikzcd} &
			\begin{tikzcd} & \gate{H} & \qw \end{tikzcd} &
			\begin{tikzcd} & \ctrl{1} & \qw \\& \targ{} & \qw \end{tikzcd}	\\
		\midrule
		\textbf{Matrix} &
			$I \equiv \begin{bmatrix} 1 & 0 \\ 0 & 1 \end{bmatrix}$ &
			$\sigma_x \equiv X \equiv \begin{bmatrix} 0 & 1 \\ 1 & 0 \end{bmatrix}$ &
			$\sigma_y \equiv Y \equiv \begin{bmatrix} 0 & -i \\ i & 0 \end{bmatrix}$ &
			$\sigma_z \equiv Z \equiv \begin{bmatrix} 1 & 0 \\ 0 & -1 \end{bmatrix}$ &
			$H \equiv \displaystyle \frac{1}{\sqrt{2}} \begin{bmatrix} 1 & 1 \\ 1 & -1 \end{bmatrix}$ &
			$\begin{bmatrix} 1 & 0 & 0 & 0 \\ 0 & 1 & 0 & 0 \\ 0 & 0 & 0 & 1 \\ 0 & 0 & 1 & 0 \end{bmatrix}$ \\
		\midrule
		\textbf{Input} & \multicolumn{5}{c|}{1 qubit} & 2 qubits\\
		\midrule
		\textbf{Operation} & Null & \multicolumn{3}{c|}{Axis Rotation} & Superposition & Entanglement\\
		\bottomrule
	\end{tabular}
\end{table*}
\subsection{Pure and Mixed States: The Density Matrix and Quantum Fidelity}
\label{sec:2.5}
Quantum states can be either \textit{pure} or \textit{mixed}. So far, we have discussed only pure states. Briefly, a pure state is a quantum state that can be described by a ket vector, i.e. it can be written in the state-vector form. This does not mean that the state-vector form has only one term: both $\ket{0}$ and $\alpha \ket{0} + \beta \ket{1}$ are pure states. 

By contrast, mixed states\footnote{For a comprehensive overview we refer the reader to \cite{VanMet-14,NieChu-11,RiePol-11}.} are not viewed as true quantum states, but rather as a way of describing a system whose state is not well defined -- it is a probabilistic mixture of well-defined pure states. In other words a mixed state is a statistical ensemble of pure states \cite{VanMet-14,RiePol-11}.
In particular, pure states give deterministic results when measured in appropriate bases, whereas mixed states give probabilistic results in all bases \cite{RiePol-11}.

The individual qubits of an EPR pair constitute examples of mixed states, since they cannot be described individually by a well-defined ket vector. However, not all the mixed states are entangled.

Mixed states inevitably arise as a consequence of quantum decoherence, as it will be discussed in Sec.~\ref{sec:5}. Mixed states are usually modeled in mathematical terms by density operators (or density matrices). Their mean can be used for the rudimentary characterization of the statistical properties of an ensemble of quantum states. More precisely, let us assume that a quantum system is in one of a number of legitimate states $\ket{\psi_i}$, where $i$ is the state-index, and the legitimate states have the respective probabilities $\{p_i\}$. In this context, $\{p_i,\ket{\psi_i}\}$ is an ensemble of pure states. The density operator (or density matrix) $\rho$ of the system is defined as \cite{NieChu-11,RiePol-11}:
\begin{equation}
	\label{eq:2.14}
	\rho= \sum_i p_i \ket{\psi_i} \bra{\psi_i}.
\end{equation}
The density operator $\rho$ is a positive operator (and hence Hermitian) with trace one, $\text{Tr}(\rho)=1$. For a pure state $\ket{\psi}$, the density matrix is equal to $ \rho= \ket{\psi} \bra{\psi}$\footnote{For a pure state it results in $\text{Tr}(\rho^2)=1$, whereas for mixed states in $\text{Tr}(\rho^2)<1$. }. 

The imperfection of mixed states can be quantified by a fundamental figure of merit known as \textit{quantum fidelity}. The fidelity $F$ of a mixed state associated with the density matrix $\rho$, with respect to a certain desired pure state $\ket{\psi}$, is a metric -- taking values between 0 and 1 -- of the distinguishability of the two quantum states, defined as \cite{Joz-94}:
\begin{equation}
\label{eq:2.16}
	F= \bra{\psi} \rho \ket{\psi}.
\end{equation}
Based on this definition, the fidelity can be conceptually described as the ``overlap'' of the mixed state with the desired state $\ket{\psi}$. The fidelity is $1$ for a pure state and it decreases as the decoherence degrades the ``quality'' of the state \cite{VanMet-14}.

Finally, we note that the postulates of quantum mechanics can be reformulated in terms of
the density operator \cite{NieChu-11}. This reformulation\footnote{As an example, the postulate related to the evolution of a closed quantum system and reported in \eqref{eq:2.13}, can be reformulated by stating that the state $\rho$ of the system at time instant $t_1$ is related to the state $\rho'$ of the system at time instant $t_0$ through a unitary operator $U$, which depends only on the time instants $t_1$ and $t_0$: $\rho=U \rho' U^\dag$.} is mathematically equivalent to the description in terms of the state vector. Nevertheless, as mentioned before, the density operator approach is extensively utilized for characterizing the quantum impairments, as discussed in Sec.~\ref{sec:5}.

\subsection{The Bloch Vector}
\label{sec:2.6}
\begin{figure}[t!]
	\centering
	\includegraphics[width=.66\columnwidth]{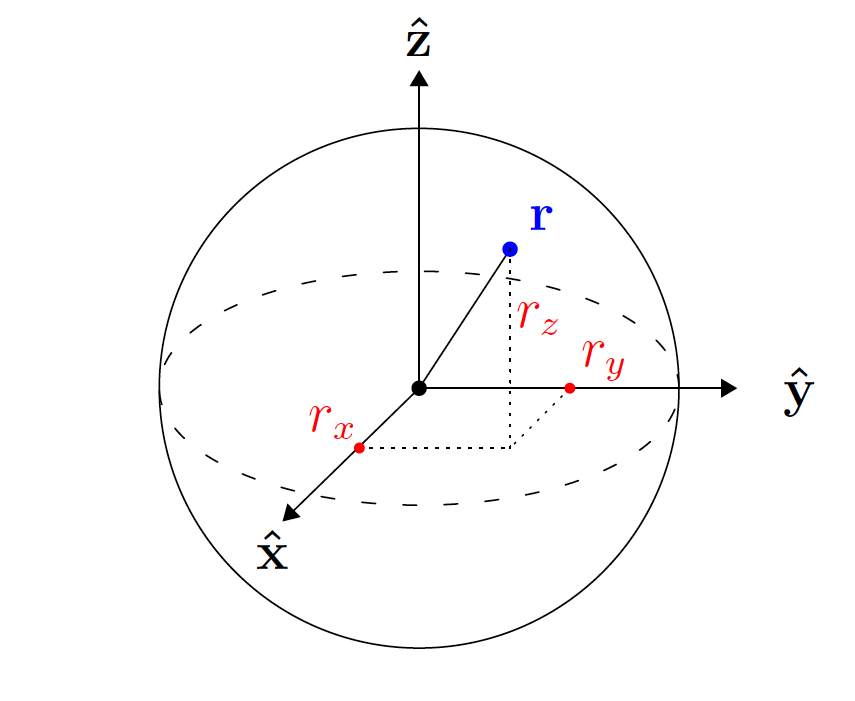}
	\caption{Bloch Vector: geometrical representation of any (pure or mixed) quantum state in Cartesian coordinates $[r_x,r_y,r_z]$, which may be contrasted to its counterpart relaying on the spherical coordinates in Fig.~\ref{Fig:01}.}
	\label{Fig:02}
\end{figure}

Geometrically, the Bloch sphere of Fig.~\ref{Fig:01} includes single-qubit mixed states. In fact, mixed states are constituted by linear combinations of pure states having non-negative weighting coefficients that sum to 1. Hence it is not surprising that single-qubit mixed states can be viewed as laying within the interior of the Bloch sphere \cite{RiePol-11}. The precise connection with the geometry relies on the fact that any density matrix of a single-qubit system, which is a $2 \times 2$ matrix, can be written as \cite{RiePol-11,NieChu-11}:
\begin{equation}
\label{eq:2.17}
	\rho = \begin{bmatrix} \rho^{00} & \rho^{01} \\ \rho^{10} & \rho^{11} \end{bmatrix} = \frac{1}{2} \left(I+r_x\sigma_x+r_y\sigma_y+r_z \sigma_z\right),
\end{equation}
where $\sigma_x, \sigma_y, \sigma_z$ represent the Pauli matrices defined in Tab.~\ref{Tab:01} and $r_x, r_y, r_z$ are the Cartesian coordinates of the quantum state considered \cite{RiePol-11}, as shown in Fig.~\ref{Fig:02}. Recall that its spherical coordinate based counterpart was shown in Fig.~\ref{Fig:01}.

Hence, there exists a one-to-one mapping between any (pure or mixed) quantum state associated with the density matrix $\rho$ and the real three-dimensional vector $\mathbf{r}=[r_x,r_y,r_z] \in \mathcal{R}^3$, known as \textit{Bloch vector}, where we have:
\begin{equation}
	\label{eq:2.18}
	\begin{aligned}
		r_x & = \rho^{01} + \rho^{10} = 2 Re(\rho^{01})\\
		r_y & = i (\rho^{01} - \rho^{10}) = 2 Im(\rho^{10})\\
		r_z & = \rho^{00} - \rho^{11},
	\end{aligned}
\end{equation}
with the norm of $\mathbf{r}$ being strictly smaller than one for mixed states, i.e. $||\mathbf{r}||<1$, while the norm being equal to one for pure states, i.e. $||\mathbf{r}||=1$. A similar relationship exists between a density matrix and the Cartesian coordinates, when multiple-qubit systems are considered.

Indeed, the one-to-one mapping of \eqref{eq:2.18} between the density matrix $\rho$ and the Bloch vector $\mathbf{r}$ insightfully visualizes the effects of the 1-qubit quantum gates of Table~\ref{Tab:01}. Specifically, commencing from \eqref{eq:2.18} and then accounting for the evolution of quantum systems in terms of their density matrices formulated as $\rho_{\text{out}} = U \rho_{\text{in}} U^\dag$, we arrive at:
\begin{equation}
	\label{eq:2.19}
	\mathbf{r}_{\text{in}} = [r_x,r_y,r_z] \xrightarrow{U} \begin{cases}
		\mathbf{r}_{\text{out}} = [r_x,r_y,r_z] & \text{if } U = I \\
		\mathbf{r}_{\text{out}} = [r_x,-r_y,-r_z] & \text{if } U = X \\
		\mathbf{r}_{\text{out}} = [-r_x,r_y,-r_z] & \text{if } U = Y \\
		\mathbf{r}_{\text{out}} = [-r_x,-r_y,r_z] & \text{if } U = Z \\
		\mathbf{r}_{\text{out}} = [r_z,-r_y,r_x] & \text{if } U = H \\
	\end{cases}.
\end{equation}

\begin{rem}
It is important to underline that -- despite the luring illusion that the Pauli-X gate affects in some way the x-coordinate of the Bloch vector -- the Pauli-X gate leaves the x-coordinate unchanged but it affects both the y- and the z-coordinate. Similar considerations hold for the Pauli-Y and Pauli-Z gates, since they leave the y- and the z-coordinate unaltered, respectively, while affecting the remaining two coordinates.
\end{rem}

\section{From Transmission to Teleportation}
\label{sec:3}

\subsection{Quantum Teleportation Overview}
\label{sec:3.1}

Let us assume that a quantum state $\ket{\psi}$ must be transmitted from a sender, say Alice, to a remote receiver, say Bob.

We assume without any loss of generality that $\ket{\psi}$ is a pure qubit, yielding:
\begin{equation}
	\label{eq:3.1}
	\ket{\psi} = \alpha \ket{0} + \beta \ket{1}.
\end{equation}

If the transmitter \textit{knows} the quantum state $\ket{\psi}$, i.e., if Alice knows $\alpha$ and $\beta$, the task can be accomplished by transmitting the values of $\alpha$ and $\beta$ to Bob and by letting Bob \textit{prepare} $\ket{\psi}$, i.e., to ``transform'' a default state into $\ket{\psi}$ with the aid of Bob's local operations. Indeed, several open questions arise in conjunction with practical quantum state preparation \cite{BenDiVSho-01,BenHayLeu-05,Gir-18}, such as the specific construction of a set of universal gates, the required depth of the quantum circuit, or the minimum required fidelity of the reconstructed state. Nevertheless, from a communications engineering perspective, the communication task can be accomplished -- at least in principle\footnote{Practical issues arise with $\alpha,\beta$ being continuous values in $\mathbb{C}$.} -- with 
the aid of classical communication resources. 

However, in the most general case\footnote{This case arises, for instance, in distributed quantum computing. Indeed, the quantum state obtained at a certain computing step of a distributed quantum algorithm is unknown by definition. Furthermore, any observation of the state before the conclusion of the quantum algorithm would imply an irreversible loss of information due to the quantum measurement postulate.}, the transmitter \textit{does not know} the quantum state $\ket{\psi}$, and the task cannot be accomplished with the aid of pure classical communication resources. In fact, the quantum measurement postulate prevents Alice from assessing $\alpha$ and $\beta$ with the aid of a quantum measurement, which would irreversibly alter the original quantum state. Furthermore, the no-cloning theorem prevents Alice from preparing multiple copies of $\ket{\psi}$ and estimating $\alpha,\beta$ by simply measuring the copies.

In other words, quantum mechanics does not allow a qubit to be copied or measured. Hence, although a photon is capable of conveying a qubit and it can be directly transmitted to a remote node -- e.g., via a fiber link -- if the traveling photon is lost due to attenuation or corrupted by decoherence, the original quantum information is definitely destroyed. As a consequence, the direct transmission of qubits via photons is not practically feasible.

\begin{figure}
	\centering
	\includegraphics[width=.9\columnwidth]{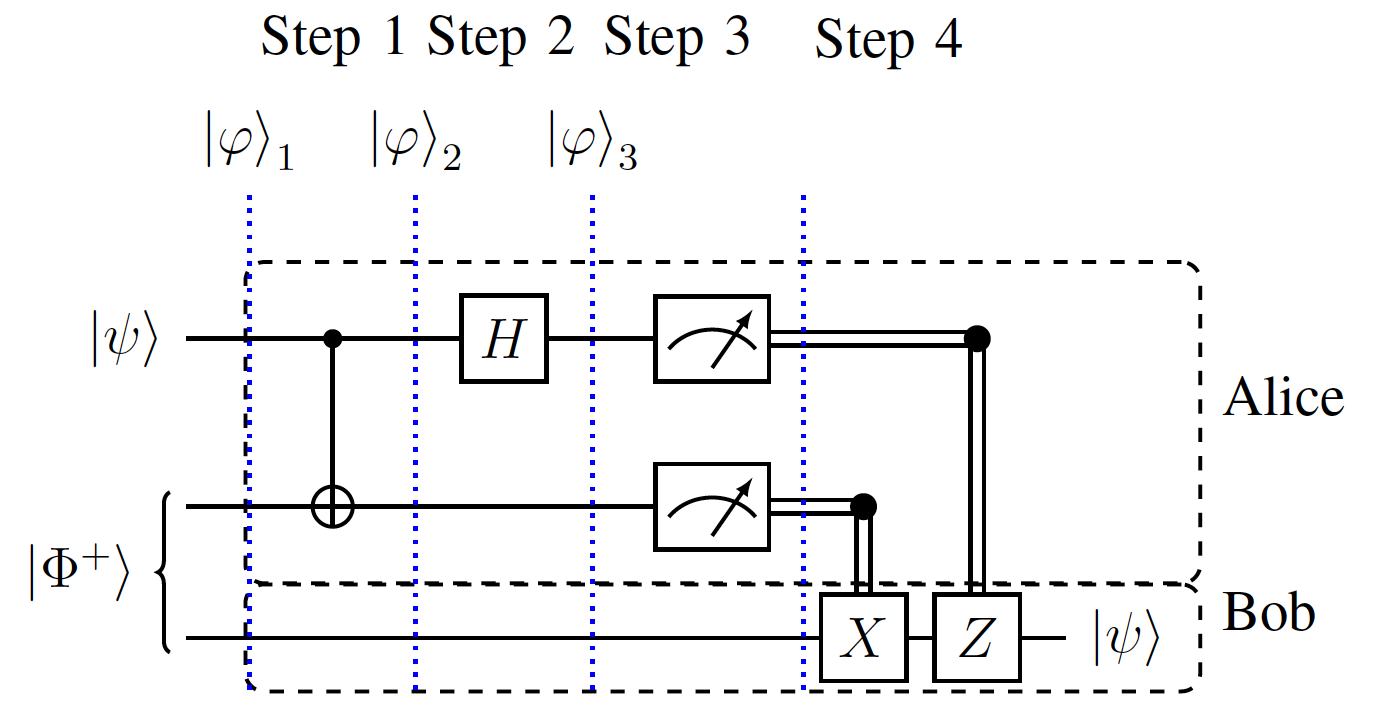}
	\caption{Quantum Teleportation Circuit, where $\ket{\psi}$ denotes the unknown state to be transmitted from Alice to Bob, while $\ket{\Phi^+}$ denotes the EPR pair generated and distributed so that one qubit is stored at Alice and another qubit is stored at Bob. Furthermore, $\ket{\varphi}_i$ denotes the global quantum state at the $i$-th step. The symbol \raisebox{.5 ex}{\protect\meterCom} denotes the measurement operation and the double-line \raisebox{.5 ex}{\protect\cwCom} represents the transmission of a classical bit from Alice to Bob.}
	\label{Fig:03}
\end{figure}

Thankfully, \textit{Quantum Teleportation} constitutes a priceless technique of transmitting qubits without either the physical transfer of the particle storing the qubit, or without the violation of the quantum mechanical principles. As shown in Figure~\ref{Fig:03}, with the aid of local operations and an EPR pair\footnote{Although multipartite-entangled states can be used for quantum teleportation, here we restrict our attention to EPR pairs for the sake of simplicity.} shared between the source and destination, quantum teleportation allows us to ``\textit{transmit}'' an unknown quantum state \cite{VanMet-14,CacCalTaf-18,CalCacBia-18,NieChu-11}, as it will be detailed in the next section.

Quantum teleportation implies the destruction of both the original qubit (encoding the quantum information to be transmitted) and the EPR member at the source, as a consequence of a measurement. Indeed, as it will be shown in Sec.~\ref{sec:3.2}, the original qubit is reconstructed at the destination, once the output of the measurement at the source -- 2 classical bits -- is received through a finite-delay classical link, obeying the speed of light in an optical link for example.

\subsection{Quantum Teleportation: Mathematical Details}
\label{sec:3.2}
In a nutshell, the teleportation process of Fig.~\ref{Fig:03} takes as its input the state $\ket{\psi}$ to be teleported and an EPR pair shared between Alice and Bob. Each of the four states $\ket{\Phi^{\pm}},\ket{\Psi^{\pm}}$ can be used for quantum teleportation, given that the state is fixed in advance by mutual agreement between Alice as well as Bob, and this mutual agreement can be achieved with the aid of a finite-delay classical link.

In the following, we assume without loss of generality that Alice and Bob share the state $\ket{\Phi^+} = \left( \ket{00} + \ket{11} \right) / \sqrt{2}$, as shown in Fig.~\ref{Fig:03}. Hence, the initial global state $\ket{\varphi_1} = \ket{\psi} \otimes \ket{\Phi^+}$ depicted in Fig.~\ref{Fig:03} is:
\begin{equation}
	\label{eq:3.2}
	\ket{\varphi_1} = \big( \alpha \ket{0} \otimes \left( \ket{00} + \ket{11} \right) + \beta \ket{1} \otimes \left( \ket{00} + \ket{11} \right) \big) / \sqrt{2}.
\end{equation}
By following the convention that the pair of leftmost qubits belongs to Alice and the rightmost qubit belongs to Bob, \eqref{eq:3.2} becomes equivalent to:
\begin{equation}
	\label{eq:3.3}
	\ket{\varphi_1} = \big( \alpha \ket{000} + \alpha \ket{011} + \beta \ket{100} + \beta \ket{111} \big) / \sqrt{2}.
\end{equation}

\textit{Step 1}. The teleportation process of Fig.~\ref{Fig:03} starts with Alice applying the CNOT gate of Table~\ref{Tab:01} to the pair of qubits at her side. By recalling that the CNOT gate maps state $\ket{10}$ into $\ket{11}$ and vice versa, the global state $\ket{\varphi_2}$ after the CNOT gate (Step 1 in Fig.~\ref{Fig:03}) becomes:
\begin{equation}
	\label{eq:3.4}
	\ket{\varphi_2} = \big( \alpha \ket{000} + \alpha \ket{011} + \beta \ket{110} + \beta \ket{101} \big) / \sqrt{2}.
\end{equation}

\textit{Step 2}. Then, as seen in Fig.~\ref{Fig:03}, Alice applies the H gate of Table~\ref{Tab:01} to the qubit to be teleported, i.e. to the leftmost qubit in \eqref{eq:3.4}. By recalling that the H gate maps $\ket{0}$ into $\frac{\ket{0}+\ket{1}}{\sqrt{2}}$ and $\ket{1}$ into $\frac{\ket{0}-\ket{1}}{\sqrt{2}}$, the global state $\ket{\varphi_3}$ after the H gate (Step 2 in Fig.~\ref{Fig:03}) is obtained from \eqref{eq:3.4} in the following form:
\begin{align}
	\label{eq:3.5}
	\ket{\varphi_3} = \big( &\alpha \ket{000} + \alpha \ket{100} + \alpha \ket{011} +\alpha \ket{111} + \nonumber \\
	& \beta \ket{010} - \beta \ket{110} + \beta \ket{001} - \beta \ket{101} \big) / 2.
\end{align}
By gathering the two leftmost qubits belonging to Alice, \eqref{eq:3.5} becomes equivalent to:
\begin{align}
	\label{eq:3.6}
	\ket{\varphi_3} = \Big( &\ket{00} \otimes \big(\alpha \ket{0} + \beta \ket{1}\big) + \ket{01} \otimes \big(\alpha \ket{1} + \beta \ket{0}\big) + \nonumber \\
	& \ket{10} \otimes \big(\alpha \ket{0} - \beta \ket{1}\big) + \ket{11} \otimes \big(\alpha \ket{1} - \beta \ket{0}\big) \Big) / 2.
\end{align}

\textit{Step 3}. Then, as indicated in Fig.~\ref{Fig:03}, Alice jointly measures the pair of qubits at her side. Remarkably, and regardless of the values of $\alpha$ and $\beta$, Alice has a $25\%$ chance of finding each of the four combinations $\ket{00}$, $\ket{01}$, $\ket{10}$ and $\ket{11}$. Alice's measurement operation instantaneously fixes Bob's qubit, regardless of the distance between Alice and Bob as a consequence of the entanglement described in Sec.~\ref{sec:2.3}\footnote{Einstein referred to this phenomenon by the light-hearted parlance of a ``\textit{spooky action at a distance}''.}. However, Bob can only recover the original qubit $\ket{\psi}$ after he correctly receives the pair of classical bits conveying the specific results of Alice's measurement. Naturally, this classical transmission has to obey the speed of light.

\textit{Step 4}. Specifically, four terms can be identified in \eqref{eq:3.6}, depending on the particular state of the two qubits at Alice. When Alice's qubits are in state $\ket{00}$, then Bob's qubit is in the state $\alpha \ket{0} + \beta \ket{1}$, i.e. Bob's qubit is identical to the original quantum state $\ket{\psi}$. Hence, if Alice obtains $00$ by jointly measuring her pair of qubits and she communicates the outcome to Bob, then Bob can directly recover the original quantum state $\ket{\psi}$ from the qubit at his side, without any quantum-domain operation. At this step, we say that the original quantum state $\ket{\psi}$ has been teleported to Bob's side.

Alternatively, if Alice's qubits are in the state $\ket{10}$, then Bob's qubit is in the state $\alpha \ket{0} - \beta \ket{1}$. Hence, if Alice obtains $10$ by jointly measuring her pair of qubits, once Bob receives the measurement outcome via a classical link, he can recover the original quantum state $\ket{\psi}$ by applying the Z gate of Table~\ref{Tab:01} (that maps $\ket{0}$ in $\ket{0}$ and $\ket{1}$ in $-\ket{1}$) to the qubit at his side, as seen in Fig.~\ref{Fig:03}. Again, at this step, we say that the original quantum state $\ket{\psi}$ has been teleported to Bob's side. Similarly \cite{NieChu-11}, if the measurement at Alice's side is $01$ or $11$, Bob recovers $\ket{\psi}$ by simply applying either the X gate of Table~\ref{Tab:01} or the X gate followed by the Z gate to the qubit at his side.

\begin{rem}
Based on the above discussions, it becomes clear that the teleportation process of a single qubit requires: i) the generation and the distribution of an EPR pair between the source and destination, ii) a finite-delay classical communication channel for conveying the pair of classical bits resulting from the measurement. Hence, it is worthwhile noting that having a tight integration between the pair of classical and quantum resources is necessary in a quantum network \cite{CacCalTaf-18}. Regarding the EPR pair, the measurement at the source destroys the entanglement. Hence, if another qubit has to be teleported, a new EPR pair must be created as well as distributed between the source and destination.
\end{rem}

\begin{figure}
	\centering
	\begin{minipage}[c]{1\linewidth}
		\centering
		\includegraphics[width=.8\columnwidth]{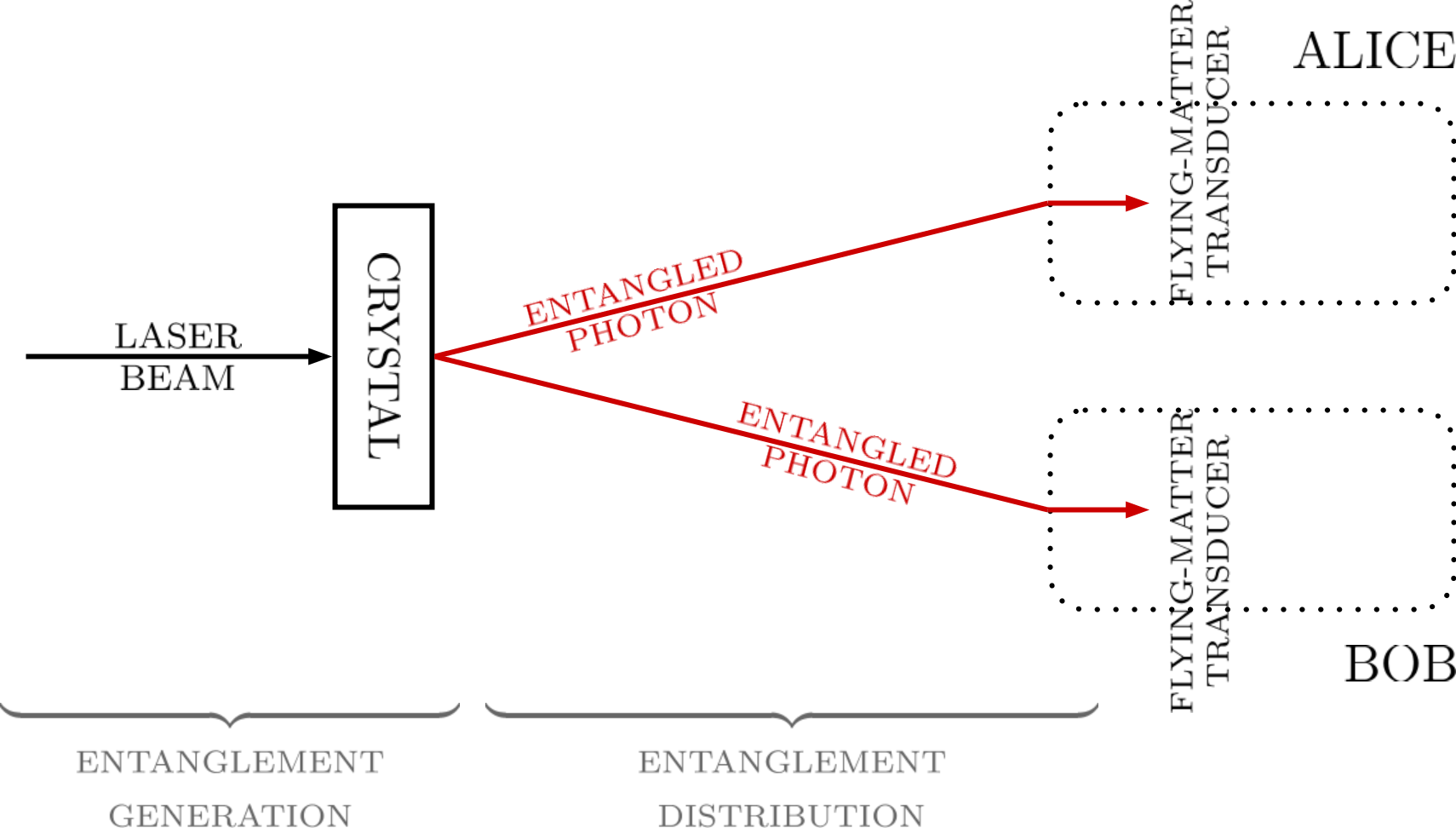}
			\subcaption{Spontaneous Parametric Down-Conversion: a laser beam is directed toward a non-linear crystal, which occasionally splits photon beams into pairs of polarization-entangled photons.}
		\label{Fig:3A}
	\end{minipage}
	\begin{minipage}[c]{1\linewidth}
		\centering
		\includegraphics[width=.8\columnwidth]{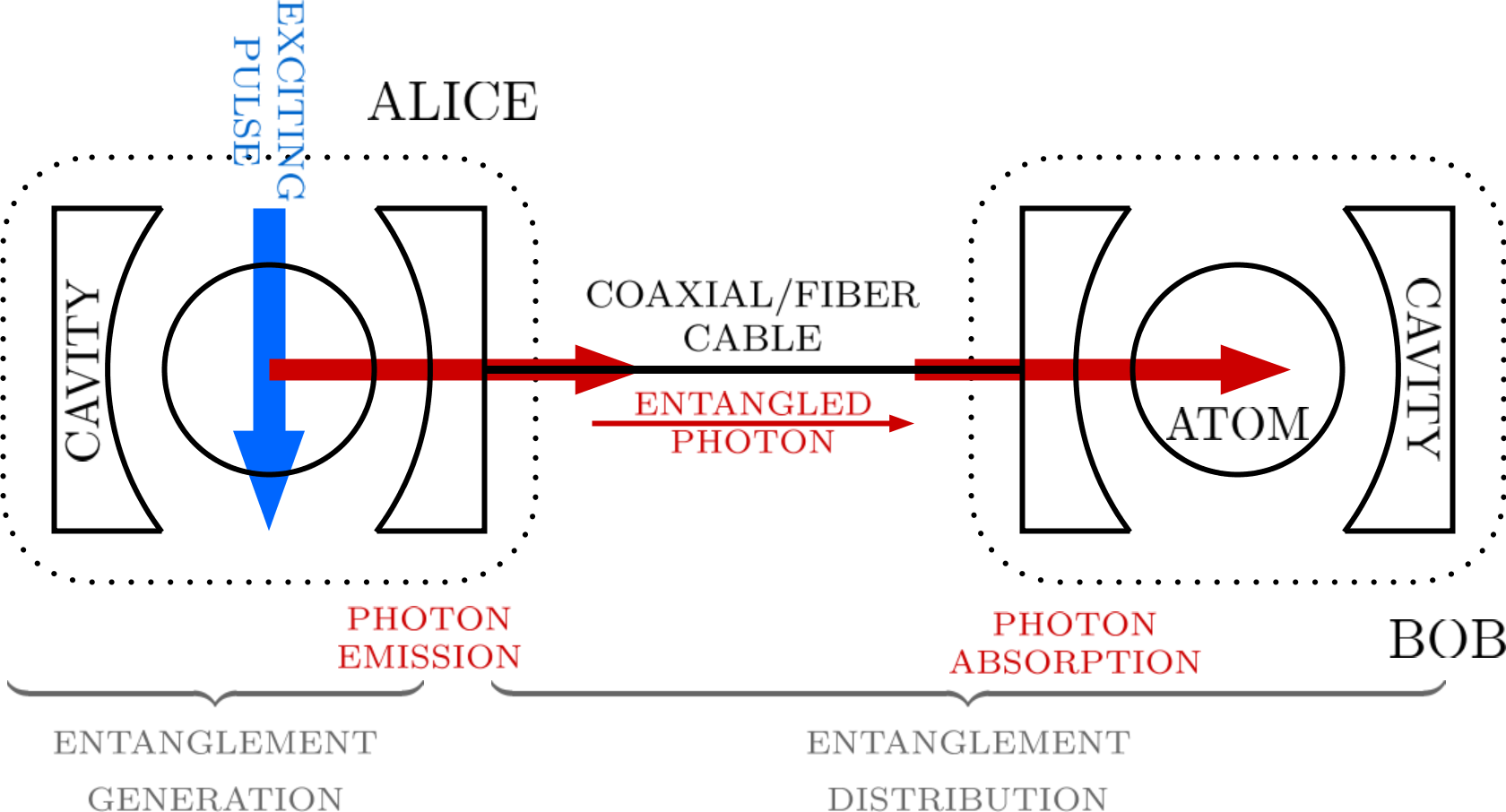}
		\subcaption{An atom strongly coupled with an optical cavity is excited by a laser beam. The resultant photon leaks out of the cavity, propagates as a wave packet through the cable, and enters an optical cavity at the second node, entangling the two remote atoms.}
		\label{Fig:3B}
	\end{minipage}
	\begin{minipage}[c]{1\linewidth}
		\centering
		\includegraphics[width=1\columnwidth]{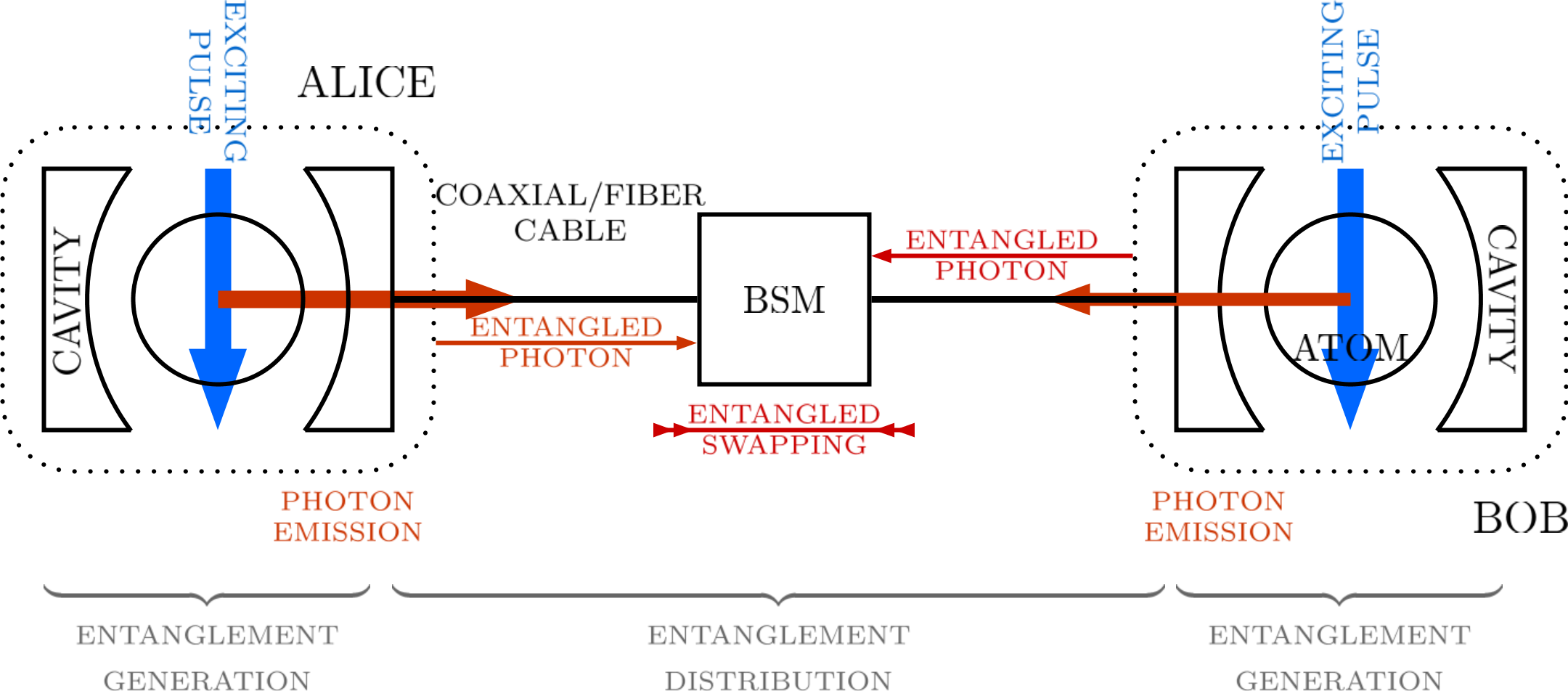}
		\subcaption{Two atoms in optical cavities are simultaneously excited with a laser pulse, leading to the emission of two atom-entangled photons. By measuring the incoming photons, the Bell State Measurement (BSM) projects the atoms into an entangled quantum state.} 
		\label{Fig:3C}
	\end{minipage}
	\caption{Practical schemes for entanglement generation and distribution. Regardless of the location of the entanglement generation functionality (\textit{at mid-point} as in Fig.~\ref{Fig:3A}, \textit{at source} as in Fig.~\ref{Fig:3B}, or \textit{at both end-points} as in Fig.~\ref{Fig:3C}), a quantum link is needed to distribute the entanglement between Alice and Bob.}
	\label{Fig:3}
\end{figure}

\subsection{Practical Entanglement Generation and Distribution}
\label{sec:3.4}

As pointed out above, the entanglement generation and distribution functionality is the key ingredient of quantum teleportation. Here, as a preliminary to the modeling to be discussed in Sec.~\ref{sec:4}, we briefly review this functionality from a practical perspective.

In a nutshell, here we gloss over many of the details, but the generation of quantum entanglement requires that two (or more) particles are in each other's \textit{spatial proximity} -- for example owing to their joint generation or due to their direct interaction -- so that the state of any of these particles cannot be described independently from the state of the other.

Since Alice and Bob represent remote nodes, the entanglement generation occurring at one side must be complemented by the entanglement distribution functionality, which ``moves'' one of the entangled particles to the other side. In this context, there is a broad consensus in the community concerning the adoption of photons as the substrate for the so-called \textit{flying qubits} \cite{NorBla-17}, i.e., as entanglement carriers. The rationale for this choice is related to the advantages provided by photons for entanglement distribution: moderate interaction with the environment leading to moderate decoherence as described in Sec.~\ref{sec:5}, convenient control with standard optical components as well as high-speed low-loss transmissions.

Indeed, one of the basic schemes conceived for entanglement generation -- namely, the Type-II spontaneous parametric down-conversion detailed in \cite{KwiMatWei-95} -- employs photons for both entanglement generation and distribution, as summarized in Fig.~\ref{Fig:3A}. By pointing a laser beam toward a non-linear crystal, two intersecting cones emerge from the crystal with a vertically polarized photon on the upper cone and a horizontally polarized photon on the lower cone. Polarization-entangled photons in one of the Bell states are generated at the two intersections of the cones. The entangled photons travel through a quantum channel to reach their destinations, namely Alice and Bob, where a \textit{transducer} is invoked at each side \cite{CacCalTaf-18} for ``transferring'' the entanglement from the flying qubit to the \textit{matter} qubit, i.e. to a qubit suitable for information processing/storage within a quantum device \cite{AfzGisDer-15}. Indeed, the transducer is needed at the destination, since the very feature that makes photons attractive for entanglement distribution -- namely, their moderate interaction with the environment -- represents a major drawback when it comes to employing photons as substrate for quantum-domain processing/storage. In fact, quantum processing/storage requires the qubits to interact with each other \cite{FlaSpa-18}.

A different scheme designed for generating and distributing the entanglement is depicted in Fig.~\ref{Fig:3B}. This scheme utilizes atoms in optical cavities \cite{CirZolKim-97,MatChaJen-06,RiNolHah-12,WelHacDai-18} linked by a photonic channel, such as a coaxial/fiber cable. Specifically, the first step is to excite an atom coupled with an optical cavity by a laser pulse, which leads to the emission of a photon into the cavity mode. The polarization of the photon is entangled with some internal state of the atom. The photon, exiting the first cavity and traveling along the quantum channel, reaches the second cavity, where it is coherently absorbed and its polarization is mapped onto the state of the remote atom. The atom-photon entanglement is thus converted into entanglement between the two remote atoms. In this scheme, the cavity acts as the matter-flying transducer described in the context of the first scheme.

Finally, a third scheme conceived for generating and distributing the entanglement is summarized in Fig.~\ref{Fig:3C}. Here, both the atoms are simultaneously excited with a laser pulse, leading to the emission of a photon in each cavity. Each photon is entangled with the emitting atom and travels along a quantum channel. Both the photons are then combined at a beam-splitter-based Bell State Measurement (BSM), which stochastically projects\footnote{This procedure is also known under the name of \textit{entanglement swapping}: the entanglement is swapped from the two original atom-photon pairs to the atom-atom pair \cite{ZukZeiHor-93,VanLadMun-09}.} the remote atoms into an entangled quantum state \cite{FerBra-10,NorBla-17,Cal-17-2,OlmMatMau-18}. This scheme has also been proposed in the context of Nitrogen-Vacancy (NV) defect centers in diamonds \cite{BerHen-13}. Finally, it has been extended to artificial atoms such as transmons \cite{KurMag-18}, which constitute one of the most popular substrates for computational qubits. Similarly to the scheme in Fig.~\ref{Fig:3B}, the cavity in this scheme also acts as a matter-flying transducer.

Although the above discussion is far from being exhaustive\footnote{Additionally, the entanglement can be distributed by literally moving stationary qubits and their associated hardware -- after entangling them at one party. However, this scheme is far from being scalable.}, some related considerations can be drawn as follows. The specific ``location'' of the entanglement generation varies among the schemes. For example, in the scheme of Fig.~\ref{Fig:3A} it is ``\textit{at mid-point}''. By contrast, it is ``\textit{at the source}'' in Fig.~\ref{Fig:3B} and ``\textit{at both end-points}'' in Fig.~\ref{Fig:3C} \cite{JonKim-16}. Nevertheless, each scheme requires a quantum channel between Alice and Bob for entanglement distribution. Furthermore, a transducer is needed\footnote{This is the case, unless the flying qubits are also used for computing as in the case of photonic-based quantum processors \cite{FlaSpa-18}, although this research is still in its infancy.} for interfacing the matter qubit with the flying qubit traveling through the quantum channel. These key features of practical entanglement generation/distribution are exploited in the next section for modeling quantum teleportation from a communications engineering perspective.

\begin{table*}[t]
	\centering
	\scriptsize
	\caption{Communication System Modeling: Classical Communications vs Quantum Teleportation}
	\begin{tabular}{|l|p{0.35\textwidth}|p{0.4\textwidth}|}
		\toprule
		\textbf{Block} & \textbf{Classical Communication} & \textbf{Quantum Teleportation} \\
		\midrule
		Information Source
			& The source output can be:
				\begin{itemize}
					\item safely read without altering the embedded information
					\item copied, and hence it can be re-transmitted whenever corrupted by noise
				\end{itemize}
			& The source output \textit{cannot} be:
				\begin{itemize}
					\item read without altering the embedded information -- the quantum measurement postulate
					\item duplicated -- the no-cloning theorem -- and hence it \textit{cannot} be re-transmitted when corrupted by noise
				\end{itemize}
			\vspace{-9pt}\\
		\midrule
		EPR Source
			& Absent
			& Entangling a certain inner state of two particles\\
		\midrule
		\multirow{2}[25]{*}{Transmitter}
			& Classical Transmitter:
				\begin{itemize}
					\item mapping classical information into a classical signal suitable for transmission over a classical channel
				\end{itemize}
			& Classical Transmitter
				\begin{itemize}
					\item mapping the (classical) output of the quantum pre-processing into a classical signal suitable for the classical channel
				\end{itemize}
			\vspace{-9pt}\\
		\cmidrule{2-3}
			& EPR Transmitter:
				\begin{itemize}
					\item absent
				\end{itemize}
			& EPR Transmitter
				\begin{itemize}
					\item mapping the entangled particle into a quantum signal suitable for the considered quantum channel, for conveying the entanglement at the remote node
				\end{itemize}
			\vspace{-9pt}\\
		\midrule
		\multirow{2}[35]{*}{Channel}
			& Classical Channel
				\begin{itemize}
					\item the communication range can be extended through classical amplify-and-forward or decode-and-forward techniques, since a classical signal can be measured without altering the encoded information
				\end{itemize}
			\vspace{-9pt}
			& Classical Channel
				\begin{itemize}
					\item medium used to transmit the classical signal from the Tele-Transmitter to the Tele-Receiver
				\end{itemize}\\
		\cmidrule{2-3}
			& Quantum Channel
				\begin{itemize}
					\item absent
				\end{itemize}
			& Quantum Channel
				\begin{itemize}
					\item conveying the entanglement to remote nodes
					\item the communication range \textit{cannot} be extended through classical amplify-and-forward or decode-and-forward techniques, due to the quantum measurement. \textit{Quantum Repeaters} should be adopted
				\end{itemize}
			\vspace{-9pt}\\
		\midrule
		\multirow{2}[25]{*}{Receiver}
			& Classical Receiver
				\begin{itemize}
					\item decoding the classical message from the received classical signal
				\end{itemize}
			\vspace{-9pt}
			& Classical Receiver
				\begin{itemize}
					\item decoding the classical input of the quantum post-processing block from the received classical signal
				\end{itemize}
			\vspace{-9pt}\\
		\cmidrule{2-3}
			& EPR Receiver
				\begin{itemize}
					\item absent
				\end{itemize}
			& EPR Receiver
				\begin{itemize}
					\item decoding the entangled input of the quantum pre-processing and post-processing blocks from the received quantum signals
				\end{itemize}
			\vspace{-9pt}\\
		\bottomrule
	\end{tabular}
	\label{tab:02}
\end{table*}

\begin{figure*}[t]
	\centering
	\includegraphics[width=1\textwidth]{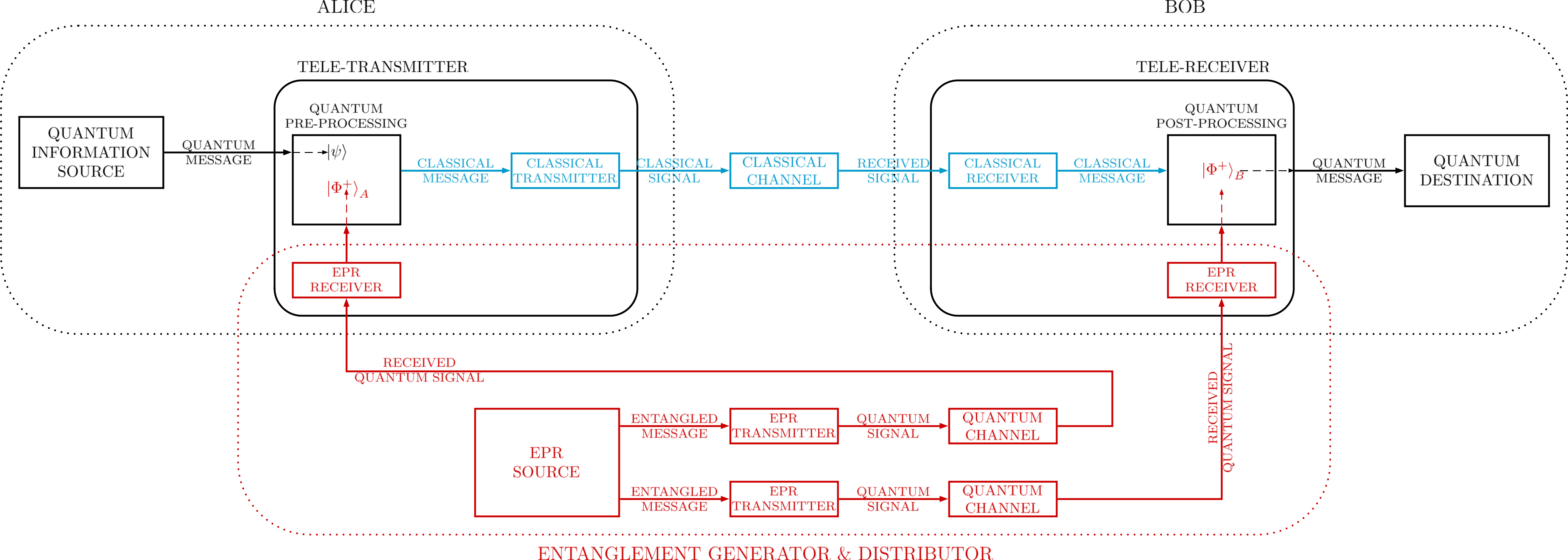}
	\caption{A Noiseless Communication System Model of Quantum Teleportation.}
	\label{Fig:4}
\end{figure*}

\begin{figure}[t]
	\centering
	\includegraphics[width=1\columnwidth]{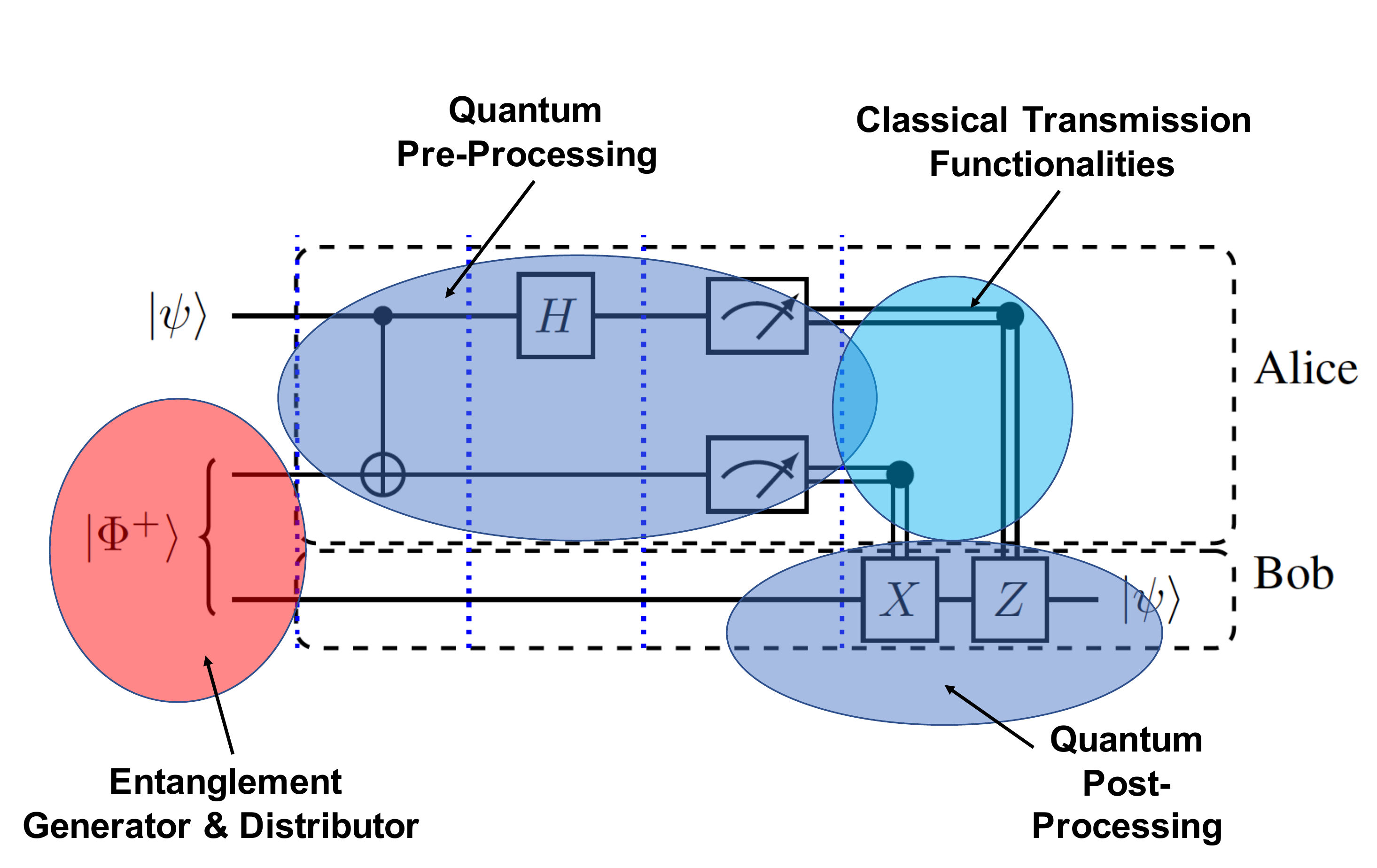}
	\caption{Quantum Teleportation Circuit of Fig.~\ref{Fig:03} interpreted in the light of Fig.~\ref{Fig:4}.}
	\label{Fig:4_bis}
\end{figure}

\section{A Noiseless Communication System Model of Quantum Teleportation}
\label{sec:4}

Based on the discussions developed in the previous section, it comes to light that a communication system model conceived for quantum teleportation relies on the tight integration of both classical/quantum operations and communications. Hence, the classical communication system model originally proposed by Shannon \cite{Sha-48} must be revised to account for the specific nature of quantum teleportation. Hence we propose the model depicted in Fig.~\ref{Fig:4}, where the different system blocks summarized in Table~\ref{tab:02} and discussed below may be readily identified. For the sake of clarity, we also highlight in Fig.~\ref{Fig:4_bis} the communication functionalities of the aforementioned blocks with reference to the quantum teleportation circuit described in Fig.~\ref{Fig:03}.

\begin{itemize}
\item[(a)] The \textit{\underline{Quantum Information Source}} of Fig.~\ref{Fig:4} provides the Tele-Transmitter with the unknown quantum message to be teleported to Bob and maps it to the qubit state $\ket{\psi}$, which again cannot be read or copied\footnote{As it will be detailed in the next section, this implies that classical error correction techniques, based on parity or repetition techniques, cannot be employed in a quantum network \cite{ChaBabNgu-18}.}.

\item[(b)] The \textit{\underline{Tele-Transmitter}} of Fig.~\ref{Fig:4} processes the quantum message mapped to $\ket{\psi}$ to produce a classical signal suitable for transmission over a classical channel. This operation is accomplished by the following sub-blocks of Fig.~\ref{Fig:4}: the \textit{Quantum Pre-Processing}, the \textit{Classical Transmitter} and the \textit{EPR Receiver}. Specifically, the \textit{EPR Receiver} is a sub-block of the \textit{Entanglement Generator \& Distributor} super-block of Fig.~\ref{Fig:4}, which supplies the Tele-Transmitter and the Tele-Receiver with the respective members of the entangled pair -- e.g., $\ket{\Phi^+}_A$ and $\ket{\Phi^+}_B $. The \textit{Entanglement Generator \& Distributor} as well as its sub-blocks are described in detail at the end of this section. The \textit{Quantum Pre-Processing} jointly operates on $\ket{\psi}$ and $\ket{\Phi^+}_A$ by applying a sequence of operations, as detailed in Sec.~\ref{sec:3.2} and further highlighted in Fig.~\ref{Fig:4_bis}. Explicitly, the CNOT gate of Table~\ref{Tab:01} is applied to both the qubits at Alice's side, followed by applying the H gate of Table~\ref{Tab:01} to the qubit to be teleported, further followed by a joint measurement applied to both qubits. Hence its output -- namely the result of the measurement -- is a classical message, which is in turn mapped by the \textit{Classical Transmitter} to a classical signal suitable for transmission to Bob over a \textit{classical channel}.

\item[(c)] The \textit{\underline{Tele-Receiver}} of Fig.~\ref{Fig:4} performs the inverse operations of the Tele-Transmitter: it reconstructs the unknown quantum message from the received classical signal. This operation is accomplished by the following sub-blocks of Fig.~\ref{Fig:4}: the \textit{Classical Receiver}, the \textit{Quantum Post-Processing} and the \textit{EPR Receiver}. Specifically, first the \textit{Classical Receiver} decodes the received classical signal into a classical message. Then, the \textit{Quantum Post-Processing} recovers $\ket{\psi}$ from $\ket{\Phi^+}_B$, which is provided by the \textit{EPR Receiver}, with the sequence of quantum operations (gates) indicated by the classical message (see Sec.~\ref{sec:3.2}) as also highlighted in Fig.~\ref{Fig:4_bis}.

\item[(d)] The \textit{\underline{Entanglement Generator \& Distributor}} of Fig.~\ref{Fig:4} is the super-block responsible for the generation and distribution of the EPR pair members $\ket{\Phi^+}_A$ and $\ket{\Phi^+}_B$ at Alice and Bob, respectively, as also highlighted in Fig.~\ref{Fig:4_bis}. Following from our discussion in Sec.~\ref{sec:3.4}, the \textit{Entanglement Generator \& Distributor} super-block is constituted by four blocks, as shown in Fig.~\ref{Fig:4}. Although the specific location and the physical implementation of each block may vary -- depending on the particulars of the practical scheme adopted for entanglement generation and distribution -- the model of Fig.~\ref{Fig:4} allows us to highlight each communication functionality required for generating and distributing the EPR pair to remote nodes.
\begin{itemize}
\item[(d1)] The \textit{EPR Source} of Fig.~\ref{Fig:4} generates the EPR pair by maximally entangling a certain inner state of two particles. The particular nature of the particles depends on the specific scheme considered -- ranging from photons through atoms to artificial atoms in superconducting circuits. The location of the EPR source may vary as well, as shown in Figs.~\ref{Fig:3A}-\ref{Fig:3C}. Nevertheless, from a communications engineering perspective, the entangled states represent the ``entangled messages'' to be transmitted to both Alice and Bob. 

\item[(d2)] The \textit{EPR Transmitter} of Fig.~\ref{Fig:4} processes the entangled message to produce a quantum signal suitable for transmission over a quantum channel, such as an optical fiber or Free-Space Optical (FSO) link. Again, the broad consensus is that of adopting photons as substrate for \textit{flying qubits} \cite{Div-00}. However, the entanglement can be mapped to a photon's different features -- such as its polarization, time-bin, etc. \cite{FlaSpa-18}. Hence, these degrees of freedom are exploited by the EPR Transmitter to produce a quantum signal suitable for transmission over the specific quantum channel.

\item[(d3)] The \textit{Quantum Channel} block of Fig.~\ref{Fig:4} represents the medium used for transmitting the quantum signal from the EPR Transmitter to the EPR Receiver. The quantum channel characteristics as well as the maximum achievable communication range vary significantly, depending on the specific choice of transmission medium -- i.e., FSO terrestrial of satellite channel or alternatively optical fiber channel. Furthermore, in contrast to classical channels, where the communication range can be extended using classical amplify-and-forward or decode-and-forward techniques, quantum channels require the adoption of \textit{quantum repeaters}\footnote{The BSM block of Fig.~\ref{Fig:3C} may be regarded as a (very) basic quantum repeater \cite{VanLadMun-09}. Please refer to Sec.~\ref{sec:7.2} for further details about quantum repeaters \cite{devitt2016high}.} \cite{BrieDurCir-98}, since the quantum signal cannot be measured without irreversibly altering the original quantum signal due to the measurement postulate \cite{VanMet-14,PirLau-17,MurLi-16,JonKim-16}.

\item[(d4)] The \textit{EPR Receiver} of Fig.~\ref{Fig:4} performs the inverse operation of the EPR Transmitter, by extracting the entangled state from the received quantum signal. It is responsible for providing the Quantum Pre-Processing and the Quantum Post-Processing with the entangled pair members $\ket{\Phi^+}_A$ and $\ket{\Phi^+}_B$.
\end{itemize}
\end{itemize}

\begin{rem}
Again, the modeling of the \textit{Entanglement Generator \& Distributor} super-block of Fig.~\ref{Fig:4} aims for providing a general portrayal of the communication functionalities required. However, the physical-counterparts of its component blocks vary, depending on the specific choices of the entanglement generation/distribution technique adopted, as described in Sec.~\ref{sec:3.4}. For instance, the schemes in Figs.~\ref{Fig:3A}-\ref{Fig:3B} employ a single EPR Source, which is located either at Alice or halfway between Alice and Bob. By contrast, the scheme in Fig.~\ref{Fig:3C} employs a pair of EPR sources, one at Alice and one at Bob. As regards to the EPR Transmitter, we have a single transmitter in Figs.~\ref{Fig:3A}-\ref{Fig:3B} (the crystal in the former, acting also as the EPR source, and the cavity in the latter, acting as the matter-flying qubit transducer) while two transmitters in Fig.~\ref{Fig:3C} (the cavities at each side, acting as transducer as well). Finally, regarding the EPR Receiver, in Fig.~\ref{Fig:3A} this functionality is performed by the matter-flying transducers located at Alice's side and at Bob's side, whereas there exists a unique physical EPR Receiver\footnote{The EPR Receiver block is ``virtual'' for Alice, since the process of ``receiving'' Alice's member of the entangled pair is fulfilled without the physical reception of the particle, by exploiting the above-mentioned ``spooky action at a distance'' determined by the interaction between the photon and the cavity.} -- the cavity -- at Bob's side in Fig.~\ref{Fig:3B}. By contrast, no physical EPR Receiver is present in Fig.~\ref{Fig:3C}, since the EPR pair members are locally generated at both Alice's and Bob's side through entanglement swapping. In other words, in Fig.~\ref{Fig:3C}, the EPR Receiver block is ``\textit{virtual}'' at both Alice's and Bob's sides, since its communication functionality, i.e., the process of ``receiving'' the member of the entangled pair is fulfilled without the physical reception of the particle, which is the phenomenon being exploited by the entanglement swapping.
\end{rem}

\begin{rem}
It is worthwhile noting that in Fig.~\ref{Fig:4} the quantum equivalent of the classical source-encoder block is absent. Specifically, the classical source-encoder is responsible for efficiently representing the source output in a sequence of binary digits with little or no redundancy \cite{Proakis-01}. However, this functionality is based on the assumption that any classical information can be read anywhere at any time, whilst this does not hold for the quantum domain. Hence, a one-to-one mapping between the classical source-encoder and a quantum-equivalent source-encoder may not be feasible. Hence further research is needed. By contrast, the classical channel-encoder block is responsible for imposing carefully controlled redundancy on the message for detecting and correcting (to some degree) the errors inflicted by the channel impairments \cite{Proakis-01}. Its quantum equivalent will be discussed in Sec.~\ref{sec:7}.
\end{rem}

\section{Imperfect Quantum Teleportation}
\label{sec:5}

The discussions of the previous section are valid under the idealized simplifying hypothesis of experiencing no decoherence. However, similarly to classical communications, the quantum communication model should account for the presence of realistic imperfections.

Again, realistic quantum systems suffer from undesired interactions with the \textit{environment}. Hence, they constitute \textit{open} rather than \textit{closed} physical systems \cite{Sch-07}. These interactions with the environment irreversibly affect any quantum state by the process of \textit{decoherence} \cite{NieChu-11,RiePol-11,LidBru-13}. Decoherence is unavoidable and it affects not only the unknown quantum state to be teleported, but also the entanglement generation and distribution process required for implementing the quantum teleportation.

However, decoherence is not the only source of impairments \cite{CacCalTaf-18}. In fact, the quantum teleportation relies on a sequence of operations applied to the quantum states, as detailed before. The contamination of these operations further aggravates the imperfections of the quantum teleportation.

\begin{figure}
	\centering
	\includegraphics[width=0.95\columnwidth]{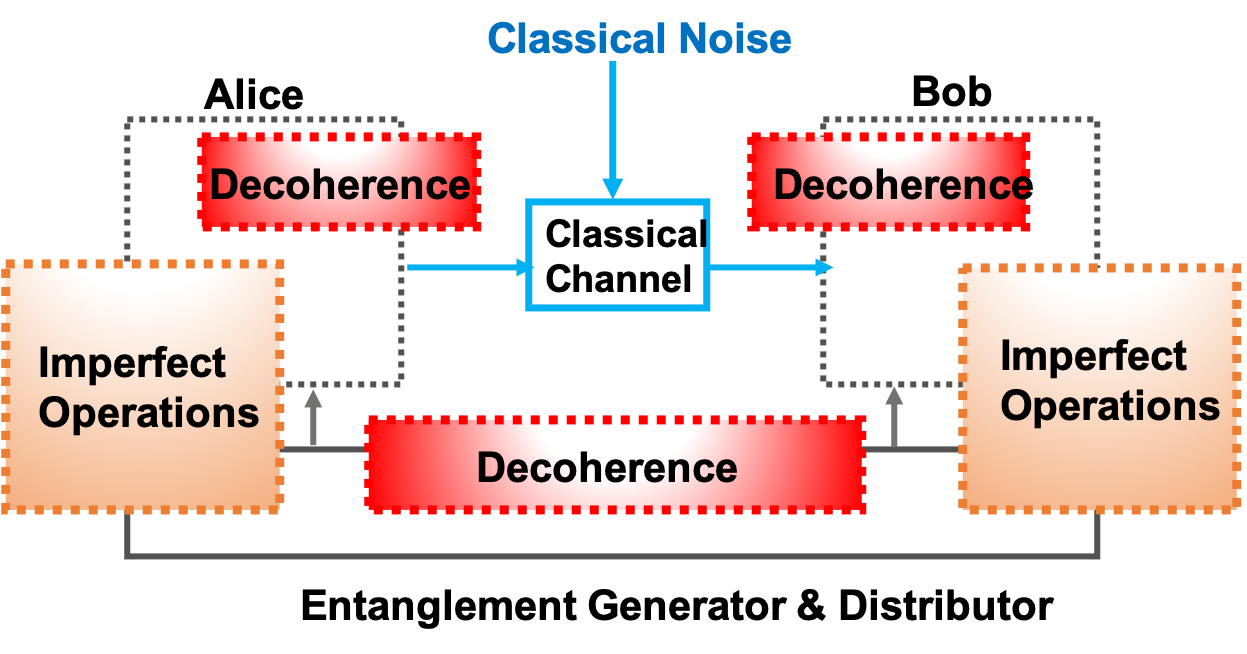}
	\caption{Imperfections Contaminating the Quantum Teleportation Process.}
	\label{Fig:5}
 \end{figure}

In Fig.~\ref{Fig:5}, we depict the relationship of different imperfections\footnote{Although the figure represents also the classical noise affecting the \textit{classical channel} block of Fig.~\ref{Fig:4} for the sake of completeness, in the following we will focus our attention on the quantum impairments only.} degrading the fidelity of the ``teleported'' qubit. Regardless of the specific cause of degradations, the effect of the quantum impairments imposed on a quantum system is that pure quantum states evolve into mixed quantum states \cite{LidBru-13}. However, despite its pivotal importance, the accurate modelling of quantum-domain impairments capable of capturing the effects of the different imperfections on the quantum teleportation process remains an open problem at the time of writing this treatise. 

With this in mind, to gain further insights into the behavior of the composite quantum impairment:
\begin{itemize} 
	\item in Secs.~\ref{sec:5.1}--\ref{sec:5.4}, we will provide a rudimentary communications-engineering perspective on the theoretical framework of characterizing the effects of quantum decoherence on an arbitrary qubit;
	\item in Sec.~\ref{sec:6}, we then complement this theoretical analysis by an experimental perspective, in which we characterize the cumulative quantum impairments affecting the quantum teleportation relying on an operational quantum chip using the IBM Q platform \cite{IBM-Q}. 
\end{itemize}
The rationale for splitting the analysis into these two steps is that the errors arising in the quantum teleportation owing to imperfect operations strongly depend on the particulars of the technology adopted for representing a qubit. As a consequence, to gain tangible insights into the behavior of the composite quantum impairment, the adoption of a specific quantum chip technology is inevitable.

Furthermore, with the analysis about to be developed in Secs.~\ref{sec:5.1}-\ref{sec:5.4} in mind we will be able to point out two distinctive features of the quantum impairments:
\begin{itemize}
	\item[i)] the quantum impairments are \textit{multiplicative} rather than being additive, hence this phenomenon might be deemed more reminiscent of the classical fading effects, rather than of the classical additive noise imposed by the Brownian motion of electrons;
	\item[ii)] the quantum impairments exhibit an asymmetric behavior, since they alter the coordinates of the Bloch vector representing the qubit differently; this phenomenon might be interpreted as a sort of \textit{spatial selectivity} in terms of Bloch vector coordinates.
\end{itemize}

These distinctive features of the quantum impairments will be confirmed by the experimental findings of Sec.~\ref{sec:6}.

\subsection{Modeling Quantum Decoherence}
\label{sec:5.1}

When an open quantum system $\mathcal{S}$ of interest -- a qubit in our case -- interacts with the environment $\mathcal{E}$, they together form the closed quantum system $\mathcal{S}\mathcal{E}$ \cite{NieChu-11,RiePol-11,GyoImrNgu-18}. This closed system $\mathcal{S}\mathcal{E}$ evolves according to a unitary transformation $U_{\mathcal{S}\mathcal{E}}$ formulated as: 
\begin{equation}
	\label{eq:4.1}
	\rho_{\mathcal{S}\mathcal{E}}(t)=U_{\mathcal{S}\mathcal{E}} \rho_{\mathcal{S}\mathcal{E}}(0) U_{\mathcal{S}\mathcal{E}}^\dagger,
\end{equation}
with $\rho_{\mathcal{S}\mathcal{E}}(\cdot)$ representing the density matrix of the closed quantum system $\mathcal{S}\mathcal{E}$, as defined in Sec.~\ref{sec:2}. 

The status of the system $\mathcal{S}$ of interest can be recovered by tracing out the environment via the partial trace operator $\text{Tr}_{\mathcal{E}}(\cdot)$ over the environment $\mathcal{E}$, which is expressed as:
\begin{equation}
	\label{eq:4.2}
	\rho_{\mathcal{S}}(t)=\text{Tr}_{\mathcal{E}}\left[\rho_{\mathcal{S}\mathcal{E}}(t)\right]=\text{Tr}_{\mathcal{E}}\left[U_{\mathcal{S}\mathcal{E}} \rho_{\mathcal{S}\mathcal{E}}(0) U_{\mathcal{S}\mathcal{E}} ^\dagger\right],
\end{equation}
where $\rho_{\mathcal{S}}(t)$ is referred to as the \textit{reduced density matrix}.

Due to the complex interactions between the system and the environment, in general $\rho_{\mathcal{S}}(t)$ may not be directly related to the initial state $\rho_{\mathcal{S}}(0)$ through a unitary transformation. Furthermore, it is quite a challenge to evaluate \eqref{eq:4.2}, since it requires us to determine the dynamics $\rho_{\mathcal{S}\mathcal{E}}(t)$ of the composite system $\mathcal{S}\mathcal{E}$. Indeed, the status of the environment is always unknown and cannot be controlled in reality. 

However, by applying some approximations, it is often possible to derive the approximate evolution of $\rho_{\mathcal{S}}(t)$ vs. time via the master equation formalism of \cite{Sch-07}. 
Accordingly, the evolution of the system $\mathcal{S}$ vs. time can be expressed in the Lindblad form\footnote{In the following, $[A,B]$ denotes the commutator between two operators and it is defined as $[A,B]=AB-BA$. Similarly, $\{A,B\}$ denotes the anti-commutator between two operators and it is defined as $\{A,B\}=AB+BA$.} as a time-local first-order differential equation system of the following form \cite{OhLeeLee-02,Sch-07}:
\begin{align}
	\label{eq:4.3}
	\frac{d}{dt}\rho_{\mathcal{S}}(t)& = \overbrace{-\frac{i}{\hbar}[H_s,\rho_{\mathcal{S}}(t)]}^{\text{unitary evolution}} + \nonumber \\
		& + \overbrace{\sum_{k}\left(L_{k}\rho_{\mathcal{S}}(t)L_{k}^{\dag}-\frac{1}{2}\{L_{k}{\dag} L_{k},\rho_{\mathcal{S}}(t)\}\right)}^{\text{non-unitary evolution}}.
\end{align} 
In \eqref{eq:4.3}, we note the presence of two components. The unitary evolution component depends both on Planck's constant $\hbar$, whose value must be experimentally determined, and on the Hermitian operator $H_s$, referred to as the \textit{Hamiltonian} of the system\footnote{In general, $H_s$ in \eqref{eq:4.3} is not identical to the unperturbed free Hamiltonian of the system that would govern the evolution of the system in the absence of any environmental effects \cite{Sch-07}. Indeed, the environmental interference typically perturbs the free Hamiltonian, leading to a re-normalization of the energy levels of the system. However, this effect (also known as Lamb-shift \cite{Sch-07}) does not contribute to the non-unitary evolution imposed by the environmental perturbations -- it only affects the unitary part of the reduced dynamics.}. In the following we assume $\hbar=1$, without loss of generality.
In the case of a closed system, the knowledge of the Hamiltonian implies having the knowledge of the entire dynamics of the system. The non-unitary evolution component follows from the non-unitary nature of the trace operation used for obtaining the reduced density matrix, and it is driven by the Lindblad operators, $L_{k}$, representing the coupling of the system to its environment. For multiple qubit based systems associated with the dimension $d$, $d^2-1$ Lindblad operators are needed in \eqref{eq:4.3}. Hence, for one-qubit systems, three Lindblad operators are required and they are given by the Pauli matrices $\{\sigma_k\}_{k=x,y,z}$ \cite{NieChu-11,OhLeeLee-02,Sch-07} reported in Table~\ref{Tab:01}:
\begin{equation}
	\label{eq:4.4}
	L_{k}=\sqrt{\gamma_k}\sigma_{k}, \quad k \in \{x,y,z\}.
\end{equation}
In \eqref{eq:4.4}, the coefficients $\{\gamma_k\}$ are referred to as decay rates, for the reasons justified in the following, and they depend on the specific interaction between the system $\mathcal{S}$ and the environment $\mathcal{E}$.

\begin{rem}
The choice of the specific Lindblad operator set is not unique and it usually depends on the particular type of quantum impairment under investigation. For example, \eqref{eq:4.4} models an important category of quantum impairments, generally termed as \textit{depolarizing phenomenon}, which imposes errors -- such as \textit{bit-flips}, \textit{phase-flips} and their combinations \cite{NieChu-11} -- typically arising in quantum computation and quantum communications. By contrast, upon setting $L_{\pm}=\sqrt{\gamma_\pm} \left(\sigma_{x} \pm i \sigma_{y} \right)$, it becomes possible to directly model another important category of quantum impairment, referred to as \textit{thermalization noise}: a qubit, if left alone for sufficiently long time, will eventually settle into some classical distribution of the basis states $\ket{0}$ and $\ket{1}$ \cite{VanMet-14,NieChu-11} -- as a consequence of energy exchange with the environment. The depolarizing phenomenon acts on a much shorter time-scale than the thermalization, hence it is the first impairment that must be considered, when modeling quantum teleportation from a communication perspective. Hence, in the following, we restrict our attention to the former. However, it is worthwhile noting that thermalization plays a crucial role in quantum networks, where the quantum states must be stored in quantum memories for fulfilling the communication needs at hand -- exemplified by waiting for reply messages from across the network. Hence, we set aside the thermalization modeling for our future work.
\end{rem}

Based on these preliminaries, in the next subsections we will review some of the different impairments from a communications engineering perspective. Without loss of generality, in the following we assume having a Hamiltonian of $H_s=\frac{\Omega}{2} \sigma_z$, which indicates that the Hamiltonian is dominated by the unperturbed qubit energy splitting $\Omega$ \cite{Sch-07,BieDohUys-11}.

\subsection{Phase Damping}
\label{sec:5.2}

One of the quantum depolarizing processes with no direct counterpart in the classical world is \textit{phase damping}, which models the erosion of quantum information without loss of energy. This is one of the most common perturbations in quantum information processing. The phase damping is described by the Lindblad operator $L_z=\sqrt{\gamma_z}\sigma_{z}$, where again, $\gamma_z$ is the decay-rate. Upon substituting it into \eqref{eq:4.3} and by accounting for $L_{z}{\dag} L_{z}=\gamma_z \,I$, the resultant time-domain evolution of the system $\mathcal{S}$ may be formulated as:
\begin{align}
\nonumber
	\label{eq:4.5}
	\frac{d}{dt}\rho_{\mathcal{S}}(t)& = -i\left[\frac{\Omega}{2} \sigma_z,\rho_{\mathcal{S}}(t)\right] + \nonumber \\
		&+\overbrace{L_{z}\rho_{\mathcal{S}}(t)L_{z}^{\dag}-\frac{1}{2}\{L_{z}{\dag} L_{z},\rho_{\mathcal{S}}(t)\}}^{\text{non-unitary evolution induced by the Phase Damping}}.
\end{align}

Solving \eqref{eq:4.5} as detailed in \cite{Pr:CacCal-19}, we find that the diagonal elements $\rho_{\mathcal{S}}^{jj}(t)$ are time-invariant, i.e. we have $\rho_{\mathcal{S}}^{jj}(t)=\rho_{\mathcal{S}}^{jj}(0) \, \forall t$ with $j=0,1$, whereas the off-diagonal elements are:
\begin{equation}
	\label{eq:4.6}
	\begin{aligned}
		\rho_{\mathcal{S}}^{01}(t) &= \rho_{\mathcal{S}}^{01}(0) e^{-(i\Omega + 2\gamma_z)t}\\
		\rho_{\mathcal{S}}^{10}(t) &=\rho_{\mathcal{S}}^{10}(0) e^{-(-i\Omega + 2\gamma_z)t} = \left[\rho_{\mathcal{S}}^{01}(t)\right]^*.
	\end{aligned}
\end{equation} 

\begin{figure}
	\centering
	\includegraphics[width=.66\columnwidth]{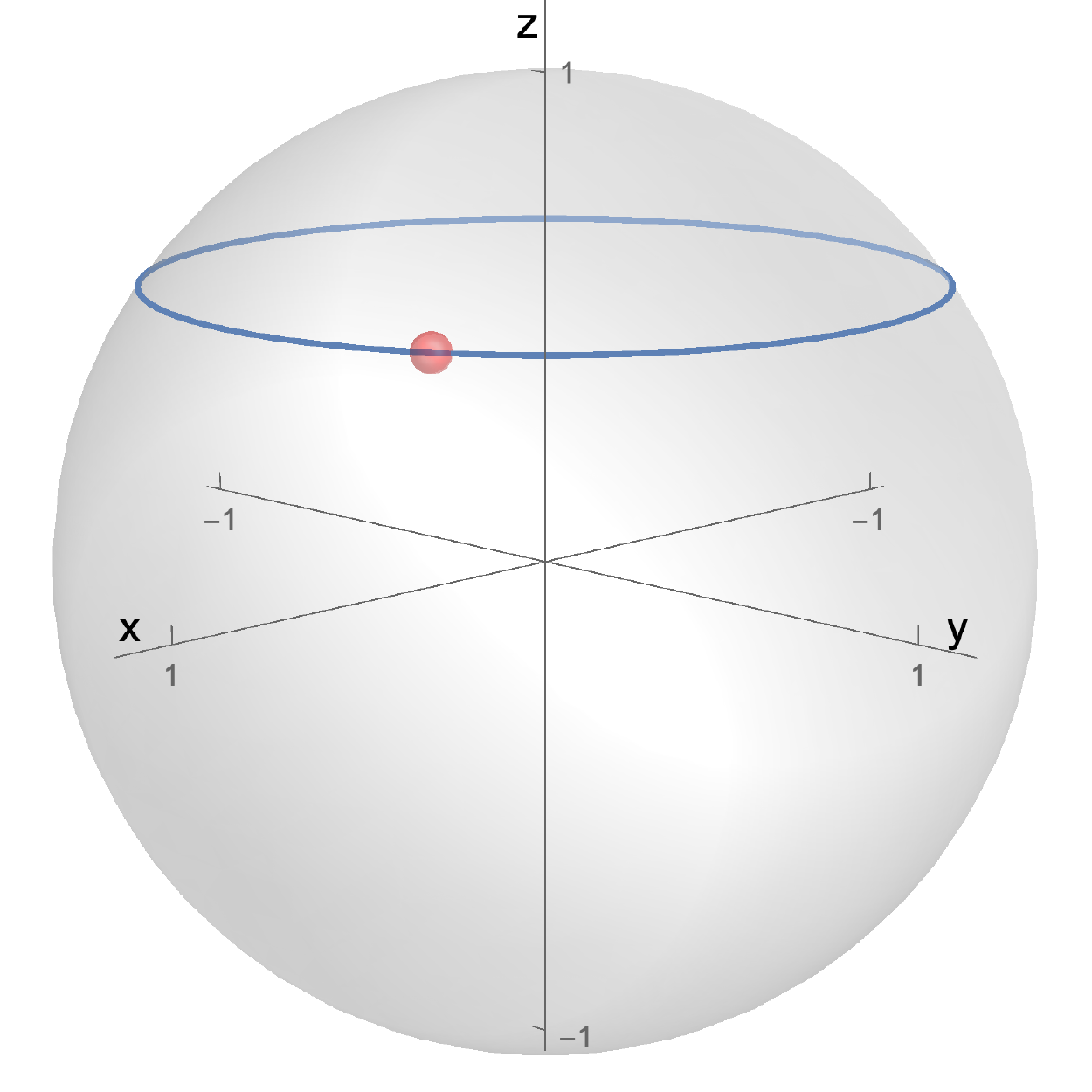}
	\caption{Bloch sphere representation of the free time-domain evolution of a qubit subject to the Hamiltonian $H_s=\Omega\sigma_z/2$, when emerging from the initial pure state of $\ket{\psi(0)} = \sqrt{\frac{1+ \sqrt{3}}{2\sqrt{3}}} \ket{0} + \frac{i + \sqrt{3}}{2 \sqrt{3+\sqrt{3}}} \ket{1}$, located at Bloch vector coordinates $\mathbf{r}(0)=\left[\frac{1}{\sqrt{2}},\frac{1}{\sqrt{6}},\frac{1}{\sqrt{3}}\right]$ and indicated by the red point within the figure.}
	\label{Fig:6}
\end{figure}

Observe from \eqref{eq:4.6} that an arbitrary qubit obeys a phase-evolution that depends on: i) the energy difference between the states $\ket{0}$ and $\ket{1}$ via the term $\Omega t$ of \eqref{eq:4.6}, which induces a rotation around the z-axis of the Bloch sphere, as shown in Fig.~\ref{Fig:6} \cite{Pr:CacCal-19}; ii) the damping decay rate via the term $2\gamma_z t$ of \eqref{eq:4.6}. Since the phase evolution imposed by $\Omega t$ can be compensated by a Phase shift gate $R_{\phi}$ associated with the opposite linear phase of $\phi = \Omega t$ \cite{Pr:CacCal-19}, it becomes possible to streamline the noise effects formulated in \eqref{eq:4.6} in the compact form of:
\begin{equation}
	\label{eq:4.8}
	\rho_{\mathcal{S}}(t) = R_{\Omega t} \rho_{\mathcal{S}}(t) R_{\Omega t}^\dagger =
		\begin{bmatrix}
			\rho_{\mathcal{S}}^{00}(0) & \rho_{\mathcal{S}}^{01}(0) e^{-2\gamma_z t}\\
			\rho_{\mathcal{S}}^{10}(0) e^{-2\gamma_z t} & \rho_{\mathcal{S}}^{11}(0)
		\end{bmatrix}.
\end{equation}

Observe in \eqref{eq:4.8} that the off-diagonal elements, $\rho_{\mathcal{S}}^{ij}(t)$ with $i\neq j$, decay exponentially vs. the time at a decay-rate of $\gamma_z$. Hence the original information embedded into the initial quantum state represented by these elements exponentially erodes vs. time owing to the phenomenon of phase damping.

In order to further augment the physical interpretation of the phase-damping phenomenon from a communications engineering perspective, let us visualize its effects on the Bloch sphere of Fig.~\ref{Fig:02} using the Cartesian coordinates of the Bloch vector $\mathbf{r}=[r_x,r_y,r_z] \in \mathcal{R}^3$ representing the quantum state \cite{RiePol-11} and introduced in Sec~\ref{sec:2.6}. Specifically, by exploiting \eqref{eq:2.17} and \eqref{eq:2.18}, after some algebraic manipulations detailed in \cite{Pr:CacCal-19}, the Bloch vector coordinates $\mathbf{r}(t)=[r_x(t),r_y(t),r_z(t)]$ of the qubit subject to phase damping may be expressed at time instant $t$ as:
\begin{equation}
	\label{eq:4.9}
	\begin{split}
		& r_x(t)=r_x(0) e^{-2 \gamma_z t}, \\
		& r_y(t)=r_y(0) e^{-2 \gamma_z t}, \\
		& r_z(t)=r_z(0),
	\end{split}
\end{equation}
where $\mathbf{r}(0)=[r_x(0),r_y(0),r_z(0)]$ represents the Bloch vector at time instant $0$.

\begin{rem}
\label{rem_Znoise}
Observe in \eqref{eq:4.9} that the phase-damping effects are multiplicative impairments imposed on the Bloch vector coordinates $r_x$ and $r_y$ of the quantum state. More explicitily, the phase damping exhibits an asymmetric behavior, since it affects the coordinates of the Bloch vector differently. Specifically, it damps both the x- and the y-coordinate, i.e. $r_x$ and $r_y$, while it leaves the z-coordinate, i.e. $r_z$, unaltered. This phenomenon might also be interpreted as a sort of \textit{spatial selectivity} in terms of the Bloch vector coordinates.
\end{rem}

\begin{rem}
\label{rem_Znoise_bis}
We further note that although the phase damping is modelled by the Pauli-Z gate of Table~\ref{Tab:01} via the Lindblad operator $L_z=\sqrt{\gamma_z}\sigma_{z}$, this should not be confused with the pure unitary evolution imposed by the Pauli-Z gate in equation \eqref{eq:2.19}, because additionally we have to take into account the non-unitary evolution induced by $L_z$ in \eqref{eq:4.5}. More explicitly, similarly to the unitary evolution imposed by the Pauli-Z gate in \eqref{eq:2.19}, the phase damping leaves the z-coordinate unaffected, but in contrast to the phase-flipping imposed by the Pauli-Z gate based unitary evolution, the non-unitary evolution represented by $L_z$ resulted in a damping of the x- and y-coordinate rather than flipping them. 
\end{rem}

\begin{figure}
	\begin{minipage}[c]{0.49\textwidth}
		\centering
		\includegraphics[width=.66\columnwidth]{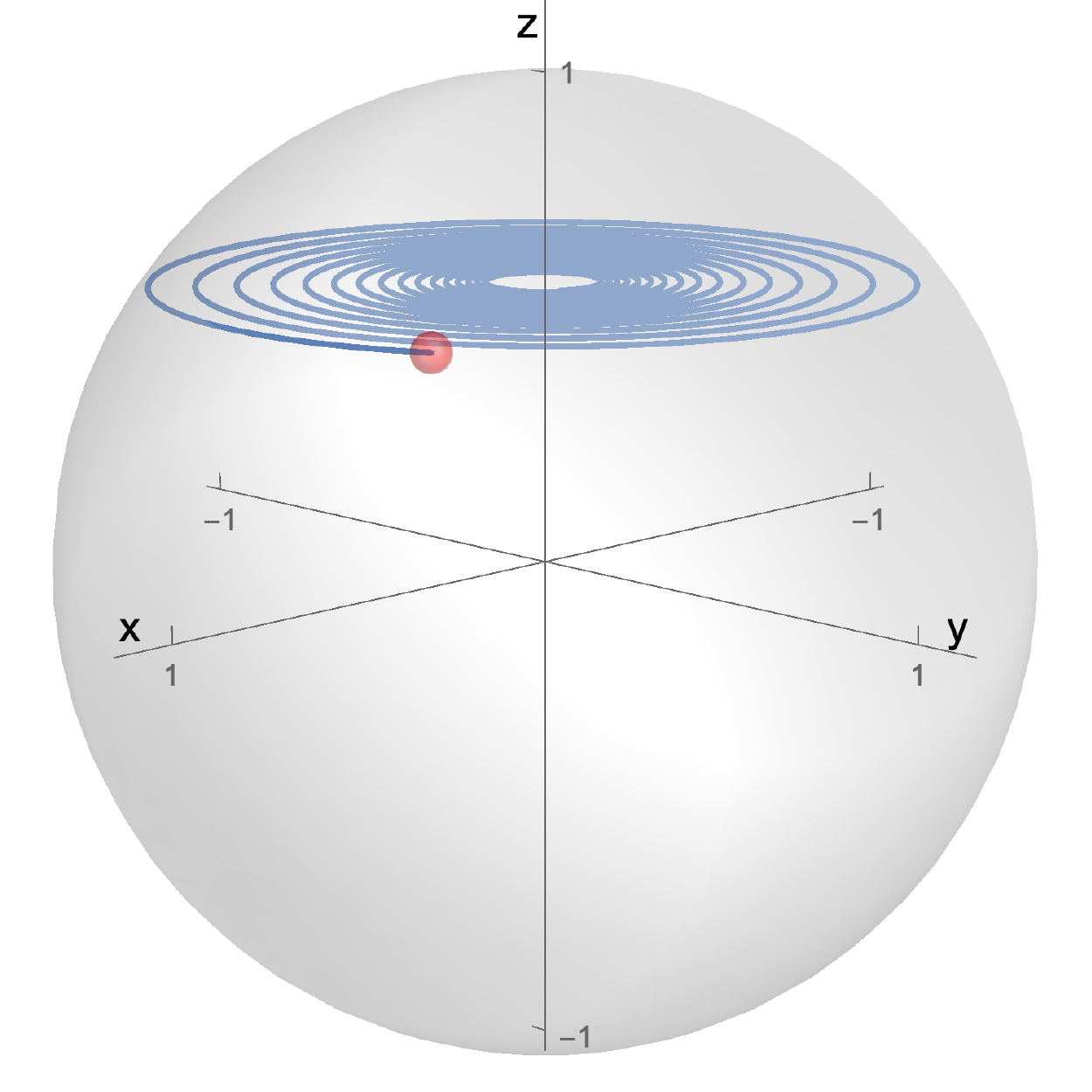}
	\end{minipage}
	\hspace{0.01\textwidth}
	\begin{minipage}[c]{0.49\textwidth}
		\centering
		\includegraphics[width=1\columnwidth]{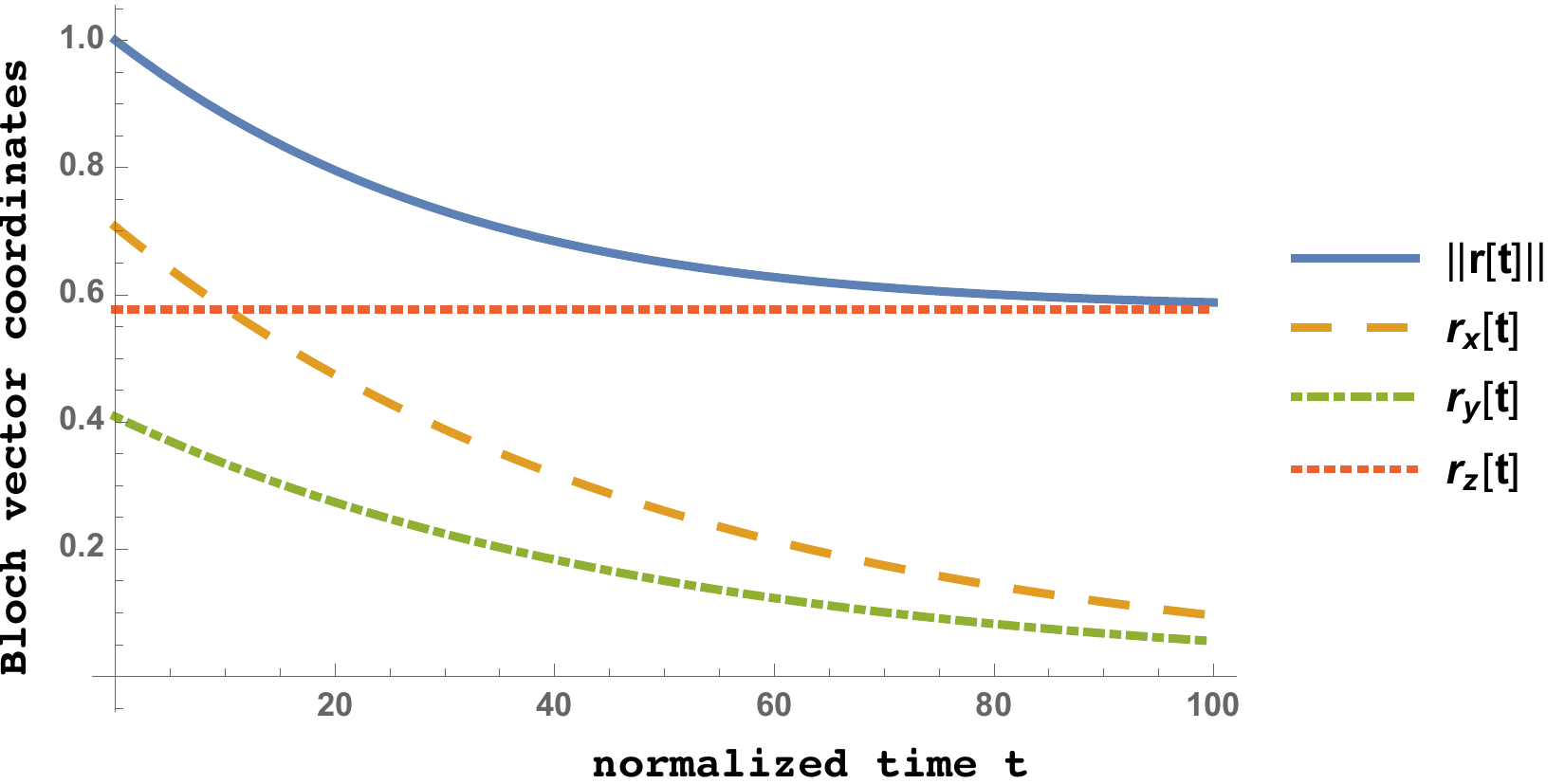}
	\end{minipage}
	\caption{Phase Damping: time-domain evolution of a qubit subject to the Hamiltonian $H_s=\Omega\sigma_z/2$, when emerging from the initial pure state of $\ket{\psi(0)} = \sqrt{\frac{1+ \sqrt{3}}{2\sqrt{3}}} \ket{0} + \frac{i + \sqrt{3}}{2 \sqrt{3+\sqrt{3}}} \ket{1}$ located at the Bloch vector coordinates $\mathbf{r}(0)=\left[\frac{1}{\sqrt{2}},\frac{1}{\sqrt{6}},\frac{1}{\sqrt{3}}\right]$ and ending up inside the sphere with unchanged z-coordinate. Left plot: representation of the qubit's time evolution with respect to the Bloch sphere. Right plot: Bloch vector coordinates $\mathbf{r}(t)$ as a function of time, with the phase evolution engendered by the energy difference $\Omega$ in $H_s$ appropriately compensated, as detailed in the text.}
	\label{Fig:7}
\end{figure}

\begin{rem}
Observe in \eqref{eq:4.9} that the initial pure state $\ket{\psi(0)}$ is transformed into a mixed state associated with $||\mathbf{r}(t)|| < 1$, as shown in the left plot of Fig.~\ref{Fig:7} and discussed in the following. Hence, the qubit at time $t>0$ is found within the interior of the Bloch sphere.
\end{rem}

To provide a graphical representation of the previous remarks, in Fig.~\ref{Fig:7}, which was originally reported in \cite{Pr:CacCal-19}, we portray the time-domain evolution of a qubit emerging from the same initial state as in Fig.~\ref{Fig:6} and experiencing phase damping.

Specifically, in the left plot of Fig.~\ref{Fig:7}, we portray the qubit's time-domain evolution commencing from the surface of the Bloch sphere and exhibiting phase damping in agreement with \eqref{eq:4.9}, where the vertical coordinate $r_z$ remains unchanged. By contrast, both the horizontal coordinates $r_x$ and $r_y$ exponentially decay toward zero. Consequently, the qubit asymptotically evolves towards the mixed state represented by the point $\mathbf{r}=[0,0,r_z(0)]$. Similarly to Fig.~\ref{Fig:6}, the characteristic orbital evolution in terms of the horizontal coordinates around the vertical axis is a consequence of the linear phase accumulation induced by the system Hamiltonian $H_s$ of \eqref{eq:4.6}. The phase-damping behavior is further characterized by the right plot of Fig.~\ref{Fig:7}, where the Bloch vector coordinates $\mathbf{r}(t)$ as function of the normalized time are portrayed, but after compensating as in \eqref{eq:4.8} the linear phase-rotation around the vertical axis as induced by $H_s$ for explicitly highlighting the spatial selectivity of the phase-damping.

\begin{figure}
	\centering
	\includegraphics[width=.6\columnwidth]{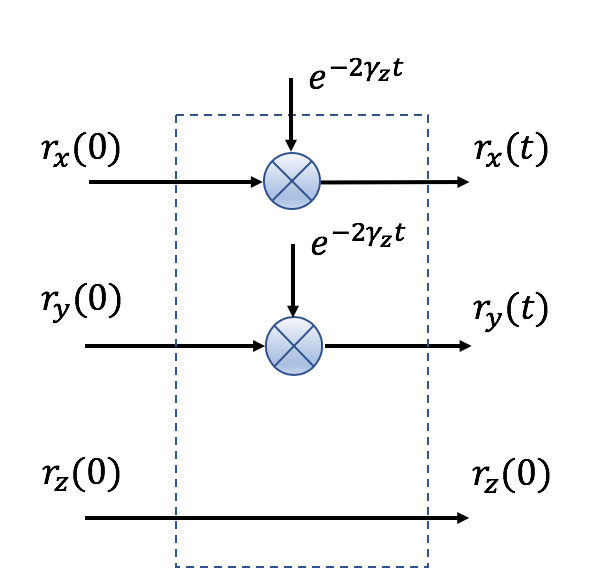}
	\caption{Phase-Damping Model.}
	\label{Fig:8}
 \end{figure}

Hence the phase-damping effects can be modeled from a communications-engineering perspective using the model of Fig.~\ref{Fig:8}. Specifically, the x- and the y-coordinate of the qubit are affected by the multiplicative damping, attenuating each coordinate according to an exponential decay governed by $\gamma_z$, whereas the z-coordinate of the qubit remains unchanged.

For historical reasons, the phase damping is also often referred to as the $T_2$ relaxation process \cite{NieChu-11}, where $T_2$ is the time it takes for a $\ket{+}$ state seen in Fig.~\ref{Fig:01} to flip to a $\ket{-}$ state with the probability of $e^{-1}$ \cite{VanMet-14}. Since the probability that the qubit is flipped is given by $\frac{1-e^{- \gamma_z t}}{2}$ \cite{NieChu-11}, the relationship between the decay rate $\gamma_z$ defined in \eqref{eq:4.4} and $T_2$ can be expressed as: 
\begin{equation}
	T_2= -\frac{1}{ \gamma_z } \ln\left\{\frac{e-2}{e}\right\}.
\end{equation}

\subsection{y-z Damping}
\label{sec:5.3}

\begin{figure}
	\begin{minipage}[c]{0.49\textwidth}
		\centering
		\includegraphics[width=.66\columnwidth]{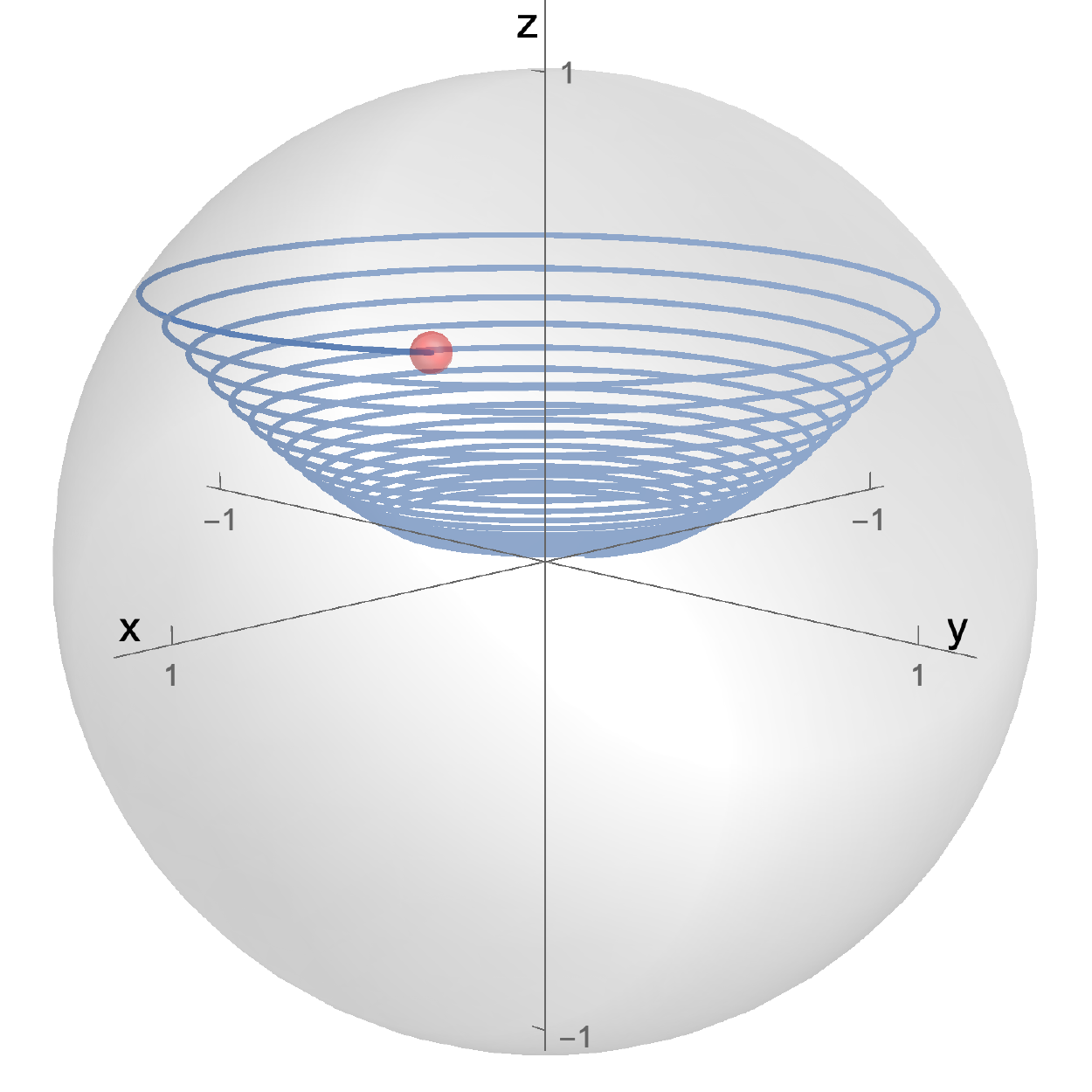}
	\end{minipage}
	\hspace{0.01\textwidth}
	\begin{minipage}[c]{0.49\textwidth}
		\centering
		\includegraphics[width=1\columnwidth]{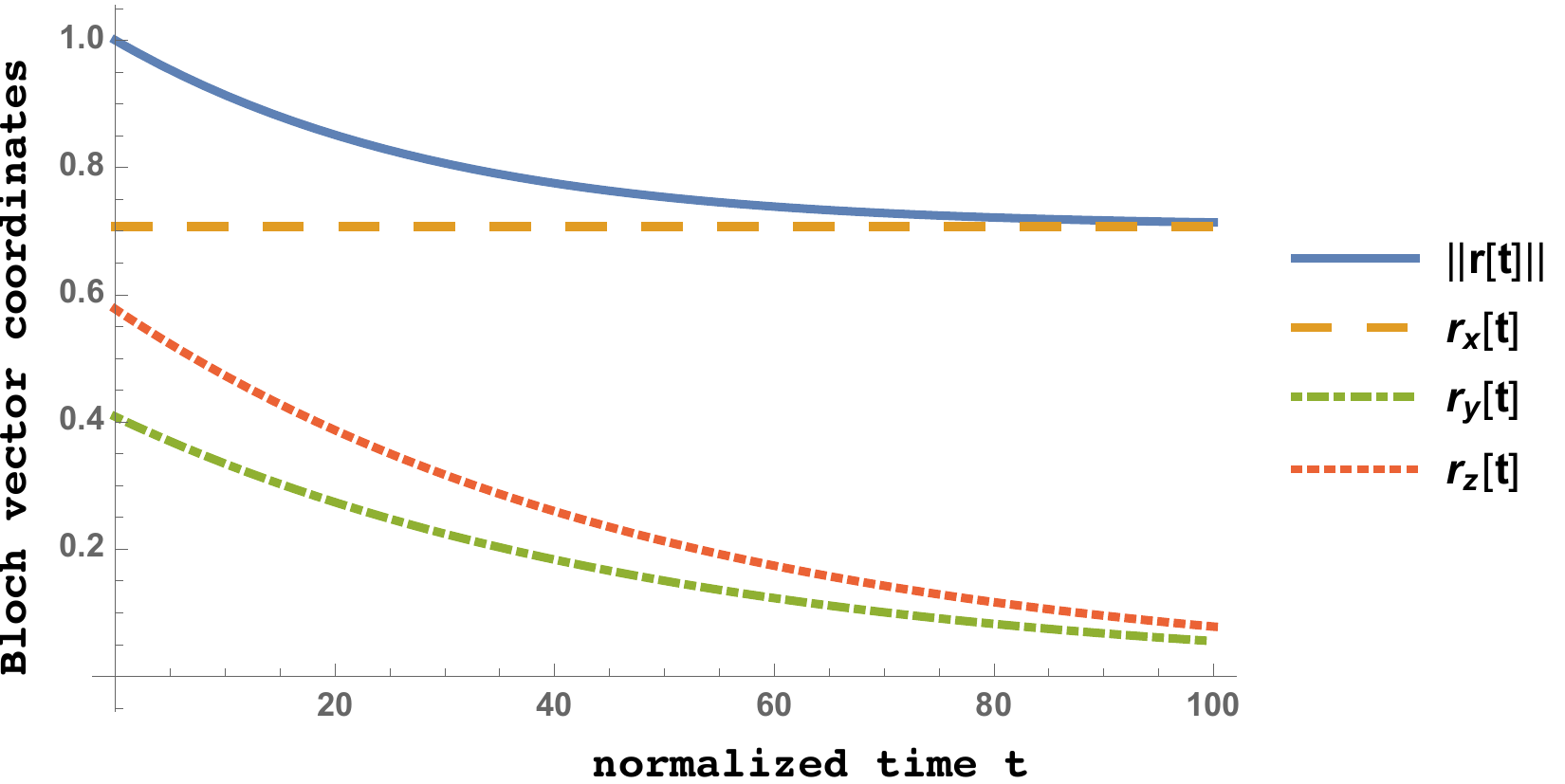}
	\end{minipage}
	\caption{y-z Damping: time-domain evolution of a qubit with Hamiltonian $H_s=\Omega\sigma_z/2$, when emerging from the initial pure state of $\ket{\psi(0)} = \sqrt{\frac{1+ \sqrt{3}}{2\sqrt{3}}} \ket{0} + \frac{i + \sqrt{3}}{2 \sqrt{3+\sqrt{3}}} \ket{1}$ located at the Bloch vector coordinates of $\mathbf{r}(0)=\left[\frac{1}{\sqrt{2}},\frac{1}{\sqrt{6}},\frac{1}{\sqrt{3}}\right]$. Left plot: representation of the qubit's time-domain evolution with respect to the Bloch sphere. Right plot: Bloch vector coordinates $\mathbf{r}(t)$ as a function of time, with the phase evolution engendered by the energy difference $\Omega$ in $H_s$ appropriately compensated, as detailed in the text.}
	\label{Fig:9}
\end{figure}

Let us now consider the scenario, when the impairments are modeled by the Lindblad operator $L_x=\sqrt{\gamma_x}\sigma_{x}$ \cite{Pr:CacCal-19}. In the following, we refer to this type of depolarizing noise process -- also known as \textit{bit-flip error process} -- as \textit{y-z damping} in analogy with Sec.~\ref{sec:5.2} and for the reasons to be further highlighted later in Remark~\ref{rem_Xnoise}.

By solving \eqref{eq:4.3} we may characterize the time-domain evolution of the system $\mathcal{S}$ as in \eqref{eq:4.10}, shown at the top of the next page. To elaborate further on \eqref{eq:4.10}, we have compensated the phase evolution engendered by $H_s$ as in Sec.~\ref{sec:5.2} for explicitly highlighting these impairments. 

\begin{figure*}[h!]
	\begin{equation}
		\label{eq:4.10}
		\rho_{\mathcal{S}}(t) = \frac{1}{2}\begin{bmatrix}
			\rho_{\mathcal{S}}^{00}(0)(1+e^{-2\gamma_x t})+\rho_{\mathcal{S}}^{11}(0)(1-e^{-2\gamma_x t}) & \rho_{\mathcal{S}}^{01}(0)(1+e^{-2\gamma_x t})+\rho_{\mathcal{S}}^{10}(0)(1-e^{-2\gamma_x t}) \\
			\rho_{\mathcal{S}}^{10}(0)(1+e^{-2\gamma_x t})+\rho_{\mathcal{S}}^{01}(0)(1-e^{-2\gamma_x t}) & \rho_{\mathcal{S}}^{11}(0)(1+e^{-2\gamma_x t})+\rho_{\mathcal{S}}^{00}(0)(1-e^{-2\gamma_x t})
		\end{bmatrix}
	\end{equation}
\end{figure*}

Similarly to phase damping, by exploiting \eqref{eq:2.17} as well as \eqref{eq:2.18}, and accounting for \eqref{eq:4.10}, after some further algebraic manipulations, the Bloch vector coordinates $\mathbf{r}(t)=[r_x(t),r_y(t),r_z(t)]$ of the qubit subject to y-z damping evolve in the time-domain according to the following relationship:
\begin{equation}
	\label{eq:4.11}
	\begin{split}
		&r_x(t)=r_x(0), \\
		&r_y(t)=r_y(0) e^{-2 \gamma_x t},\\
		&r_z(t)=r_z(0) e^{-2 \gamma_x t},
	\end{split}
\end{equation}
where $\mathbf{r}(0)=[r_x(0),r_y(0),r_z(0)]$ represents the Bloch vector at time instant $0$.

\begin{rem}
\label{rem_Xnoise}
Similarly to phase damping, observe in \eqref{eq:4.11} that the y-z damping effects are also multiplicative impairments imposed on the Bloch vector coordinates of the quantum state. Furthermore and similarly to the phase damping, the y-z damping exhibits an asymmetric behavior, since it affects the individual Bloch vector coordinates differently. Specifically, it damps exponentially both the y- and the z-coordinate, i.e. $r_y$ and $r_z$, while it leaves the x-coordinate, i.e. $r_x$, unaffected. This phenomenon might be interpreted again as a sort of \textit{spatial selectivity} in terms of the Bloch vector coordinates. Moreover, similarly to phase damping, the initial pure state $\ket{\psi(0)}$ is transformed into a mixed state associated with $||\mathbf{r}(t)|| < 1$ for $t>0$. 
\end{rem}

\begin{rem}
\label{rem_Xnoise_bis}
We note furthermore that although the y-z damping is modelled by the Pauli-X gate of Table~\ref{Tab:01} via the Lindblad operator $L_x=\sqrt{\gamma_x}\sigma_{x}$, we should not confuse the unitary evolution imposed by the Pauli-X gate and highlighted in equation \eqref{eq:2.19} with the non-unitary evolution induced by $L_x$ in \eqref{eq:4.3}. More explicitly, similarly to the unitary evolution imposed by the Pauli-X gate in \eqref{eq:2.19}, the y-z damping leaves the x-coordinate unaffected, but in contrast to the bit-flipping imposed by the Pauli-X gate based unitary evolution, the non-unitary evolution represented by $L_x$ resulted in an exponential damping of the y- and the z-coordinate rather than flipping them. 
\end{rem}

To visualize the above discussion, in Fig.~\ref{Fig:9} we portray the time-evolution of a qubit emerging from the same initial state as in Figs.~\ref{Fig:6} as well as \ref{Fig:7} and subject to y-z damping. 

Specifically, in the left plot, we portray the qubit's time-domain evolution commencing from the surface of the Bloch sphere. But then as time elapses, we observe that all the coordinates decay to zero, which seems to contradict \eqref{eq:4.11}. However, this contradiction is only illusory, since in the left plot, we do not compensate the orbital evolution of the x- and y-coordinate around the z-axis induced by $\Omega$ in $H_s$. This is confirmed by the right plot of Fig.~\ref{Fig:9}, where we have appropriately compensated this phase evolution as in \eqref{eq:4.8} for explicitly highlighting the spatial selectivity of the y-z damping. In agreement with \eqref{eq:4.11}, the x-coordinate remains unaltered, whereas the y- and the z-coordinates exponentially decay towards zero with a decay-rate of $\gamma_x$. Hence, the qubit asymptotically evolves towards the mixed state represented by the point $\mathbf{r}=[r_x(0),0,0]$.

 \begin{figure}
	\centering
	\includegraphics[width=0.6\columnwidth]{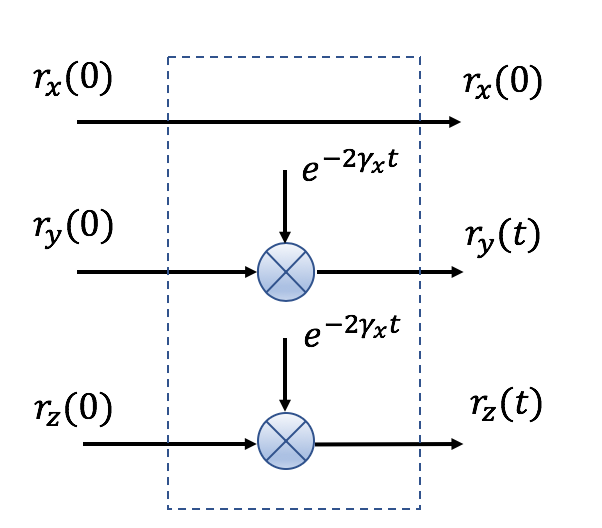}
	\caption{y-z Damping Model.}
	\label{Fig:10}
 \end{figure}
 
Based on the above discussions, the y-z damping effects can be modeled from a communications-engineering perspective using the model of Fig.~\ref{Fig:10}. Specifically, the y- and the z-coordinate of the qubit are affected by a multiplicative damping, attenuating each coordinate according to an exponential decay governed by $\gamma_x$, whereas the x-coordinate of the qubit remains unchanged.

\subsection{Combined y-z-Phase Damping}
\label{sec:5.4}

Based on the discussions of Secs.~\ref{sec:5.3} and \ref{sec:5.4}, the spontaneous question arises: ``is the multiplicative nature of the quantum impairments modeled in \eqref{eq:4.3} general?''

To gain further insights concerning this question, we analyze the behavior of the composite impairments, namely when the impairments are modeled by a combination of the two Lindblad operators $L_x=\sqrt{\gamma_x}\sigma_{x}$ and $L_z=\sqrt{\gamma_z}\sigma_{z}$. In the following, we refer to this type of depolarizing process as \textit{combined y-z-phase damping} in analogy with Secs.~\ref{sec:5.2}-\ref{sec:5.3}.

As before, by compensating for the phase evolution induced by the Hamiltonian $H_s=\frac{\Omega}{2} \sigma_z$, commencing from \eqref{eq:4.3}, after some algebraic manipulations we arrive at:
\begin{equation}
	\label{eq:4.12}
	\begin{split}
		&r_x(t)=r_x(0) e^{-2 \gamma_z t}, \\
		&r_y(t)=r_y(0) e^{-2 (\gamma_x+\gamma_z) t}, \\
		&r_z(t)=r_z(0) e^{-2 \gamma_x t}.
	\end{split}
\end{equation}

\begin{rem}
\label{rem_XZnoise}
Similarly to the previous scenarios, observe in \eqref{eq:4.12} that the combined y-z-phase damping effects impose multiplicative impairments on the Bloch vector coordinates of the quantum state. Furthermore, the combined y-z-phase damping exhibits an asymmetric behavior, since it affects the individual coordinates of the Bloch vector differently. Specifically, it damps all the three coordinates exponentially, but at different decay rates. This phenomenon might again be interpreted as a \textit{spatial selectivity} in terms of Bloch vector coordinates, with the initial pure state $\ket{\psi(0)}$ being transformed into a mixed state. 
\end{rem}

\begin{rem}
\label{rem_XZnoise_bis}
Similarly to the previous subsections, we should avoid confusing the unitary evolution imposed by a sequence of the Pauli-X and Pauli-Z gates with the non-unitary evolution induced by a combination of the two Lindblad operators $L_x$ and $L_z$ present in \eqref{eq:4.3}.
\end{rem}

\begin{figure}
	\begin{minipage}[c]{0.49\textwidth}
		\centering
		\includegraphics[width=.66\columnwidth]{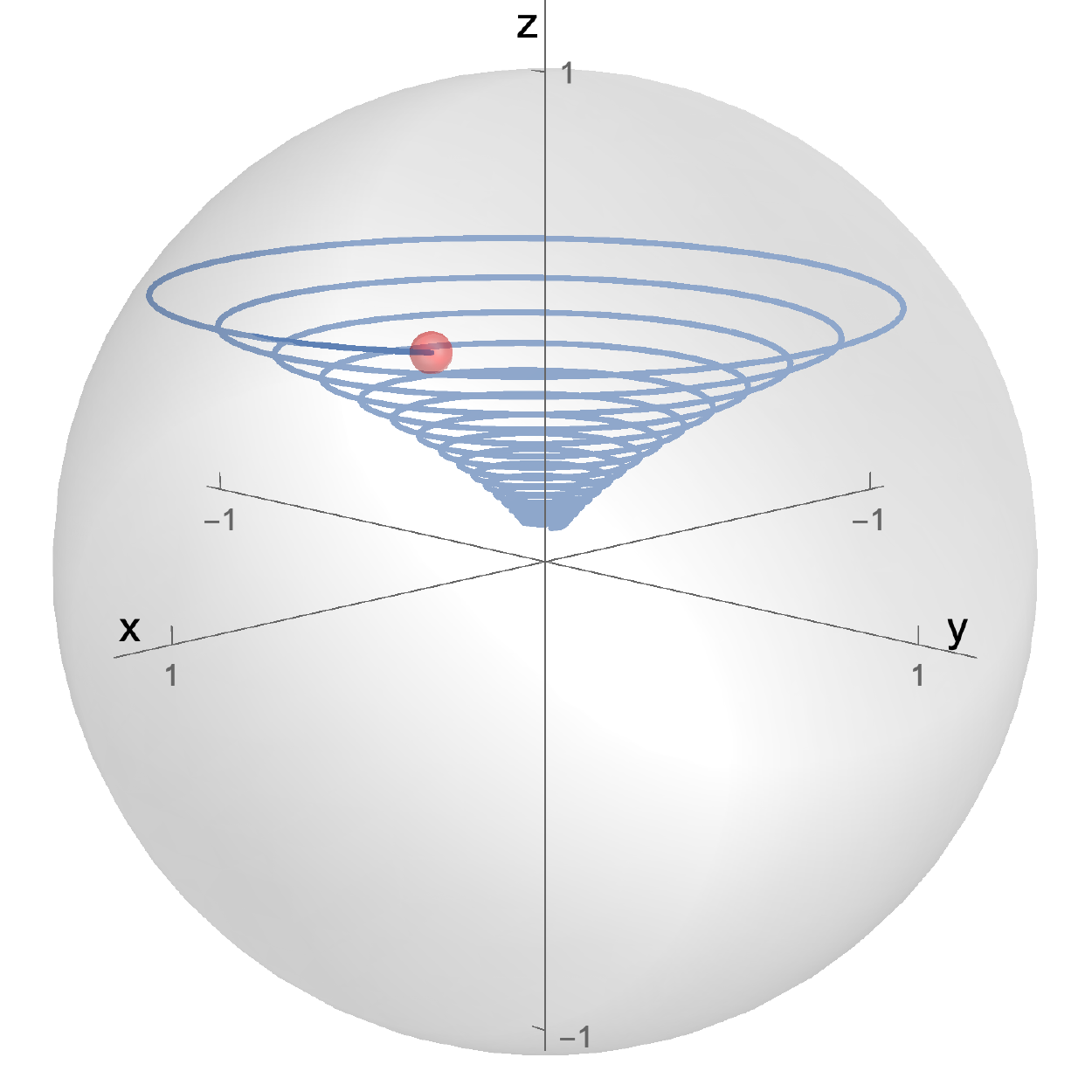}
	\end{minipage}
	\hspace{0.01\textwidth}
	\begin{minipage}[c]{0.49\textwidth}
		\centering
		\includegraphics[width=1\columnwidth]{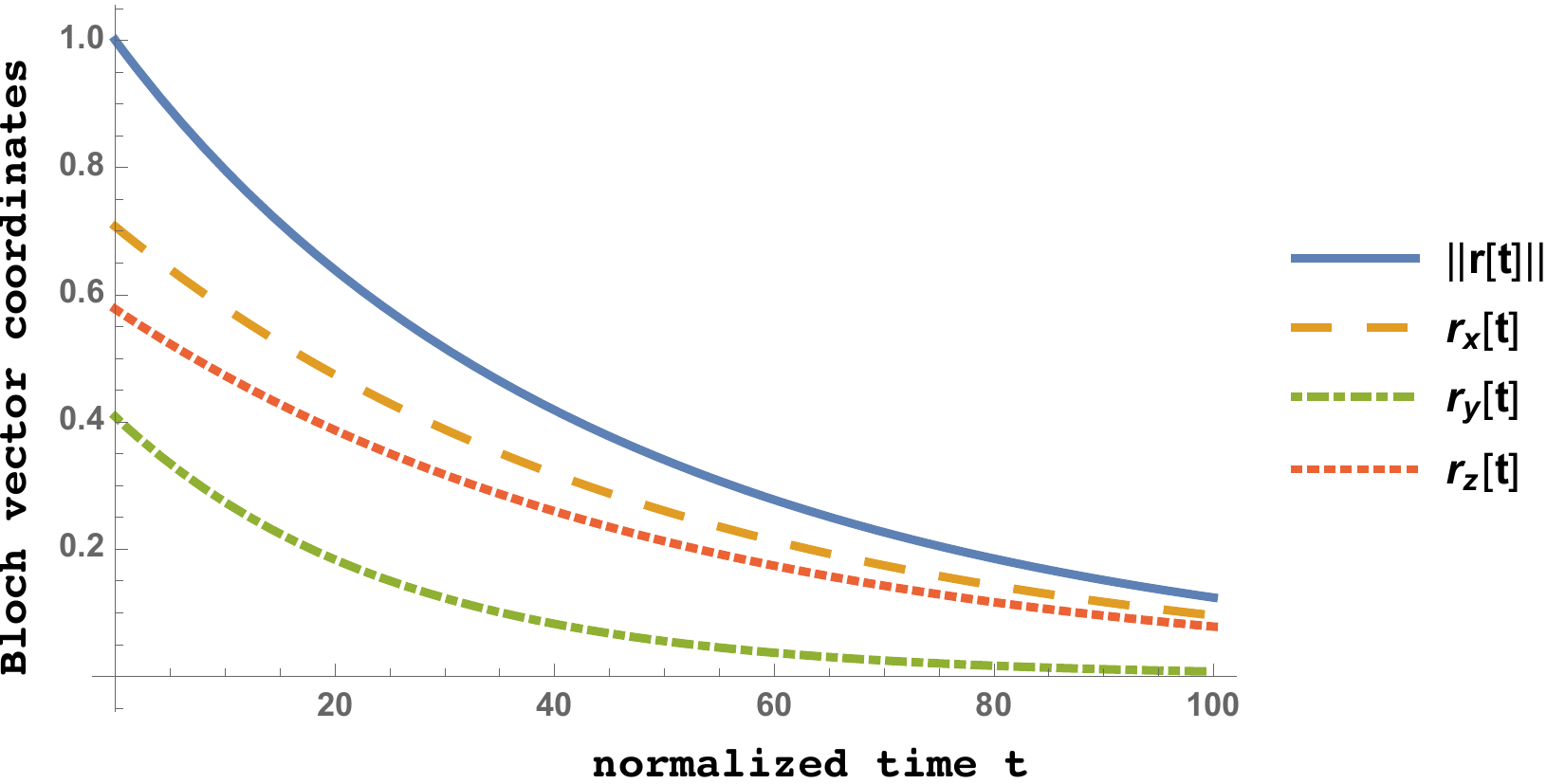}
	\end{minipage}
	\caption{Combined y-z-phase damping: time-domain evolution of a qubit subject to the Hamiltonian $H_s=\Omega\sigma_z/2$. when emerging from initial pure state of $\ket{\psi(0)}= \sqrt{\frac{1+ \sqrt{3}}{2\sqrt{3}}} \ket{0} + \frac{i + \sqrt{3}}{2 \sqrt{3+\sqrt{3}}} \ket{1}$ located at the Bloch vector coordinates $\mathbf{r}(0)=\left[\frac{1}{\sqrt{2}},\frac{1}{\sqrt{6}},\frac{1}{\sqrt{3}}\right]$. Left plot: representation of the qubit's time evolution with respect to the Bloch sphere. Right plot: Bloch vector coordinates $\mathbf{r}(t)$ as a function of time, with the phase evolution engendered by the energy difference $\Omega$ in $H_s$ appropriately compensated, as detailed in the text.}
	\label{Fig:11}
\end{figure}

The multiplicative nature of the combined impairments is further confirmed by Fig.~\ref{Fig:11}, showing the time evolution of a qubit subject to combined y-z-phase damping. Specifically, as done in the previous subsections, we consider a qubit emerging from the initial pure state of $\ket{\psi(0)}$ whose Bloch vector coordinates are $\mathbf{r}(0)=\left[\frac{1}{\sqrt{2}},\frac{1}{\sqrt{6}},\frac{1}{\sqrt{3}}\right]$. The left plot shows the representation of the qubit's time-domain evolution starting from the surface of the Bloch sphere. Similarly to the left plot of Fig.~\ref{Fig:9}, all the Bloch vector coordinates evolve towards zero. However, by comparing the two figures, we recognize that in the y-z-phase damping scenario, the evolution of the quantum state towards the center of the sphere is faster. This is in agreement with \eqref{eq:4.12}. In fact, as shown by the right plot in Fig.~\ref{Fig:11}, by appropriately compensating the phase evolution induced by $H_s$, all the Bloch coordinates exponentially decay toward zero, with a decay-rate of: i) either $\gamma_z$ or $\gamma_x$ for the x- and z-coordinate, respectively; ii) the sum of $\gamma_x$ and $\gamma_z$ for the y-coordinate.

\begin{figure}
	\begin{minipage}[c]{0.49\textwidth}
		\centering
		\includegraphics[width=0.6\columnwidth]{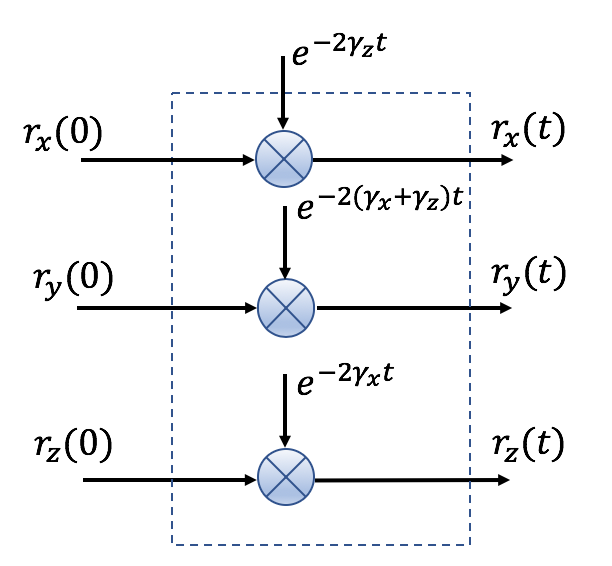}
		\caption{Combined y-z-Phase Damping Model.}
		\label{Fig:12}
	\end{minipage}
	\hspace{0.01\textwidth}
	\begin{minipage}[c]{0.49\textwidth}
		\centering
		\includegraphics[width=0.6\columnwidth]{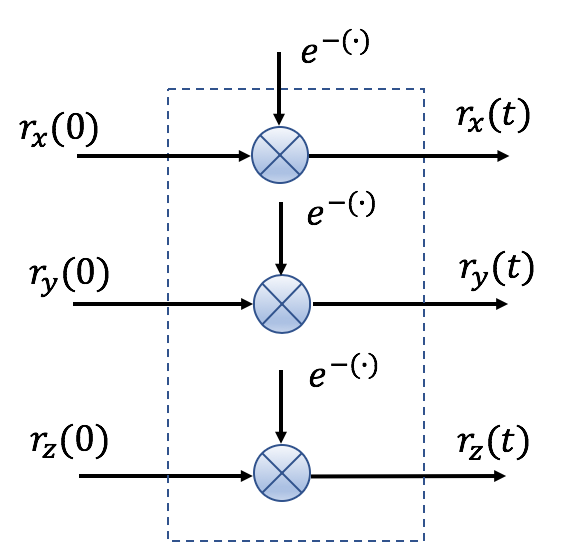}
		\caption{Arbitrary Decoherence Process Model.}
		\label{Fig:13}
	\end{minipage}
\end{figure}

Based on the above discussions, the combined y-z-phase damping effects can be modeled from a communications-engineering perspective using the model of Fig.~\ref{Fig:12}. Specifically, all the qubit coordinates are affected by the multiplicative damping, with an exponential decay rate that differs among the different coordinates.

Based on the reasoning above, it may be readily shown that a quantum impairment modeled by the combination of all the three Lindblad operators -- $L_x=\sqrt{\gamma_x} \sigma_x$, $L_y=\sqrt{\gamma_y} \sigma_y$ and $L_z=\sqrt{\gamma_z} \sigma_z$ -- changes the Bloch vector coordinates of a qubit according to the following relationship:
\begin{equation}
	\label{eq:4.11_bis}
	\begin{split}
		&r_x(t)=r_x(0) e^{-2 (\gamma_y+\gamma_z) t}, \\
		&r_y(t)=r_y(0) e^{-2 (\gamma_x+\gamma_z) t}, \\
		&r_z(t)=r_z(0) e^{-2 (\gamma_x+\gamma_y) t}.
	\end{split}
\end{equation}
Consequently, all the previous considerations continue to hold.

\begin{rem}
The above analysis revels that the decoherence affects the three spatial directions by a similar mechanism, namely by multiplying the Bloch coordinates of the qubit with an exponential function. However, the exponent is different for the three coordinates, dictated by the complex interaction between the system and the environment. From a communications engineering perspective, the decoherence may be represented as in Fig.~\ref{Fig:13}. 
\end{rem}


\section{IBM Q Experimental Results}
\label{sec:6}
In this section we report the results of teleportation experiments relying on a real quantum computer through the IBM Q platform \cite{Cas-17}. Our objective is to gain experimental insights concerning the composite quantum impairments affecting the teleported qubit.

Specifically, we perform quantum process tomography \cite{ChoGamCor-12,AltJefKwi-05,Blu-10,MahRoz-13,OkaSat-16} for fully characterizing the dynamics of teleportation. At the time of writing the IBM Q project does not allow us to account for the channel effects within the entanglement distribution. Nevertheless, the experiments allow us to evaluate the deleterious effects of cumulative impairments encountered during entanglement generation, as well as by imperfect quantum gates at both the source and destination, and finally the decoherence effects.

For the quantum circuit depicted in Figure~\ref{Fig:03}, we perform over 8 million quantum process tomography\footnote{Quantum process tomography is sensitive to the so called state preparation and measurement (SPAM) errors. Hence, the experiment results discussed in the following are inevitably affected also by SPAM errors.} experiments using the 5-qubits IBM Tenerife \textit{ibmqx4} quantum processor during the period of January 14th to January 19th 2019. The quantum state to be teleported was placed in the chip-qubit $q[0]$ and the EPR pair was created\footnote{By applying a H gate on $q[0]$, followed by a CNOT gate on $q[1]$ with $q[0]$ as control.} between chip-qubits $q[1]$ and $q[2]$. 

\subsection{Teleporting a basis state}
\label{sec:6.1}

\begin{figure*}
	\begin{minipage}[c]{0.32\textwidth}
		\centering
		\includegraphics[width=1\columnwidth]{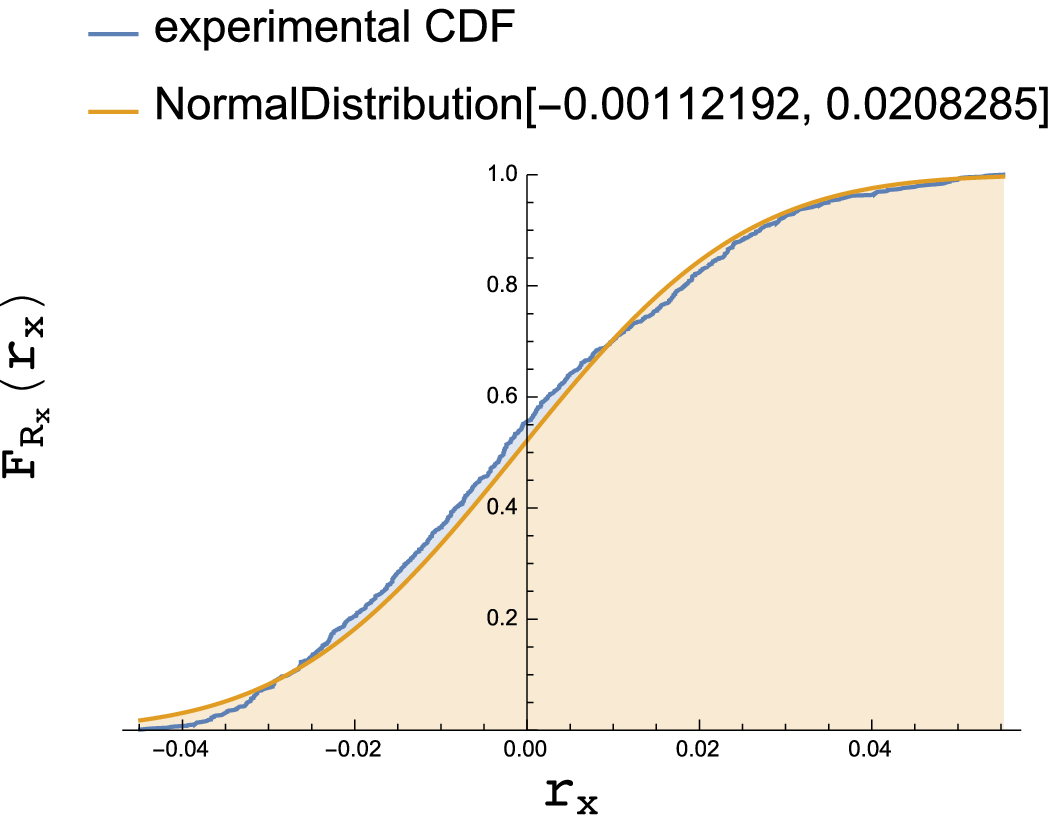}
		\subcaption{Marginal CDF $F_{R_x}(r_x)$ of the Bloch x-coordinate.}
		\label{Fig:B1}
	\end{minipage}
	\hspace{0.005\textwidth}
	\begin{minipage}[c]{0.32\textwidth}
		\centering
		\includegraphics[width=1\columnwidth]{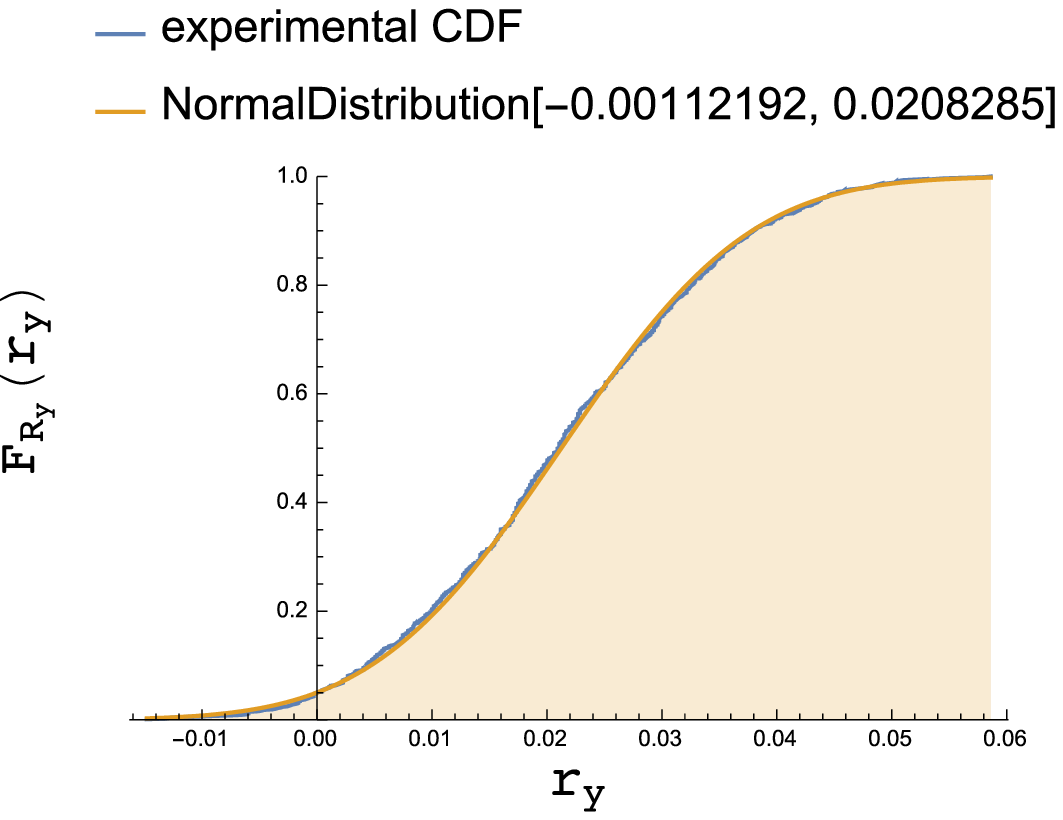}
		\subcaption{Marginal CDF $F_{R_y}(r_y)$ of the Bloch y-coordinate.}
		\label{Fig:B2}
	\end{minipage}
	\hspace{0.005\textwidth}
	\begin{minipage}[c]{0.32\textwidth}
		\centering
		\includegraphics[width=1\columnwidth]{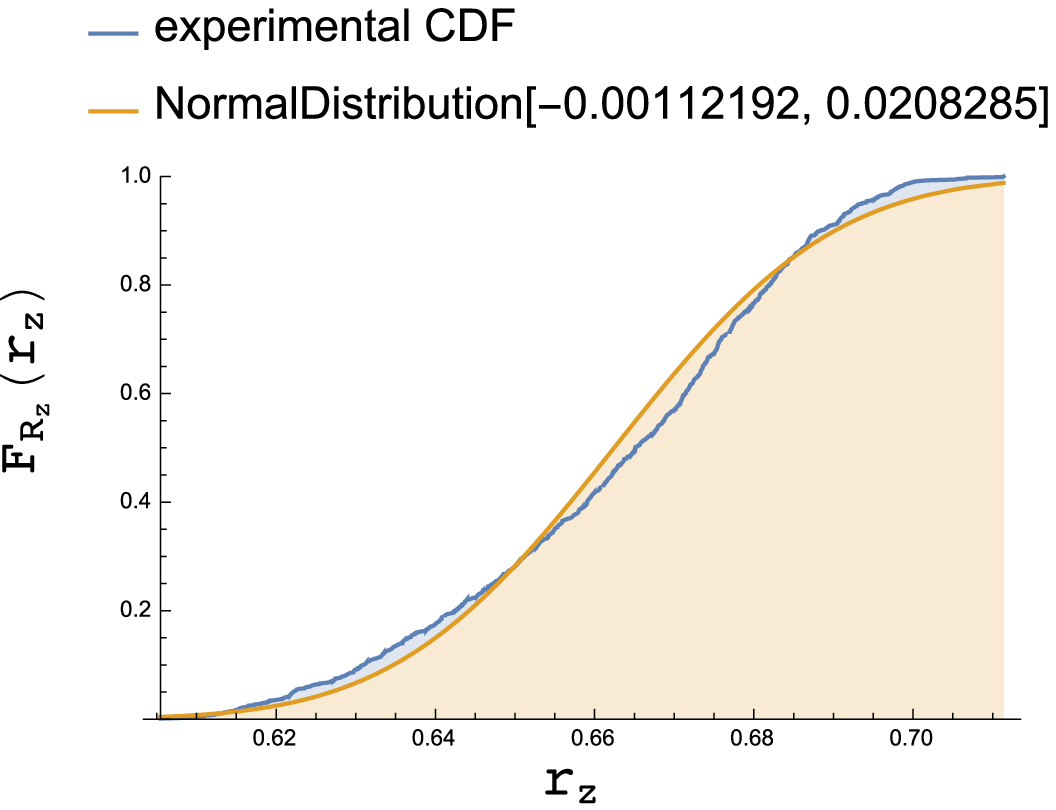}
		\subcaption{Marginal CDF $F_{R_z}(r_z)$ of the Bloch z-coordinate.}
		\label{Fig:B3}
	\end{minipage}\\
	\begin{minipage}[c]{0.49\textwidth}
		\centering
		\includegraphics[width=.98\columnwidth]{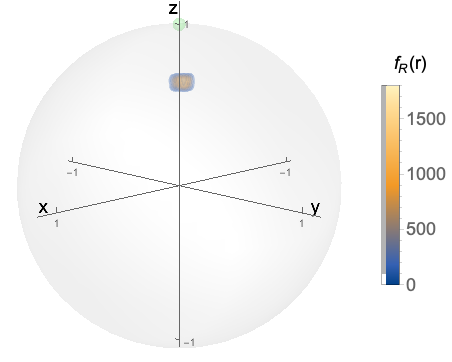}
		\subcaption{Density plot of the experimental joint PDF $f_{\mathbf{R}}(\mathbf{r})$ of the Bloch vector $\mathbf{r}=[r_x,r_y,r_z]$.}
		\label{Fig:B4}
	\end{minipage}
	\hspace{0.01\textwidth}
	\begin{minipage}[c]{0.49\textwidth}
		\centering
		\includegraphics[width=1\columnwidth]{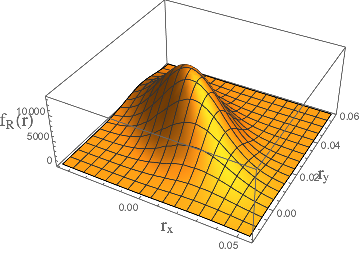}
		\subcaption{Density plot of the experimental joint PDF $f_{\mathbf{R}}(\mathbf{r})$ of the Bloch vector $\mathbf{r}=[r_x,r_y,r_z]$ when $r_z = E[R_z]$.} 
		\label{Fig:B5}
	\end{minipage}
	\caption{Teleportation with 5-qubits IBM Tenerife \textit{ibmqx4} quantum processor between chip-qubits $q[0]$ and $q[2]$. Initial pure state $\ket{\psi} = \ket{0}$ located at the Bloch vector coordinates $\mathbf{r}=\left[0,0,1\right]$ (green dot in Fig.~\ref{Fig:B4}).}
	\label{Fig:B}
\end{figure*}

In Figure~\ref{Fig:B}, we evaluate the effects of the cumulative quantum impairments when the qubit to be teleported is the pure state $\ket{\psi} = \ket{0}$. In the absence of impairments, the teleported qubit would coincide with the original qubit $\ket{0}$, and it will be placed at the Bloch vector coordinates of $[0,0,1]$ (green dot in Fig.~\ref{Fig:B4}). However, as discussed in Sec.~\ref{sec:5}, real quantum systems constitute open physical systems -- where decoherence arises due to the interactions with the environment. Hence, the teleported qubit differs from the original qubit $\ket{0}$ and, in agreement with the theoretical results of Sec.~\ref{sec:5}, the teleported qubit is no longer in a pure state, but it is rather in a mixed state laying inside the Bloch sphere.

Specifically, we characterize the effects of the cumulative impairments using the marginal Cumulative Distribution Functions (CDFs) of the Bloch vector coordinates of the teleported state, as seen in Figs.~\ref{Fig:B1}-\ref{Fig:B3}.

Regarding the Bloch x-coordinate, Fig.~\ref{Fig:B1} shows its marginal CDF $F_{R_x}(r_x)$, where, by definition, $F_{R_x}(r_x) \eqdef P\left( R_x \leq r_x\right)$, i.e., $F_{R_x}(r_x)$ is the probability of the x-coordinate of the teleported state being smaller than or equal to $r_x$. We observe that, with probability roughly equal to one, the x-coordinate assumes values within the interval $(-0.044, 0.056)$, with an average value $\mu_x$ around $0$. Furthermore, the figure also shows the distribution that better fits with the experimental data, obtained with the aid of the Mathematica \textit{FindDistribution} and \textit{EstimatedDistribution} packages. Observe that the x-coordinate of the teleported qubit can be roughly modeled by the normal distribution $\mathcal{N}(\mu_x,\sigma_x)$ associated with $\mu_x \simeq 0$ and $\sigma_x \simeq 0.021$. However, we note that rigorous distribution fitting theory would result in a low confidence metric, because there are intervals, where there is a consistent shape-deviation between the theory and measurement. 

Regarding the Bloch y-coordinate, whose CDF is shown in Fig.~\ref{Fig:B2}, the results show that it can also be roughly modeled by a normal distribution with a variance of $\sigma_y \simeq 0.013$. However, in this case, there is a slight drifting toward positive values, with an average value $\mu_y$ of around $0.02$.

Finally, regarding the Bloch z-coordinate whose CDF is shown in Fig.~\ref{Fig:B3}, we observe that the z-coordinate of the teleported qubit shrinks from the original value of $1$ to an average value $\mu_z \simeq 0.66$. Furthermore, with a probability of approximately zero, the z-coordinate assumes values outside the interval $(0.605, 0.712)$ and it can be modeled by a normal distribution having a variance of $\sigma_z \simeq 0.022$.

The drift of the qubit from the original pure state $\ket{0}$ to a mixed state laying in the interior of the Bloch sphere becomes evident in Fig.~\ref{Fig:B4}, where we report the density plot of the experimental joint Probability Density Function (PDF) of the Bloch vector coordinates of the teleported qubit. Specifically, the experimental data indicate that the teleported qubit lays within a sphere roughly centered at coordinates of $[\mu_x=0,\mu_y=0.02,\mu_z=0.66]$ with radius around $0.1$. The probability of obtaining a value within such a sphere is clearly given by the joint PDF depicted in Fig.~\ref{Fig:B4}.

We note furthermore that the effect of the quantum impairments on the different coordinates is not independent. Indeed, as shown in Fig.~\ref{Fig:B5}, there exists a correlation between the effects inflicted upon the x- and y-coordinate. Specifically, a drift toward positive values of the x-coordinate is coupled with a drift toward lower values of the y-coordinate.

\begin{rem}
\label{rem:Exp_basis}
This first set of experiments seems to suggest, in agreement with the theoretical analysis developed in Sec.~\ref{sec:5}, that the cumulative impairments are multiplicative, resulting in a pure state being transformed into a mixed state. However, to gain more general insights into the composite quantum impairments, in the next sub-section we carry out a quantum process tomography experiment for characterizing the dynamics of the teleportation of superposed quantum states.
\end{rem}

\subsection{Teleporting superposed states}
\label{sec:6.2}

\begin{figure*}
	\begin{minipage}[c]{0.32\textwidth}
		\centering
		\includegraphics[width=1\columnwidth]{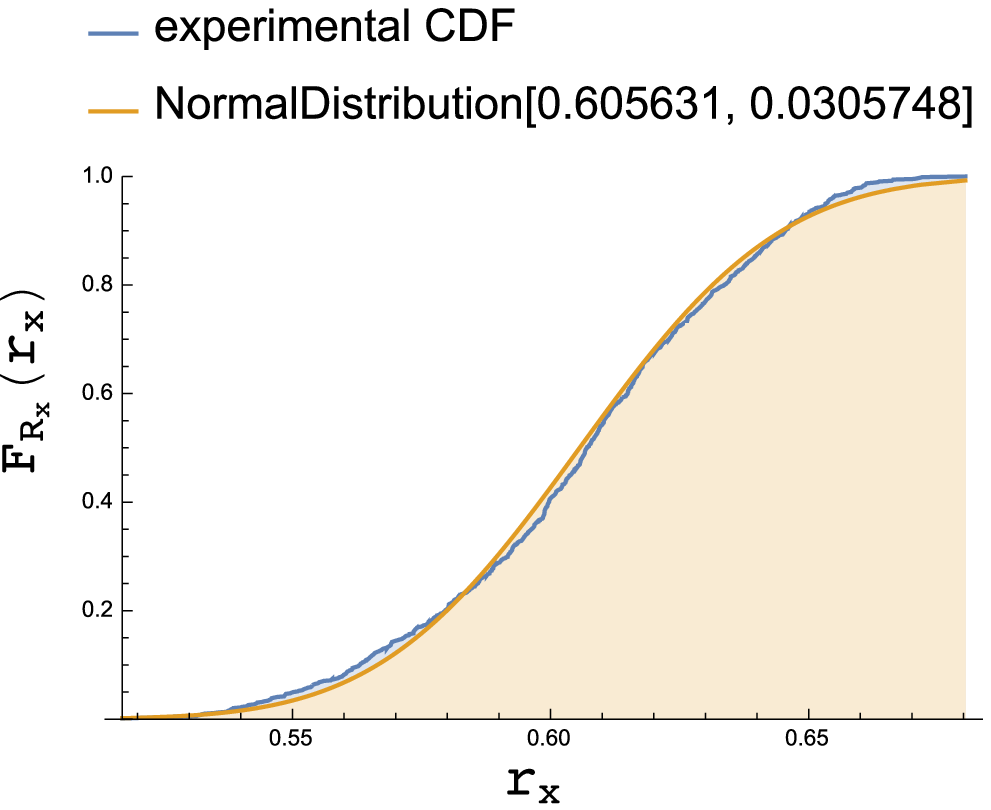}
		\subcaption{Marginal CDF $F_{R_x}(r_x)$ of the Bloch x-coordinate.}
		\label{Fig:C1}
	\end{minipage}
	\hspace{0.005\textwidth}
	\begin{minipage}[c]{0.32\textwidth}
		\centering
		\includegraphics[width=1\columnwidth]{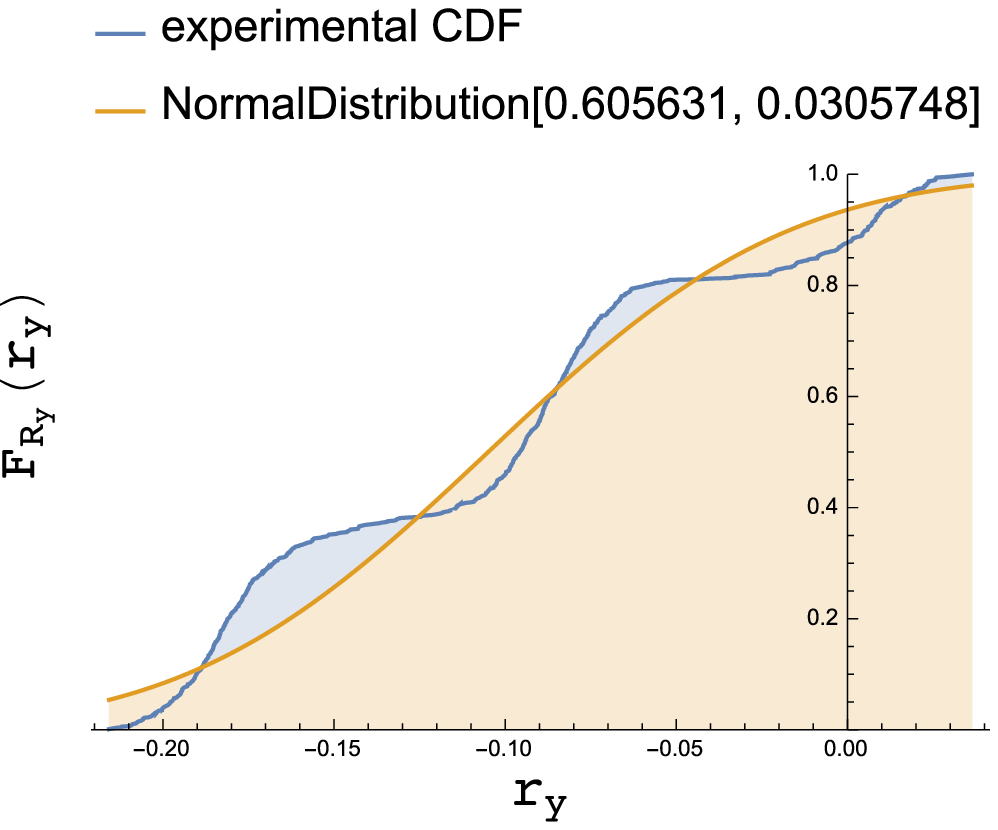}
		\subcaption{Marginal CDF $F_{R_y}(r_y)$ of the Bloch y-coordinate.}
		\label{Fig:C2}
	\end{minipage}
	\hspace{0.005\textwidth}
	\begin{minipage}[c]{0.32\textwidth}
		\centering
		\includegraphics[width=1\columnwidth]{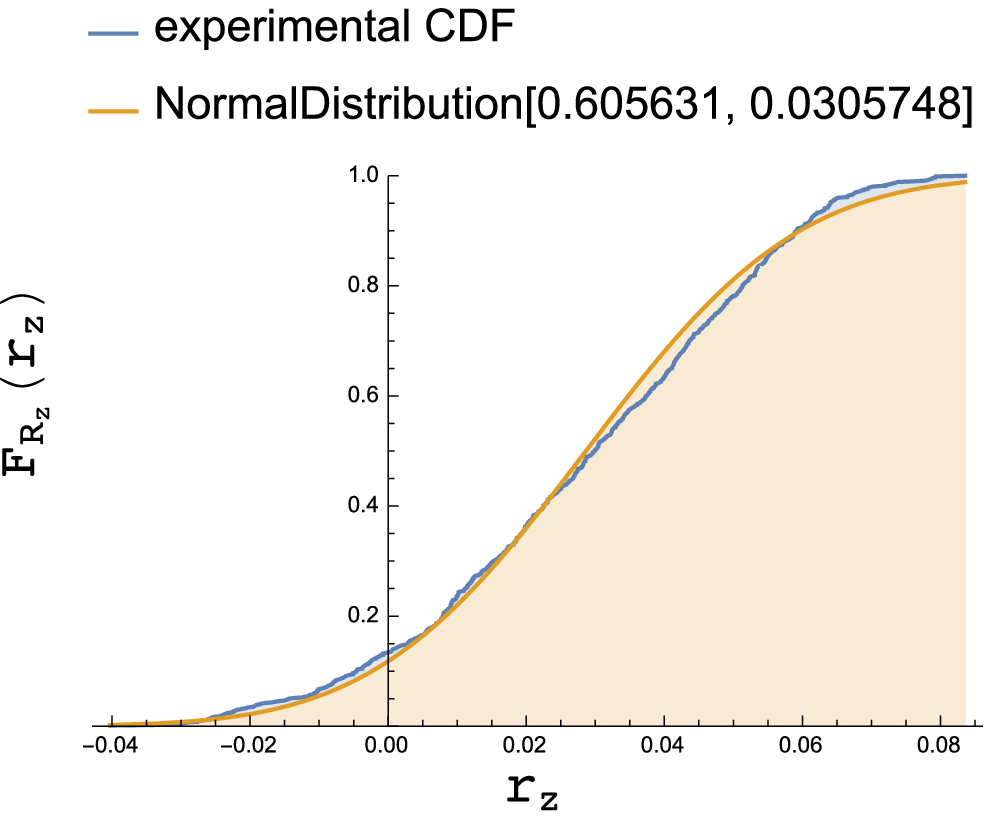}
		\subcaption{Marginal CDF $F_{R_z}(r_z)$ of the Bloch z-coordinate.}
		\label{Fig:C3}
	\end{minipage}\\
	\begin{minipage}[c]{0.49\textwidth}
		\centering
		\includegraphics[width=.98\columnwidth]{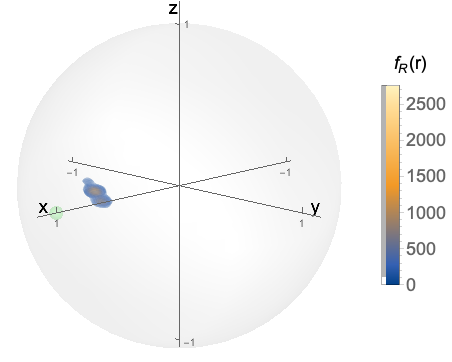}
		\subcaption{Density plot of the experimental joint PDF $f_{\mathbf{R}}(\mathbf{r})$ of the Bloch vector $\mathbf{r}=[r_x,r_y,r_z]$.}
		\label{Fig:C4}
	\end{minipage}
	\hspace{0.01\textwidth}
	\begin{minipage}[c]{0.49\textwidth}
		\centering
		\includegraphics[width=1\columnwidth]{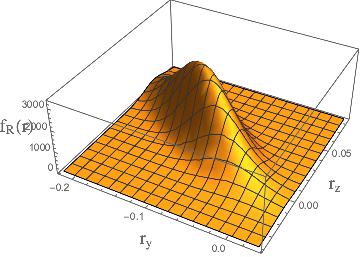}
		\subcaption{Density plot of the experimental joint PDF $f_{\mathbf{R}}(\mathbf{r})$ of the Bloch vector $\mathbf{r}=[r_x,r_y,r_z]$ when $r_x = E[R_x]$.} 
		\label{Fig:C5}
	\end{minipage}
	\caption{Teleportation with 5-qubits IBM Tenerife \textit{ibmqx4} quantum processor between chip-qubits $q[0]$ and $q[2]$. Initial pure state $\ket{\psi} = \ket{+}$ with Bloch vector coordinates $\mathbf{r}=\left[1,0,0\right]$ (green dot in Fig.~\ref{Fig:C4}).}
	\label{Fig:C}
\end{figure*}

In Figure~\ref{Fig:C}, we evaluate the effects of the cumulative quantum impairments when the qubit to be teleported is in the pure state of $\ket{\psi} = \ket{+} = (\ket{0} + \ket{1})/\sqrt{2}$. Ideally, the teleported qubit would coincide with the original qubit $\ket{+}$ placed at the Bloch vector coordinates of $[1,0,0]$ (green dot in Fig.~\ref{Fig:C4}). However, due to the impairments and in agreement with the theoretical results of Sec.~\ref{sec:5}, the teleported qubit is corrupted into a mixed state laying inside the Bloch sphere.

Let us now focus our attention on Figs.~\ref{Fig:C1}-\ref{Fig:C3}, where we characterize the cumulative impairments through the marginal CDFs of the Bloch vector coordinates of the teleported qubit and the results of Sec.~\ref{sec:6.1} are confirmed. Specifically, Fig.~\ref{Fig:C1} also provides clear evidence of the multiplicative nature of the cumulative impairments: the x-coordinate of the teleported qubit shrinks from the original value of $1$ to an average value of about $\mu_x=0.61$. Furthermore, the probability that the x-coordinate of the teleported qubit assumes values outside the interval $(0.517, 0.681)$ is approximately zero. To elaborate a little further, similarly to Sec.~\ref{sec:6.1}, the experimental distributions of the x-coordinate and the z-coordinate seen in Fig.~\ref{Fig:C3} are reminiscent of normal distributions, with parameters $(\mu_x=0.61, \sigma_x=0.03)$ and $(\mu_z=0.03, \sigma_x=0.02)$, respectively.

Regarding the y-coordinate, whose CDF is shown in Fig.~\ref{Fig:C2}, we observe a slight drifting toward negative values and the distribution exhibits three local peaks roughly at $-0.18$, $-0.09$, and $0.01$. We believe the peaks are induced by the quantum device calibration procedures \cite{Blu-19}. Indeed, the experimental campaign lasted six days, and roughly once a day the IBM chip went off-line for full calibration. If we limit the analysis to the data collected between two consecutive calibrations, we might surmise that each coordinate, including the y-coordinate, can be roughly modeled by a normal distribution. Furthermore, it is worthwhile observing that, according to the Mathematica \textit{FindDistribution} and \textit{EstimatedDistribution} packages, the normal distribution has the best fit with the experimental data, given the limited set of hypothesis-distributions, even though the rigorous goodness-of-fit metrics are low.  
Further research is needed to fully understand the effects of the calibration procedures on the cumulative impairments.

The peaks along the y-coordinate can be easily spotted in Fig.~\ref{Fig:C4}, where we portray the density plot of the experimental joint PDF of the Bloch vector coordinates of the teleported state. Furthermore, similarly to Sec.~\ref{sec:6.1}, the effects of the cumulative impairments on the different coordinates are not independent. For instance, as shown in Fig.~\ref{Fig:C5}, there exists a correlation between the y- and z-coordinate, where the positive values of the z-coordinate are more likely to be associated with lower values of the y-coordinate.

\begin{rem}
This second set of experiments confirms Remark~\ref{rem:Exp_basis}. Specifically, we observe also for a superposed state that the cumulative impairments are multiplicative, which is in agreement with the theoretical analysis developed in Sec.~\ref{sec:5}. Nevertheless, to provide further insights, in the following we consider a superposed quantum state with all the three Bloch vector coordinates being different from zero.
\end{rem}

\begin{figure*}
	\begin{minipage}[c]{0.32\textwidth}
		\centering
		\includegraphics[width=1\columnwidth]{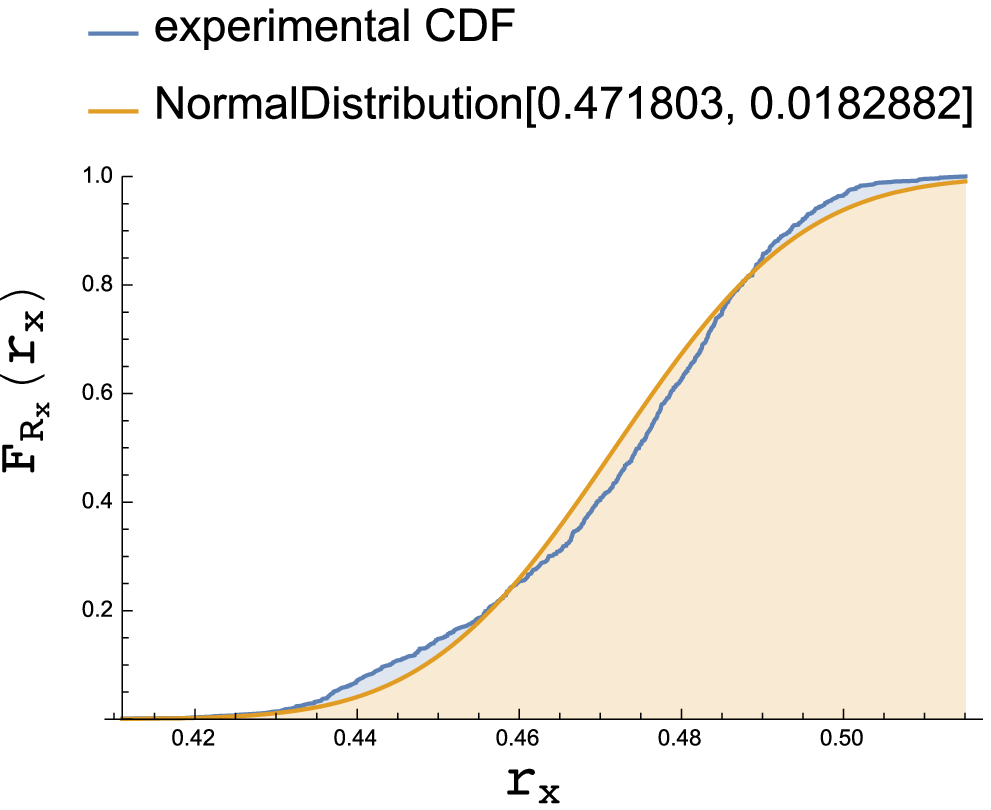}
		\subcaption{Marginal CDF $F_{R_x}(r_x)$ of the Bloch x-coordinate.}
		\label{Fig:D1}
	\end{minipage}
	\hspace{0.005\textwidth}
	\begin{minipage}[c]{0.32\textwidth}
		\centering
		\includegraphics[width=1\columnwidth]{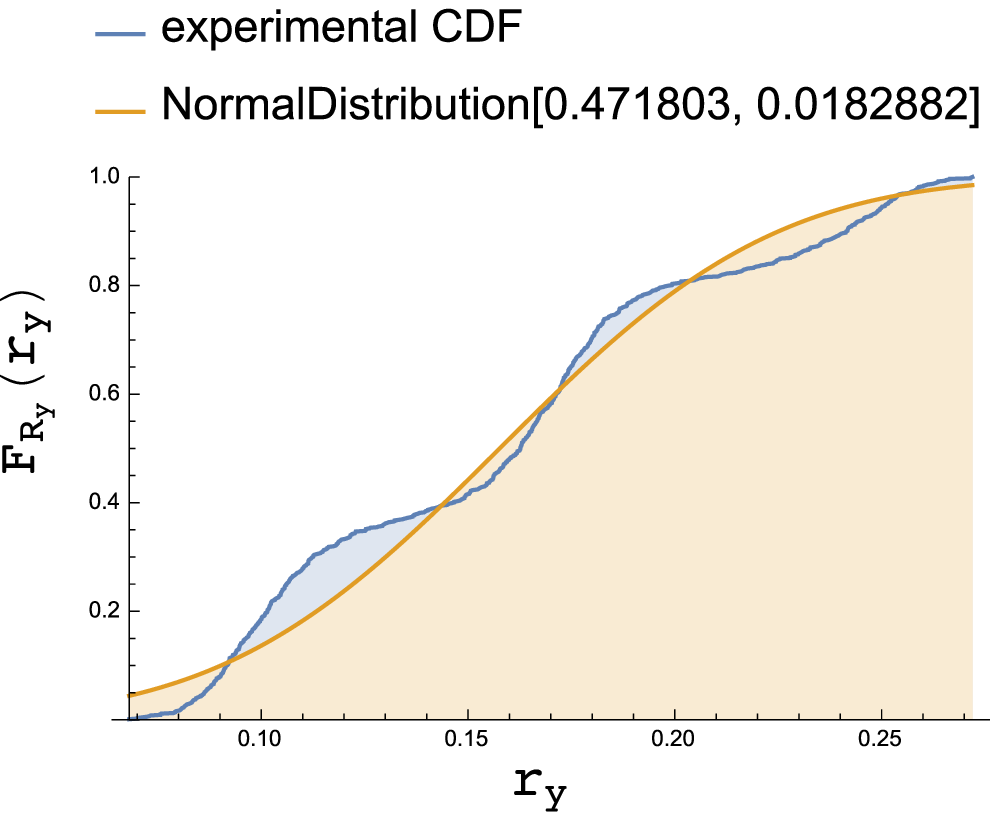}
		\subcaption{Marginal CDF $F_{R_y}(r_y)$ of the Bloch y-coordinate.}
		\label{Fig:D2}
	\end{minipage}
	\hspace{0.005\textwidth}
	\begin{minipage}[c]{0.32\textwidth}
		\centering
		\includegraphics[width=1\columnwidth]{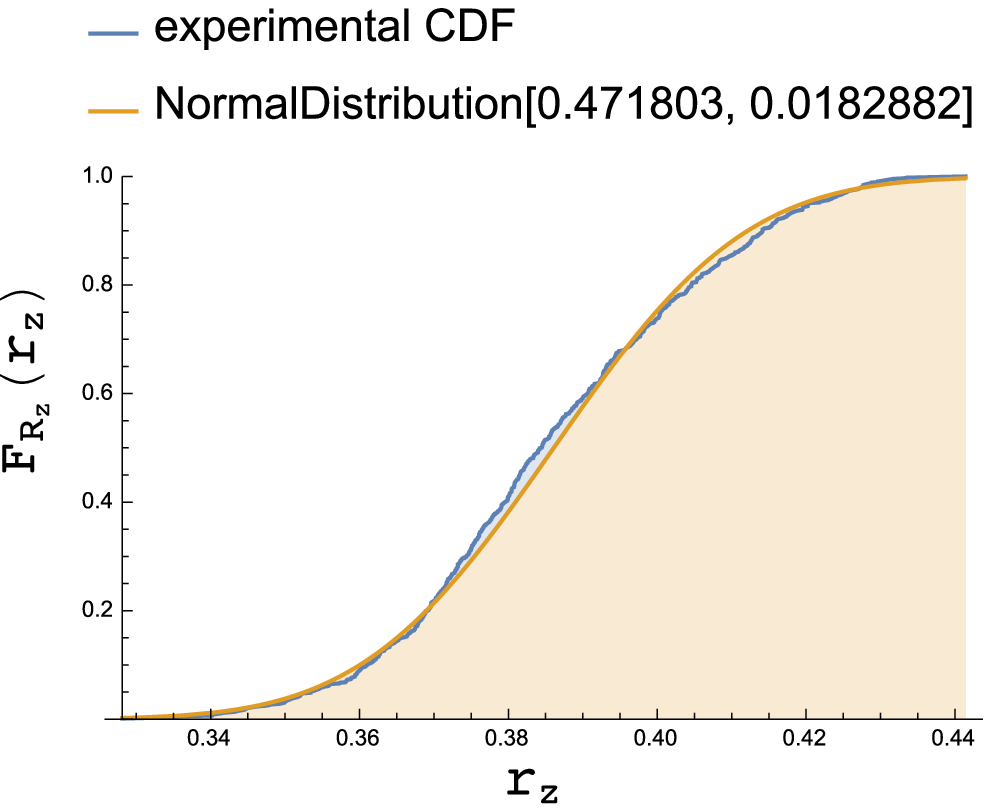}
		\subcaption{Marginal CDF $F_{R_z}(r_z)$ of the Bloch z-coordinate.}
		\label{Fig:D3}
	\end{minipage}\\
	\begin{minipage}[c]{0.49\textwidth}
		\centering
		\includegraphics[width=1\columnwidth]{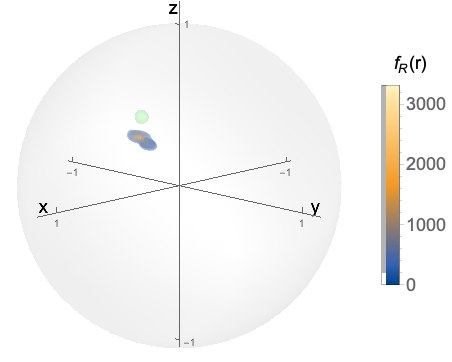}
		\subcaption{Density plot of the experimental joint PDF $f_{\mathbf{R}}(\mathbf{r})$ of the Bloch vector $\mathbf{r}=[r_x,r_y,r_z]$.}
		\label{Fig:D4}
	\end{minipage}
	\hspace{0.01\textwidth}
	\begin{minipage}[c]{0.49\textwidth}
		\centering
		\includegraphics[width=1\columnwidth]{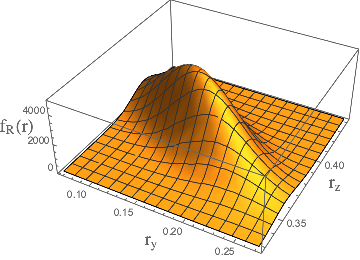}
		\subcaption{Density plot of the experimental joint PDF $f_{\mathbf{R}}(\mathbf{r})$ of the Bloch vector $\mathbf{r}=[r_x,r_y,r_z]$ when $r_x = E[R_x]$.} 
		\label{Fig:D5}
	\end{minipage}
	\caption{Teleportation with 5-qubits IBM Tenerife \textit{ibmqx4} quantum processor between chip-qubits $q[0]$ and $q[2]$. Initial pure state $\ket{\psi} = \sqrt{\frac{1+ \sqrt{3}}{2\sqrt{3}}} \ket{0} + \frac{i + \sqrt{3}}{2 \sqrt{3+\sqrt{3}}} \ket{1}$ with Bloch vector coordinates $\mathbf{r}=\left[\frac{1}{\sqrt{2}},\frac{1}{\sqrt{6}},\frac{1}{\sqrt{3}}\right]$ (green dot in Fig.~\ref{Fig:D4}).}
	\label{Fig:D}
\end{figure*}

In Figure~\ref{Fig:D}, we evaluate the effects of the cumulative quantum impairments, when the qubit to be teleported is the pure state $\ket{\psi}$, placed at the Bloch vector coordinates of $\left[r_x^{in}=1/\sqrt{2},r_y^{in}=1/\sqrt{6},r_z^{in}=1/\sqrt{3}\right]$ (green dot in Fig.~\ref{Fig:D4}), as in Sec.~\ref{sec:5}. As in the previous experiments, due to the composite impairments, the teleported qubit is found in a mixed state.

Finally, in Figs.~\ref{Fig:D1}-\ref{Fig:D3}, we characterize the cumulative effects through the marginal CDFs of the Bloch vector coordinates of the teleported qubit. The results of the previous experiments are confirmed: i) reasonable fitting between the experimental distribution and the Gaussian one; ii) multiplicative nature of the quantum impairments.

However, with this last set of experiment we gain some additional insights concerning the theoretical system model. 

Specifically, we observe from Figs.~\ref{Fig:D1}-\ref{Fig:D3} that the x-coordinate, y-coordinate and z-coordinate of the teleported qubit shrink from the original values of $r_x^{in},r_y^{in},r_z^{in}$ to average values of about $\mu_x =0.47,\mu_y =0.16,\mu_z=0.39$.

By analyzing the average attenuations experienced by the x- and the z-coordinate, we note that they are similar, $\mu_x/r_x^{in} \simeq \mu_z/r_z^{in} \simeq 0.67$, which is also in line with the first two experiments. Explicitly, in Fig.~\ref{Fig:B3} we have $\mu_z/(r_z^{in}=1)\simeq 0.66$ for the z-coordinate and in Fig.~\ref{Fig:C1} $\mu_x/(r_x^{in}=1)\simeq 0.61$ for the x-coordinate.

As regards to the y-coordinate, we observe that it is subjected to a stronger average attenuation, i.e. $\mu_y/r_y^{in}\simeq 0.39$, about twice that of the x- and z-coordinate.

\begin{rem}
\label{rem:compositenoise}
This third experiment suggests that the cumulative impairments behave in agreement with the y-z-phase damping model of Sec.~\ref{sec:5.4} as summarized in Fig.~\ref{Fig:12}, in conjunction with equal decay rates, i.e. $\gamma_x=\gamma_z$, as also confirmed by the first two experiments. However, further research is needed to confirm this hypothesis, along with the specific values of the decay rates and their relationship with the statistical parameters governing the joint distribution of the teleported Bloch vector coordinates. The correlation of the teleported Bloch vector coordinates also has to be further investigated in order to characterize the decay rates.
\end{rem}

\section{Outlook}
\label{sec:7}

Quantum teleportation is the core functionality of the Quantum Internet, which facilitates the ``transfer'' of qubits without either the physical transfer of the particle storing the qubit or the violation of the quantum mechanical principles. Unfortunately, the quantum teleportation process is gravely affected by the quantum impairments, as analyzed in Sec.~\ref{sec:5}. 

To this aim, Secs.~\ref{sec:5} and ~\ref{sec:6} provided a first step toward the modeling of the quantum impairments arising during the quantum teleportation process from a communications engineering perspective. Both the theoretical analysis and the experimental campaign allowed us to gain important insights into the behavior of the composite quantum impairments. Specifically, they revealed that the impairments are multiplicative and they also exhibit an asymmetric behavior, affecting the Bloch vector coordinates of a qubit differently. Moreover, the composite quantum impairments obey the theoretical y-z-phase damping model analyzed in Sec.~\ref{sec:5.4}, although this has to be further investigated.

At this stage a natural question arises: ``how can we generalize the communication system model proposed in Fig.~\ref{Fig:4} for the quantum teleportation in order to account for all quantum impairments?''

To answer to this question, we first summarize some of the considerations developed in the previous sections. Specifically, according to the communication system model given in Fig.~\ref{Fig:4}, the noise acts on:
\begin{itemize}
	\item[i)] the \textit{Entanglement Generator \& Distributor} super-block
	\item[ii)] the \textit{Quantum Pre/Post-Processing} blocks.
\end{itemize}

The impairment imposed at the \textit{Entanglement Generator \& Distributor} affects both the generation and the distribution of the entangled pair at Alice and Bob. This in turn implies that the imperfect generation and/or distribution processes provide eventually Alice and Bob with a pair of ``\textit{imperfect}'' (i.e. mixed-state) entangled qubits -- rather than a pair of maximally entangled qubits.

Furthermore, even if we assume having a perfect \textit{Entanglement Generator \& Distributor} super-block, contamination is still present at the \textit{Quantum Pre/Post-Processing} blocks due to decoherence and/or imperfect quantum operations.

In the following, we discuss the open problems to be circumvented within these communication system blocks. 

\subsection{Mitigating Impairments in \textit{Quantum Pre/Post-Processing}}
\label{sec:7.1}
Intuitively, to enhance the \textit{Quantum Pre/Post-Processing} blocks, we can modify the system model of Fig.~\ref{Fig:4} by introducing a system block at Alice's side that imposes some redundancy into the quantum information sequence. This redundancy can be exploited at Bob's side by a complementary system block to mitigate the impairments. These two system blocks are the well-known \textit{channel-encoder} and \textit{channel-decoder} pair within a classical communication system model \cite{Proakis-01}.

However, a one-to-one mapping between a classical channel-encoder (decoder) and its quantum equivalent cannot be taken for granted. In fact, the no-cloning theorem prevents the adoption of classical error correction techniques relying on the capability of copying the classical information. Hence, specifically designed Quantum Error Correction (QEC) techniques have to be used \cite{Sho-95,ChaBabNgu-18}.

\subsection{Mitigating Impairments at the \textit{Entanglement Generator \& Distributor}}
\label{sec:7.2}
We can also modify the quantum communication system model of Fig.~\ref{Fig:4} by introducing a system block for mitigating the impairments imposed on the entanglement resources.

Indeed, provided that the contamination of the entangled qubits is below a certain threshold, it is possible to \textit{purify} multiple imperfectly entangled pairs into a single ``almost-maximally entangled'' pair, albeit only at the price of additional processing. This strategy is known as \textit{entanglement distillation} or \textit{purification} \cite{Ben-96}, and it has been lavishly documented in the literature \cite{DurBri-07}.

Furthermore, the entanglement distribution rate decays exponentially with the distance between Alice and Bob \cite{TakGuh-14,PirLau-17}. As highlighted in Table~\ref{tab:02}, no classical strategies such as amplify-and-forward or decode-and-forward can be employed. Instead, \textit{quantum repeaters} have to been employed \cite{BrieDurCir-98,DurBriCir-99,VanLadMun-09,VanTou-13,MurLi-16}, which are devices implementing \textit{entanglement swapping} \cite{ZukZeiHor-93}, which allows us to entangle qubits over a long link by generating and distributing entanglement through shorter links. In practice, two EPR pairs are generated and distributed. One pair between Alice and an intermediate node -- namely the quantum repeater -- and another pair between the intermediate node and Bob. By performing a BSM on the two particles at the quantum repeater, entanglement is created between the particles at Alice and Bob \cite{VanMet-14,CacCalTaf-18}. Clearly, the decoherence-contaminated entanglement swapping procedure itself also introduces errors \cite{MunVanLou-08,LiaTayNem-09}, which can be tackled either by entanglement purification or by QEC techniques \cite{MurLi-16}.

However, further research is needed. In fact, the quantum encoder/decoder blocks at the Quantum Pre/Post-Processing blocks and the ones at the \textit{Entanglement Generator \& Distributor} can be separately designed, although a joint -- but much more difficult -- design is also conceivable. However, it is unclear whether the joint design provides superior performance.

\vspace{18pt}

In conclusion, a substantial amount of frontier-research is required for tackling the challenges and open problems associated with the Quantum Internet.

However, the excitement in contributing to this research area is intoxicating, since the Quantum Internet might pave the way for the Internet of the future, such as Arpanet had paved the way for today's Internet. This is an exciting era for quantum communications and signal processing.


\bibliographystyle{IEEEtran}
\bibliography{quantum-06}

\end{document}